\documentclass[12pt,prd, aps, showpacs, superscriptaddress,floatfix]{revtex4-2}

\usepackage{graphicx}
\usepackage{bm}
\usepackage{multirow}
\usepackage{epsfig}
\usepackage{psfrag}
\usepackage{soul}
\usepackage{multirow}
\usepackage{mathptmx}
\usepackage{mathrsfs}
\usepackage{amsmath, amssymb,amsbsy}
\usepackage{dcolumn}
\usepackage{epstopdf}
\usepackage{color}
\usepackage{hhline}
\usepackage{changebar}
\usepackage{placeins}

\usepackage{slashed}

\usepackage{etoolbox} 

\usepackage{flexisym,breqn,gensymb}

\usepackage{subcaption}
\usepackage[colorlinks=true, linkcolor=blue, filecolor=blue, urlcolor=blue, citecolor=blue,plainpages=false]{hyperref}

\newcommand\riken{RIKEN-BNL Research Center, Brookhaven National
  Laboratory, Upton, NY 11973, USA}
\newcommand\bnl{Brookhaven National Laboratory, Upton, NY 11973, USA}
\newcommand\edinb{SUPA, School of Physics, The University of
  Edinburgh, Edinburgh EH9 3JZ, UK}

\newcommand\cu{Physics Department, Columbia University, New York,
  NY 10027, USA}

\newcommand\uconn{Physics Department, University of Connecticut,
  Storrs, CT 06269-3046, USA}
\newcommand\soton{School of Physics and Astronomy, University of
  Southampton,  Southampton SO17 1BJ, UK}

\newcommand\regensburgtwo{Universit{\"a}t Regensburg, Fakult{\"a}t f{\"u}r Physik, 93040, Regensburg, Germany}

\newcommand{\mituni}{Center for Theoretical Physics, Massachusetts
Institute of Technology, Boston, MA 02139, USA}
\newcommand{\cern}{Theoretical Physics Department, CERN, 1211 Geneve 23, Switzerland} 

\newcommand{\ba}{\begin{eqnarray}}
\newcommand{\ea}{\end{eqnarray}}
\newcommand{\bas}{\begin{eqnarray*}}
\newcommand{\eas}{\end{eqnarray*}}
\newcommand{\be}{\begin{equation}}
\newcommand{\ee}{\end{equation}}
\newcommand{\bes}{\begin{equation*}}
\newcommand{\ees}{\end{equation*}}
\newcommand{\bi}{\begin{itemize}}
\newcommand{\ei}{\end{itemize}}

\newcommand{\bcentre}{\begin{center}}
\newcommand{\ecentre}{\end{center}}

\font\tenmsb=msbm10 scaled\magstep1
\font\sevenmsb=msbm7 scaled\magstep1
\font\fivemsb=msbm5 scaled\magstep1
\newfam\msbfam
\textfont\msbfam=\tenmsb
\scriptfont\msbfam=\sevenmsb
\scriptscriptfont\msbfam=\fivemsb

\def\msbar{\overline{\mathrm{MS}}}

\def\su3{SU(3)}

\def\tD{\mbox{D}\kern-0.65em\raise0.15ex\hbox{/}\kern0.15em} 
\def\sD{\mbox{\scriptsize D}\kern-0.5em\raise0.15ex\hbox{\scriptsize/}}
\def\ssD{\mbox{\tiny D}\kern-0.42em\raise0.15ex\hbox{\tiny/}}

\def\dslash{\hbox{\(\partial\)}\kern-0.5em\raise0.15ex\hbox{/}} 

\def\su3{SU(3)}

\def\nicefrac#1#2{\leavevmode\kern.1em\raise.5ex\hbox{\the\scriptfont0 #1}\kern-
.1em/\kern-.15em\lower.25ex\hbox{\the\scriptfont0 #2}}

\newcommand{\bea}{\begin{eqnarray}}
\newcommand{\eea}{\end{eqnarray}}

\newcommand{\qslash}{\slashed{q}}

\newcommand{\colvectwo}[2]{\left(\begin{array}{c}#1\\#2\end{array}\right)}

\newcommand{\prop}{\mathcal{G}}

\newcommand{\psibar}{\overline{\psi\vphantom{d}}}

\newcommand{\snk}{{\rm snk}}

\newcommand{\sep}{{\rm sep}}

\newcommand{\SMOMqq}{{\rm SMOM}(\slashed{q},\slashed{q})}
\newcommand{\SMOMgg}{{\rm SMOM}(\gamma^\mu,\gamma^\mu)}
\newcommand{\re}{{\rm Re}}
\newcommand{\im}{{\rm Im}}

\date{\today}

\begin{document}
\bibliographystyle{apsrev}

\title{Direct CP violation and the $\pmb{\Delta I=1/2}$ rule in $\pmb{ K\to\pi\pi}$ decay from the Standard Model}

\author{R.~Abbott}\affiliation{\cu}
\author{T.~Blum}\affiliation{\uconn}\affiliation{\riken}
\author{P.A.~Boyle}\affiliation{\bnl}\affiliation{\edinb}
\author{M.~Bruno}\affiliation{\cern}
\author{N.H.~Christ}\affiliation{\cu}
\author{D.~Hoying}\affiliation{\riken}\affiliation{\uconn}
\author{C.~Jung}\affiliation{\bnl}
\author{C.~Kelly}\affiliation{\bnl}
\author{C.~Lehner}\affiliation{\regensburgtwo}\affiliation{\bnl}
\author{R.D.~Mawhinney}\affiliation{\cu}
\author{D.J.~Murphy}\affiliation{\mituni}
\author{C.T.~Sachrajda}\affiliation{\soton}
\author{A.~Soni}\affiliation{\bnl}
\author{M.~Tomii}\affiliation{\uconn}
\author{T.~Wang}\affiliation{\cu}

\collaboration{RBC and UKQCD Collaborations}
%
%
\mbox{}\hfill\noaffiliation{CERN-TH-2020-058, MIT-CTP/5197}

\maketitle

\centerline{ABSTRACT}
We present a lattice QCD calculation of the $\Delta I=1/2$, $K\to\pi\pi$ decay amplitude $A_0$ and $\varepsilon'$, the measure of direct CP-violation in $K\to\pi\pi$ decay, improving our 2015 calculation~\cite{Bai:2015nea} of these quantities.  Both calculations were performed with physical kinematics on a $32^3\times 64$ lattice with an inverse lattice spacing of $a^{-1}=1.3784(68)$ GeV.  However, the current calculation includes nearly four times the statistics and numerous technical improvements allowing us to more reliably isolate the $\pi\pi$ ground-state and more accurately relate the lattice operators to those defined in the Standard Model.  We find ${\rm Re}(A_0)=2.99(0.32)(0.59)\times 10^{-7}$ GeV and ${\rm Im}(A_0)=-6.98(0.62)(1.44)\times 10^{-11}$ GeV, where the errors are statistical and systematic, respectively. The former agrees well with the experimental result ${\rm Re}(A_0)=3.3201(18)\times 10^{-7}$ GeV.   These results for $A_0$ can be combined with our earlier lattice calculation of $A_2$ ~\cite{Blum:2015ywa} to obtain ${\rm Re}(\varepsilon'/\varepsilon)=21.7(2.6)(6.2)(5.0) \times 10^{-4}$, where the third error represents omitted isospin breaking effects, and Re$(A_0)$/Re$(A_2) = 19.9(2.3)(4.4)$.   The first agrees well with the experimental result of ${\rm Re}(\varepsilon'/\varepsilon)=16.6(2.3)\times 10^{-4}$.  A comparison of the second with the observed ratio Re$(A_0)/$Re$(A_2) = 22.45(6)$, demonstrates the Standard Model origin of this ``$\Delta I = 1/2$ rule'' enhancement.


\newpage
\section{Introduction}
\label{sec:Introduction}
A key ingredient to explaining the dominance of matter over antimatter in the observable universe is the breaking of the combination of charge-conjugation and parity (CP) symmetries. The amount of CP violation (CPV) in the Standard Model is widely believed to be too small to explain the dominance of matter over antimatter, suggesting the existence of new physics not present in the Standard Model. CPV in the Standard Model is highly constrained, requiring the presence of all three quark-flavor doublets and described by a single phase~\cite{Kobayashi:1973fv}. These properties imply that the ``direct'' CPV in $K\to\pi\pi$ decays is a highly suppressed ${\cal O}(10^{-6})$ effect in the Standard Model, making it a quantity which is especially sensitive to the effects of new physics in general, and new sources of CPV in particular.

Direct CPV was first observed in $K\to\pi\pi$ decays by the NA48 (CERN) and KTeV (FermiLab) experiments~\cite{Batley:2002gn,PhysRevD.83.092001} in the late 1990s, and the most recent world average of its measure is ${\rm Re}(\epsilon'/\epsilon)=16.6(2.3)\times 10^{-4}$~\cite{PhysRevD.98.030001}, where $\epsilon$ is the measure of indirect CPV (\,$|\epsilon|=2.228(11)\times 10^{-3}$\,). However, despite the impressive success of these experiments, it was only recently that a reliable, first-principles Standard Model determination of $\epsilon'$ that could be compared to the experimental value became available. This is due to the presence of low-energy QCD effects that are difficult to model reliably.

Lattice QCD is the only known technique for determining the properties of low-energy QCD from first principles with systematically improvable errors. In this regime the high-energy physics is precisely captured by the $\Delta S=1$ weak effective Hamiltonian, 
\begin{equation}
H_W = \frac{G_F}{\sqrt{2}}V_{us}^*V_{ud}\sum_{i=1}^{10} \bigl[z_i(\mu) + \tau y_i(\mu)\bigr] Q_i(\mu)\,,
\label{eq-H_W}
\end{equation}
where the Fermi constant $G_F = 1.166 \times 10^{-5}$ GeV$^{-2}$, $V_{q'q}$ is the Cabibbo-Kobayashi-Maskawa matrix element connecting the quarks $q'$ and $q$ and $\tau =  -V_{ts}^* V_{td}/V_{us}^* V_{ud}$. The quantities $z_i$ and $y_i$ are the Wilson coefficients that encapsulate the high-energy behavior, and which have been computed to next-to-leading-order (NLO) in QCD perturbation theory and to leading order (with some important NLO terms) in electroweak perturbation theory~\cite{Buchalla:1995vs}, in the $\msbar$ scheme as a function of the scale $\mu$. The task of the lattice calculation is to determine the matrix elements $\langle \pi\pi|Q_i(\mu)|K^0\rangle$ of the weak effective operators $Q_i$ renormalized in a scheme which can be defined non-perturbatively. A further perturbative calculation is subsequently necessary to match such matrix elements to those in the $\msbar$ scheme. Conventionally, as shown in Eq.~\eqref{eq-H_W}, the weak Hamiltonian is expressed in terms of 10 operators $\{Q_i\}_{1\le i \le 10}$ (as defined, for example, in Eqs. (4.1) -- (4.5) of Ref.~\cite{Buchalla:1995vs} ) that are linearly dependent due to the Fierz symmetry.  More convenient for our purposes is a second, 7-operator ``chiral'' basis~\cite{Blum:2001xb} $\{Q'_j\}_{j=1,2,3,5,6,7,8}$ in which the operators are linearly independent and transform as irreducible representations of $SU(3)_L\times SU(3)_R$. The relationship between these bases is discussed in more detail in Sec.~\ref{sec-renorm_phys_matelems}.

For an isospin-symmetric lattice calculation it is convenient to formulate the $K\to\pi\pi$ matrix elements in terms of two amplitudes, $A_0$ and $A_2$, where $A_I = \langle (\pi\pi)_{I}|H_W|K^0\rangle$ and the subscript indicates the isospin representation of the two-pion state. These correspond to $\Delta I=1/2$ and $\Delta I=3/2$ decays, respectively. From these amplitudes, $\epsilon'$ can be obtained directly as
\begin{equation}
\frac{\varepsilon'}{\varepsilon}\ = \frac{i\omega e^{i(\delta_2-\delta_0)}}{\sqrt{2}\varepsilon}\left[ \frac{\im(A_2)}{\re(A_2)}-\frac{\im(A_0)}{\re(A_0)} \right] \label{eq-epsilonprimefromA}
\end{equation}
where $\delta_I$ are the $\pi\pi$ scattering phase shifts and $\omega = {\rm Re}(A_2)/{\rm Re}(A_0)$. Note that the effects of isospin breaking and electromagnetism are not included in our simulation and are instead treated as systematic errors as discussed in Sec.~\ref{sec:epsilonp}.

In 2015 the RBC \& UKQCD collaborations published~\cite{Bai:2015nea} the first lattice calculation of $A_0$ using 216 lattice configurations with a $32^3\times 64$ volume, an inverse lattice spacing of $a^{-1}=1.3784(68)$ GeV, and with physical kinematics. We found Re$(A_0) = 4.66(1.00)(1.26) \times 10^{-7}$ GeV and Im$(A_0) = -1.90(1.23)(1.08) \times 10^{-11}$ GeV, where the parentheses contain the statistical and systematic errors, respectively. Within the uncertainties, the former agrees with the experimental result of Re$(A_0) = 3.3201(18) \times 10^{-7}$ GeV, and the latter, combined with the experimental value of Re$(A_0)$ and the result of our previous calculation of $A_2$~\cite{Blum:2015ywa}, gives Re$(\varepsilon'/\varepsilon)=1.38(5.15)(4.59)\times10^{-4}$, which is $2.1\sigma$ below the experimental value.

In order to obtain on-shell kinematics, i.e. to ensure that $E_{\pi\pi}$, the energy of the two-pion final state, satisfies $E_{\pi\pi} =m_K$, we exploit the possibility of choosing appropriate spatial boundary conditions. With periodic boundary conditions for all the quarks, the ground state of the two-pion final state corresponds to $E_{\pi\pi}=2m_\pi$, with each of the pions at rest, and the state with $E_{\pi\pi}=m_K$ appears as an excited state. We would therefore need to resort to multi-state fits to rather noisy data in order to obtain the physical amplitudes. The change in the finite-volume corrections induced by modifying the boundary conditions is exponentially small~\cite{Sachrajda:2004mi,Christ:2019sah} or else accounted for by the L\"uscher and Lellouch-L\"uscher~\cite{Luscher:1990ux,Lellouch:2000pv} prescriptions with minor alterations~\cite{Yamazaki:2004qb}.

In our calculation of $A_2$~\cite{Blum:2015ywa,Blum:2012uk,Blum:2011ng} we employ antiperiodic spatial boundary conditions (APBC) for the down quark in some or all directions, which results in the charged pions also obeying corresponding antiperiodic boundary conditions. The momenta of the charged pions are therefore discretized in odd-integer multiples of $\pi/L$ in these directions, where the spatial volume of the lattice is $V=L^3$. Since only the spectrum of the charged pions is changed by the APBC, we compute $K^+\to\pi^+\pi^+$ matrix elements of operators which change $I_z$, the third component of isospin, by 3/2 and then use the Wigner-Eckart theorem to obtain the physical $K^+\to\pi^+\pi^0$ amplitude which is proportional to $A_2$. Note that in order to ensure that $E_{\pi\pi}=m_K$, $L$ must be appropriately tuned.

The technique of using APBC on the down quark naturally breaks the isospin symmetry. For the $\Delta I=3/2$ calculation this symmetry breaking does not pose an issue because the final state of the measured $K^+\to\pi^+\pi^+$ matrix element is the only doubly-charged two-pion state and therefore cannot mix with other $\pi\pi$ states due to charge conservation. However, the final state in the $\Delta I=1/2$ matrix elements has isospin 0 and is a linear combination of $|\pi^+\pi^-\rangle$ and $|\pi^0\pi^0\rangle$ states. Thus the breaking of isospin symmetry at the boundaries results in different energies for the $|\pi^+\pi^-\rangle$ and $|\pi^0\pi^0\rangle$ states and the APBC technique cannot be used. As a result, for the calculation of the $\Delta I=1/2$ amplitude we instead utilize G-parity boundary conditions (GPBC). G-parity is the combination of charge conjugation and a $180^\circ$ isospin rotation about the y-axis, $\hat G = \hat C e^{\pi\hat I_y}$. The charged and neutral pions are both odd eigenstates of this operation, hence when applied as a boundary condition all pion states again become antiperiodic in the spatial boundary. While more general than the APBC approach, the use of GPBC introduces a number of technical and computational difficulties that we discuss in Ref.~\cite{Christ:2019sah} and below.

Note that due to the $\pi\pi$ interaction being repulsive in the $I=2$ channel but attractive in the $I=0$ channel, the finite-volume $\pi\pi$ energies in these two representations differ at fixed lattice size and it is therefore not possible to use the same ensemble to measure both the $\Delta I=3/2$ and $\Delta I=1/2$ decay amplitudes with on-shell kinematics.  In this document we present a detailed update of the calculation of $A_0$ and will combine it with the results for $A_2$ given in  Ref.~\cite{Blum:2015ywa}.

Among the necessary ingredients in the lattice calculation of the $K\to\pi\pi$ matrix elements are the two-pion energies $E_{\pi\pi}$ and the amplitudes $\langle \pi\pi|{\cal O}_{\pi\pi}|0\rangle$, where ${\cal O}_{\pi\pi}$ is an interpolating operator that can create the required two-pion state from the vacuum. These quantities are determined by correlation functions describing the propagation of the two-pion state. The matrix elements $\langle \pi\pi\,|Q_i|K^0\rangle$ are obtained from $K\to\pi\pi$ correlation functions in which the Euclidean time dependence is exponential in $E_{\pi\pi}$, and the amplitudes corresponding to the annihilation of the two-pion state (and the creation of the kaon state) have to be removed. From the measurement of $E_{\pi\pi}$ and using the L\"uscher formula~\cite{Luscher:1990ux}, we also determine the s-wave isospin 0 $\pi\pi$-phase shift, $\delta_0(m_K)$, which enters the expression relating the matrix elements to $(\varepsilon'/\varepsilon)$, Eq.~\eqref{eq-epsilonprimefromA}. The derivative of the phase-shift with respect to the energy is additionally required to determine the power-like (i.e. non-exponential) finite-volume corrections through the Lellouch-L\"uscher formula~\cite{Lellouch:2000pv} (cf. Sec.~\ref{sec-llfactor}). In the 2015 calculation we obtained $\delta_0(E_{\pi\pi}^{\rm lat}\approx m_K)=23.8(4.9)(1.2)^\circ$, substantially smaller than the dispersive result~\cite{Colangelo:2001df}.



The observation of a discrepancy from the predicted phase shift increased our motivation to extend the earlier calculation by increasing the statistics and using more sophisticated methods to better analyze the $I=0$ two-pion system. In Ref.~\cite{Bai:2015nea} we observed excellent stability in the determination of the ground-state two-pion energy $E_{\pi\pi}$; the result was consistent between one- and two-state fits to our data (i.e. whether we assumed that just the ground-state was propagating or allowed for a contribution from an excited state) and was also independent, within the uncertainties, of the time separation between the insertion of the creation and annihilation operators (the ${\cal O}_{\pi\pi}$ introduced above). Nevertheless, we considered the best explanation for the discrepancy to be contamination from one or more excited states whose contribution with increasing time is masked by the rather rapid reduction in the signal-to-noise of our data. Therefore, in addition to increasing our statistics by more than a factor of 3, we have introduced two additional $\pi\pi$ interpolating operators. For our original calculation we used a $\pi\pi$ operator comprising two quark bilinear operators that create back-to-back moving pions of a particular momentum. Alongside this operator, which we label $\pi\pi(111)$, we have now added a scalar operator $\sigma=\frac{1}{\sqrt{2}}(\bar u u + \bar d d)$, and an operator creating pions with larger relative momenta that we label $\pi\pi(311)$. Here the number appearing in the parentheses of the $\pi\pi(\ldots)$ operators is related to the components of the pion momentum in lattice units: $(xyz)\to (\pm x,\pm y,\pm z)\pi/L$ (the total $\pi\pi$ momentum is zero in all cases). Here and for the remainder of this document we will assume the lattice size $L$ to be in lattice units unless otherwise stated. All three operators, once suitably projected onto a state that is symmetric under cubic rotations, have the same quantum numbers as the $s$-wave $I=0$ two-pion state of interest and as such project onto the same set of QCD eigenstates, albeit with different coefficients. 

In Ref.~\cite{pipi-paper} we demonstrate that a simultaneous fit to the $3\times 3$ matrix of $\pi\pi$ two-point correlation functions in which the two-pion states are created or annihilated by one of these three operators, results in a substantial reduction in the statistical and systematic errors. We find that, once the excited states are taken into account, the resulting $I = 0$ $\pi\pi$-scattering phase shift at $E_{\pi\pi}^{\rm lat}=479.5$ MeV is $\delta_0(E_{\pi\pi}^{\rm lat})=32.3(1.0)(1.8)^\circ$, where the errors are statistical and systematic, respectively.   This significant increase in our result for $\delta_0(E_{\pi\pi}^{\rm lat})$ brings us into much closer agreement with the dispersive prediction, which at our present value of $ E_{\pi\pi}^{\rm lat}$ is $\delta_0(E_{\pi\pi}^{\rm lat})_{\rm disp}=35.9^\circ$, obtained using Eqs. 17.1-17.3 of  Ref.~\cite{Colangelo:2001df} with $m_\pi=139.6$ MeV. (We refer the reader to Ref.~\cite{Colangelo:2001df} for estimates of the error on the dispersive prediction.) In this paper we present results for the $\Delta I=1/2$ $K\to\pi\pi$ matrix elements obtained from our expanded data set of 741 measurements, using all three $\pi\pi$ interpolating operators.


In this analysis we also include an improved non-perturbative determination of the renormalization factors relating the bare matrix elements in the lattice discretization to those of operators renormalized in the RI-SMOM scheme (see Sec.~\ref{sec:NPR}). Perturbation theory is then required to match the operators renormalized in the RI-SMOM scheme to those in the $\msbar$ scheme in which the Wilson coefficients have been computed. This calculation utilizes step-scaling to raise the matching scale from 1.53 GeV to 4.01 GeV, significantly reducing the systematic error associated with the perturbative matching. 

Throughout this document results are presented in lattice units unless otherwise stated.

While the current paper is intended to be self-contained it should be viewed as the third in a series of three closely related papers.  The first of these is Ref~\cite{Christ:2019sah} which gives a detailed discussion of the implementation and properties of lattice calculations which impose G-parity boundary conditions.  The second paper is Ref.~\cite{pipi-paper} in which the same ensemble of gauge configurations and many of Green's functions used in the current paper are analyzed to study $\pi\pi$ scattering.  This second paper contains the two-pion, finite-volume energy eigenvalues from which the  $I=0$ and $I=2$ $\pi\pi$ scattering phase shifts are derived as well as the matrix elements of the $\pi\pi$ interpolating operators between the corresponding energy eigenstates and the vacuum which are used in the current paper.

For the convenience of the reader we summarize the primary results of this work in Tab.~\ref{tab-summary}. For further discussion we refer the reader to Sec.~\ref{sec:finalresults}. It is important to stress that the results and uncertainties in Tab.~\ref{tab-summary} have been obtained by combining a number of elements. The major direct contribution from this work is the evaluation of the matrix elements  $\langle (\pi\pi)_{I=0}|Q_i|K^0\rangle$ in isosymmetric QCD, with the operators renormalized in the RI-$\SMOMqq$ scheme (see Tab.~\ref{tab-renorm-matrix-elems-qq-full}), with the lattice systematic uncertainties carefully estimated (see Sec.~\ref{sec:syserrs}). These matrix elements are combined with the perturbatively-calculated Wilson coefficients in the $\msbar$ scheme and the perturbative matching of the matrix elements from the RI-SMOM to $\msbar$ schemes with estimates of the corresponding systematic uncertainties. If and when these perturbative uncertainties, as well as those in the CKM matrix elements and isospin breaking, are reduced then the matrix elements in Tab.~\ref{tab-renorm-matrix-elems-qq-full} can be used to improve the precision in the determination of $A_0$.

\begin{table}[tb]
\begin{tabular}{c|c}
\hline\hline
Quantity & Value \\
\hline
${\rm Re}(A_0)$ & 2.99(0.32)(0.59)$\times 10^{-7}\ {\rm GeV}$ \\
${\rm Im}(A_0)$ & -6.98(0.62)(1.44)$\times 10^{-11}\ {\rm GeV} $\\
${\rm Re}(A_0)/{\rm Re}(A_2)$ & 19.9(2.3)(4.4) \\
${\rm Re}(\epsilon'/\epsilon)$ & 0.00217(26)(62)(50) 
\end{tabular}
\caption{A summary of the primary results of this work. The values in parentheses give the statistical and systematic errors, respectively. For the last entry the systematic error associated with electromagnetism and isospin breaking is listed separately as a third error contribution.\label{tab-summary} }
\end{table}

The layout of the remainder of this paper is as follows: In Sec.~\ref{sec:MeasTechniques} we introduce our lattice ensemble and give a general overview of our measurement techniques. In Sec.~\ref{sec:TwoPointResults} we discuss and present results from fits to the single-pion, two-pion and kaon two-point correlation functions, the values of which are required as inputs to the fits of the $K\to\pi\pi$ three-point correlation functions from which the matrix elements of the bare lattice operators are determined. In Sec.~\ref{sec:ThreePointResults} we discuss the measurement of these three-point functions and provide the results from the fits. In Sec.~\ref{sec:NPR} we discuss our procedure for the non-perturbative renormalization of the operators $Q_i$, the results of which are combined with the matrix elements of the bare lattice operators and other inputs to determine $A_0$ and $\epsilon^\prime/\epsilon$ in Sec.~\ref{sec:resultsA0epsprime}. We follow this by a detailed discussion of the systematic errors in Sec.~\ref{sec:syserrs} and present our final results for the matrix elements, decay amplitudes, and $\epsilon^\prime/\epsilon$, together with a discussion of the $\Delta I=1/2$ rule, in Sec.~\ref{sec:finalresults}. Finally we present our conclusions in Sec.~\ref{sec:conclusions}. There are two technical appendices in which we present the Wick contractions of some of the correlation functions used in this project.


\section{Overview of measurements}
\label{sec:MeasTechniques}
In this section we provide an overview of the calculation, including information on the ensemble and the measurement techniques.

\subsection{Gauge ensemble}

For this calculation we employ a single lattice of size $32^3\times 64$. We utilize $2+1$ flavors of M\"obius domain wall fermions with $L_s=12$ and M\"obius parameters $b+c=32/12$ and $b-c=1$ and light and strange quark masses of $1\times 10^{-4}$ and 0.045, respectively. We use the Iwasaki+DSDR gauge action with $\beta=1.75$, corresponding to an inverse lattice spacing of $a^{-1}=1.3784(68)$ GeV. The {\it dislocation suppressing determinant ratio} (DSDR) term~\cite{Renfrew:2009wu} is a modification of the gauge action that suppresses the dislocations, or tears in the gauge field that enhance chiral symmetry breaking at coarse lattice spacings. This enables the calculation to be performed with larger lattice spacings, and hence larger physical volumes, at fixed computational cost, ensuring good control over finite-volume systematic errors. We use G-parity boundary conditions (GPBC) in three spatial directions in order to obtain nearly physical kinematics for our $K\to\pi\pi$ decays.

The lattice parameters are equal to those of the 32ID ensemble documented in Refs.~\cite{Arthur:2012yc, Blum:2014tka} except for the boundary conditions and that we now simulate with a lighter, physical pion mass of 142 MeV versus the 170 MeV pion mass of the 32ID ensemble. This allows the use of existing measurements such as the lattice spacing, and also enables the computation of the non-perturbative renormalization factors in a regime free of the complexities associated with GPBC. 

The ensemble used for our 2015 calculation comprised 864 molecular dynamics (MD) time units (after thermalization), upon which 216 measurements were performed separated by 4 MD time units. Subsequent to the calculation, it was discovered~\cite{Bai:2016ocm} that an error existed in the generation of the random numbers used to set the conjugate momentum at the start of each trajectory, which gave rise to small correlations between widely separated lattice sites. While the resulting effects were determined to be two-to-three orders of magnitude smaller than our statistical errors, we nevertheless do not include these configurations in the present calculation. 

In the period following our previous publication, we have dramatically increased the number of measurements. Configurations were generated on seven independent Markov chains originating from widely seperated configurations of our original ensemble. Subsequent algorithmic improvements, particularly the introduction of the exact one-flavor algorithm (EOFA)~\cite{Chen:2014hyy,Chen:2014bbc,Jung:2017xef} further enhanced our rate of generation such that we have completed over 5000 additional MD time units to date.

Continuing with a measurement separation of 4 MD time units, we can potentially perform almost 1300 measurements in total. For this analysis, we include measurements on ${\sim}60\%$ of the available configurations, totaling 741. We aim to provide updated results containing measurements on the remaining portion in a future publication. For further information on the ensemble properties, generation algorithms and details of the configurations used for this analysis we refer the reader to Ref.~\cite{pipi-paper}.

\subsection{Goodness-of-fit and error estimation}
\label{sec-fitstrategy}

Aside from the central values of our fit parameters we must also estimate the standard error and the goodness-of-fit. These are obtained via bootstrap resampling, specifically the {\it non-overlapping block bootstrap} variant~\cite{carlstein1986} which allows us to account for mild autocorrelation effects observed in our data. A block size of 8 is used.

The bootstrap measurement of the goodness-of-fit is a technique developed specifically for this and our companion work~\cite{pipi-paper}, and is detailed in Ref.~\cite{Kelly:2019wfj}. To summarize, the goodness-of-fit is typically parameterized by a p-value that represents the likelihood that the data agrees with the model, allowing only for statistical fluctuations. The p-value is computed by first measuring
\begin{equation}
q^2 = \sum_{i,j} \left(\bar x_i - f(\vec p, i)\right) ({\rm cov})^{-1}_{ij} \left(\bar x_j - f(\vec p, j)\right)\,.\label{eq-q2def}
\end{equation}
where $\bar x_i$ are the ensemble-means of the data at coordinate $i$, $\vec p$ the fitting parameters, $f$ the model function, and $(\rm cov)_{ab}$ is the covariance matrix. The value obtained for $q^2$ is then compared to the null distribution that describes how this quantity varies between independent experiments if only statistical fluctuations are allowed around the model. The null distribution is typically assumed to be the $\chi^2$ distribution, but this is inappropriate when the fluctuations in the covariance matrix between experiments become significant, as is the case for our $\pi\pi$ measurements~\cite{Kelly:2019wfj}. In that work we demonstrate that the null distribution can be estimated directly from the data through a simple bootstrap procedure, allowing for a more reliable p-value that is free from assumptions. This procedure also has the benefit of allowing us to neglect the autocorrelations in the determination of the covariance matrix on each bootstrap ensemble, which dramatically improves the statistical error but changes the definition of $q^2$ in a subtle way that cannot be accounted for by traditional methods.

\subsection{Measurement technique}

Measurements are performed using the all-to-all (A2A) propagator technique of Ref.~\cite{Foley:2005ac}, whereby the quark propagator is decomposed into an exact low-mode contribution obtained from a set of, in our case 900, predetermined eigenvectors, and a stochastic approximation to the high-mode contribution. This allows for the maximal translation of correlation functions in order to take full advantage of each configuration, as well as easy implemention of arbitrarily smeared source and sink operators. We perform full spin, color, flavor and time dilution such that the stochastic source is required only to produce a delta-function in the spatial location.

For all quantities we use smeared meson sources with an exponential ($1s$ hydrogen wavefunction-like) structure,
\begin{equation}
\Theta(|\vec x - \vec y|) = \exp(-|\vec x - \vec y|/\alpha)\label{eq-smearing}
\end{equation}
where $\alpha=2$ is the smearing radius and $\vec x$ and $\vec y$ are the spatial coordinates of the two quark operators. Several technicalities must be considered when using G-parity boundary conditions, including limitations on the allowed quark momenta which has implications for the cubic rotational symmetry, the preservation of which is essential for producing an operator that projects onto the rotationally-symmetric (s-wave) $\pi\pi$ state. These are detailed in Ref.~\cite{pipi-paper}.

More specific details of the various measurements are provided in the following sections.


\section{Results from two-point correlation functions}
\label{sec:TwoPointResults}
In order to compute the $K\to\pi\pi$ matrix elements it is necessary to measure the energies and amplitudes of the pion, kaon and $\pi\pi$ two-point Green's functions. In this section we present results for the kaon two-point function and summarize the results of Ref.~\cite{pipi-paper} for the pion and $\pi\pi$ two-point functions. We also detail the determination of the energy dependence of the phase shift at the kaon mass scale, which is used to obtain the Lellouch-L\"uscher~\cite{Lellouch:2000pv} finite volume correction to the matrix elements.

\subsection{Notation}
\label{sec-GPnotation}

G-parity boundary conditions mix quark flavor at the boundary, introducing additional Wick contractions in which a quark propagates through the boundary and is annihilated by an operator of the opposite quark flavor. In Ref.~\cite{Christ:2019sah} we introduced a notation whereby the quark field and its G-parity partner are placed in a two-component vector,
\begin{dmath}
\psi_l = \colvectwo{d}{C\bar u^T}\,,
\end{dmath}
where $C$ is the charge conjugation matrix. We will refer to the index of these vectors as a ``flavor index''. In this notation the propagator becomes a $2\times 2$ ``flavor matrix'', and Pauli matrices inserted appropriately describe the flavor structure. In this notation the Wick contractions assume an almost identical form to those of the periodic case.

The strange quark is introduced into the G-parity framework as a member of an isospin doublet that includes a fictional degenerate partner, $s'$, into which the strange quark transforms at the boundary. The corresponding field operator is
\begin{dmath}
\psi_h = \colvectwo{s}{C\bar s^{\prime\,T}}\,.
\end{dmath}
With the introduction of this extra quark flavor a square-root of the $s/s'$ determinant is required in order to generate a 2+1 flavor ensemble~\cite{Christ:2019sah}. 

\begin{table}[t]
\begin{tabular}{c|cccc}
\hline\hline
State & Fit Range & $A$ & $E$ & p-value \\
\hline
Kaon & 10-29 & $4.5964(48)\times 10^{6}$ & 0.35587(10) & 0.88 \\
Pion & 14-29 & $6.194(11)\times 10^{6}$ & 0.19893(13) & 0.99
\end{tabular}
\caption{Fit results in lattice units, fit ranges and p-values for the pion and kaon states. Here $E$ is the energy of the state in question, which for the kaon is equal to the kaon mass, $m_K$. \label{tab-pik2ptfit} }
\end{table}

\subsection{Kaon two-point function}
\label{sec-kaon2pt}

Following Ref.~\cite{Christ:2019sah}, a stationary (G-parity even) kaon-like state can be constructed as 
\begin{dmath}
|\tilde K^0\rangle = \frac{1}{\sqrt{2}}\left( |K^0\rangle + |K^{\prime\,0}\rangle\right)\,,
\end{dmath}
where $K^0$ is the physical kaon and $K^{\prime\,0}$ a degenerate partner with quark content $\bar s' u$. This $|\tilde K^0\rangle$ state can be created using the following operator
\begin{dmath}
{\cal O}_{\tilde K^0}(t) = \frac{i}{\sqrt{2}} \sum_{\vec x, \vec y}e^{i\vec p\cdot (\vec x-\vec y)} \psibar_l(\vec x,t)\gamma^5 \Theta(|\vec x-\vec y|) \frac{1}{2}(1+\sigma_2)\psi_h(\vec y, t)
\end{dmath}
where $\vec p=(1,1,1)\frac{\pi}{2L}$ is the quark momentum and $\Theta$ is defined in Eq.~\eqref{eq-smearing}. Note that in the above equation and for the other operators presented in this document, the projection operators $\frac{1}{2}(1\pm \sigma_2)$ appear; these are necessary to define quark field operators that are eigenstates of translation and hence have definite momentum~\cite{Christ:2019sah}.


The two-point function
\begin{dmath}
C_K(t_1, t_2) = \langle 0|{\cal O}^\dagger_{\tilde K^0}(t_1) {\cal O}_{\tilde K^0}(t_2)|0\rangle
\end{dmath}
is measured for all $t_1$ and $t_2$, and subsequently averaged over $t_2$ at fixed $t= t_1-t_2$. The data are folded in $t$, {\it i.e.} data with $t=t_1-t_2$ are averaged with those with $t=L_T-(t_1-t_2)$, where $L_T$ is the lattice temporal extent, to improve statistics.  We perform correlated fits to the following function,
\begin{dmath}
C_K(t) = A_K \left(e^{-m_K t} + e^{-m_K(L_T-t)}\right)\,,
\end{dmath}
where the second term accounts for the state propagating backwards in time through the lattice temporal boundary. The chosen fit range, p-value and the results of the fit are given in Tab.~\ref{tab-pik2ptfit}. In physical units our kaon mass is 490.5(2.4) MeV, which is within 2\% of the physical neutral kaon mass.


\subsection{Pion two-point function}

The isospin triplet of pion states can be constructed from the operators listed in Sec. V.A. of Ref.~\cite{Christ:2019sah}. Due to the isospin symmetry the resulting two-point functions all have the same Wick contractions, and are most conveniently generated with the neutral pion operator,
\begin{dmath}
{\cal O}_\pi(\vec p_\pi, t) = \frac{-i}{\sqrt{2}} \sum_{\vec x, \vec y}e^{i(\vec p_1\cdot\vec x + \vec p_2\cdot\vec y)} \psibar_l(\vec x,t)\gamma^5 \sigma_3 \Theta(|\vec x-\vec y|){\cal P}_{\vec p_\pi} \psi_l(\vec y, t)\,,
\end{dmath}
where $\vec p_\pi = \vec p_1+\vec p_2$ is the total pion momentum and ${\cal P}_{\vec p_\pi}$ is a flavor projection operator of the form $\frac{1}{2}(1\pm \sigma_2)$ whose sign depends on the particular choice of the quark momentum, per the discussion in Sec. IV.G. of Ref.~\cite{Christ:2019sah}. We measure the two-point function with four different momentum orientations related by cubic transformations in order to improve the statistical error: $\vec p_\pi \in \left\{ (1,1,1), (-1,1,1), (1,-1,1), (1,1,-1)\right\}$ in units of $\pi/L$. The corresponding choices of quark momentum are given in Ref.~\cite{pipi-paper}. The two-point function 
\begin{dmath}
C_\pi(\vec p_\pi; t_1, t_2) = \langle 0|{\cal O}^\dagger_\pi(\vec p_\pi, t_1) {\cal O}_\pi(\vec p_\pi, t_2)|0\rangle
\end{dmath}
is again averaged over all source timeslices and also over all four momentum orientations, and the data folded to improve statistics. Correlated fits are performed to the function,
\begin{dmath}
C_\pi(t) = A_\pi \left(e^{-E_\pi t} + e^{-E_\pi(L_T-t)}\right)\,,
\end{dmath}
where $t=t_1-t_2$ as before. The chosen fit range, p-value and the results of the fit are also given in Tab.~\ref{tab-pik2ptfit}. In physical units, and assuming the continuum dispersion relation, our pion mass is 142.3(8) MeV, approximately 5\% larger than the physical value of 135 MeV. The small effect of this difference on our final results is expected to be negligible in comparison to our other errors.

\subsection{$\pmb{I=0}$ $\pmb{\pi\pi}$ two-point function}
\label{sec-pipi2pt}

\begin{table}[tb]
\begin{tabular}{c|c|c}
\hline\hline
Parameter & \multicolumn{2}{c}{Value}\\
\hline
                  	& 2-state fit	 	& 3-state fit \\
\hline
Fit range 		&    6-15 		&    4-15 \\
\hline
$A_{\pi\pi(111)}^0$ 	& $ 0.3682(31)$		& $ 0.3718(22)$\\
$A_{\pi\pi(311)}^0$ 	& $ 0.00380(32)$	& $ 0.00333(27)$\\
$A_{\sigma}^0$ 		& $-0.0004309(41)$	& $-0.0004318(42)$\\
$E_0$ 			& $ 0.3479(11)$		& $ 0.35030(70)$\\
$A_{\pi\pi(111)}^1$	& $ 0.1712(91)$		& $ 0.1748(67)$\\
$A_{\pi\pi(311)}^1$ 	& $-0.0513(27)$		& $-0.0528(30)$\\
$A_{\sigma}^1$ 		& $ 0.000314(17)$	& $ 0.000358(13)$\\
$E_1$ 			& $ 0.568(13)$		& $ 0.5879(65)$\\
$A_{\pi\pi(111)}^2$ 	& ---			& $ 0.116(29)$\\
$A_{\pi\pi(311)}^2$ 	& ---			& $ 0.063(10)$\\
$A_{\sigma}^2$ 		& ---			& $ 0.000377(94)$\\
$E_2$ 			& ---			& $ 0.94(10)$\\
\hline
p-value 		& 0.314			& 0.092\\
\end{tabular}    
\caption{Fit parameters in lattice units and the p-values for multi-operator fits to the $I=0$ $\pi\pi$ two-point functions. Here $E_i$ are the energies of the states and $A^i_\alpha$ represents the matrix element of the operator $\alpha$ between the state $i$ and the vacuum, given in units of $\sqrt{1\times 10^{13}}$. The second column gives the parameters for our primary fit which uses two-states and three operators. The third column shows a fit with the same three operators and one additional state that is used to probe the systematic effects of this third state on the $K\to\pi\pi$ matrix element fits.
\label{tab-pipi2ptfit-both}
}
\end{table}

Details of the strategy for measuring the $I=0$ $\pi\pi$ two-point function can be found in Ref.~\cite{pipi-paper}. In summary, we construct three operators with the quantum numbers of the $I=0$ $\pi\pi$ state: The first and second operators, labeled $\pi\pi(111)$ and $\pi\pi(311)$, comprise two single-pion operators carrying equal and opposite momenta separated by $\Delta=4$ timeslices in order to reduce the overlap with the vacuum state. The pion momenta in the former reside in the set $(\pm 1,\pm 1, \pm 1)\pi/L$, and those of the latter in the set $(\pm 3,\pm 1, \pm 1)\pi/L$ and permutations thereof. We average over all non-equivalent directions of the pion momentum in order to project onto the rotationally symmetric state. The final, $\sigma$ operator corresponds to the scalar two-quark operator $\frac{1}{\sqrt{2}}(\bar u u + \bar d d)$. As mentioned previously, the pion and $\sigma$ bilinear operators are smeared with a hydrogen wavefunction (exponential) smearing function of radius $2$ lattice sites in order to improve their overlap with the lowest-energy states.

Two-point correlation functions are constructed from pairs of source and sink operators thus,
\begin{dmath}
C^{\pi\pi}_{\alpha\beta}(t_1, t_2) = \langle 0| {\cal O}^\dagger_\alpha(t_1) {\cal O}_\beta(t_2)|0\rangle - \langle 0| {\cal O}^\dagger_\alpha(t_1)|0\rangle \langle 0|{\cal O}_\beta(t_2)|0\rangle\,,
\end{dmath}
where we include an explicit vacuum subtraction. Here $t_1$ specifies the earliest time in which any fermion operator appearing in the annihilation operator ${\cal O}^\dagger_\alpha$ is evaluated, and likewise $t_2$ is the latest time appearing in the creation operator ${\cal O}_\beta$, such that $t=t_1-t_2$ is the time of propagation of the shortest-lived pion state. We average over many $t_2$ at fixed $t=t_1-t_2$ and the data are folded to improve statistics as follows:
\begin{dmath}
C^{\pi\pi}_{\alpha\beta}(t) \to \frac{1}{2}\left[ C^{\pi\pi}_{\alpha\beta}(t) + C^{\pi\pi}_{\alpha\beta}(L_T-\Delta_i-\Delta_j- t)\right]
\end{dmath}
where $\Delta_i=4$ for the $\pi\pi(111)$ and $\pi\pi(311)$ operators and zero for the $\sigma$ operator. To the matrix of correlation functions we perform simultaneous correlated fits to the functions,
\begin{dmath}
C^{\pi\pi}_{\alpha\beta}(t) =  \sum_{i=0}^{i_{\rm max}} A^i_\alpha A^i_\beta \left(e^{-E_i t} + e^{-E_i(L_T-\Delta_i-\Delta_j- t)}\right)\,.
\end{dmath}

We will use the result obtained by uniformly fitting to the temporal range $6-15$ with all three $\pi\pi$ source/sink operators and allowing for two intermediate states ($i_{\rm max}=1$), which represents the ``best fit'' in Ref.~\cite{pipi-paper}. The results, reproduced from that work are given in the second column of Tab.~\ref{tab-pipi2ptfit-both} for the convenience of the reader. Note that our $\pi\pi$ and kaon energies differ by 2.2(3)\%, where the error is statistical only, and as such our $K\to\pi\pi$ calculation is not precisely energy-conserving. The effect of this difference is incorporated as a systematic error on our final result, as discussed in Sec.~\ref{sec-syserr-unphyskin}.

It is interesting to compare the statistical errors of our $\pi\pi$ ground-state fit parameters to those of our 2015 analysis, which was performed using a single operator ($\pi\pi(111)$) and the same $t_{\rm min}=6$ as our present analysis. Previously we obtained
\begin{dmath}\begin{array}{rl}
A_{\pi\pi(111)}^0 &= 0.3923(60) \\
E_0 &= 0.3606(74)\,.
\end{array}\end{dmath}
Comparing these to the results of this work in Tab.~\ref{tab-pipi2ptfit-both} we find that the error on the ground-state amplitude has reduced by a factor of $1.9$ and the energy by a factor of 6.7. The former is compatible with the expected $\sqrt{741/216}=1.9$ reduction in errors due to the increased statistics, but the latter has improved by a far greater amount. In Ref.~\cite{pipi-paper} we demonstrate that this improvement in the errors is a result of the additional operators, in particular the $\sigma$ operator, which vastly enhance the resolution on the ground-state energy.

The $I=0$ $\pi\pi$ scattering phase shift is obtained via L\"uscher's method~\cite{Luscher:1986pf,Luscher:1990ux} and has the value,
\begin{dmath}
\delta_0 = 32.3(1.0)(1.8)^\circ\,,\label{eq-delta0value}
\end{dmath}
where the errors are statistical and systematic, respectively. The procedures by which we estimate our errors are detailed in Ref.~\cite{pipi-paper}.

Our decision to fit the $\pi\pi$ two-point function with two states limits the number of states that we can include in our $K\to\pi\pi$ matrix element analysis. In order to study the possibility of residual contamination from a third state we repeat the analysis of the $\pi\pi$ two-point function with 3 states, the results of which are given in the third column of Tab.~\ref{tab-pipi2ptfit-both}. For a stable fit to the $\pi\pi$ data we found it necessary to use $t_\min=4$, which is lower than the $t_\min=6$ used for the primary fit and which exposes the result to enhanced excited state contamination. However comparing the results between the second and third columns of Tab.~\ref{tab-pipi2ptfit-both} we find little relative difference in the parameters associated with the ground-state, suggesting any such effects on the $K\to\pi\pi$ matrix elements are small.

\subsection{Phase-shift derivative at the kaon mass}
\label{sec-phase-shift-deriv}

As detailed in Sec.~\ref{sec-llfactor}, the Lellouch-L\"uscher finite volume correction to the $K\to\pi\pi$ matrix elements requires the evaluation of the derivative of the phase-shift with respect to the $\pi\pi$ energy evaluated at the kaon mass scale, or more specifically with respect to the variable $q=kL/2\pi$ where $k^2 = (E_{\pi\pi}/2)^2 - m_\pi^2$ is the square of the interacting pion momentum. This derivative cannot presently be obtained experimentally at this energy scale, and therefore an interpolating ansatz or direct lattice measurement is required.

In Ref.~\cite{pipi-paper}, alongside the stationary state examined above, we also compute the $\pi\pi$ energy at several non-zero center-of-mass momenta, allowing us to obtain the phase-shift at two values of the rest-frame energy that are lower than the kaon mass as well as a threshold determination of the scattering length. These results are also close to their corresponding dispersive predictions, albeit with somewhat larger excited-state systematic errors. Using these results we can directly measure the derivative of the phase-shift with respect to the energy using a finite-difference approximation, for which we obtain
\begin{dmath}
\frac{{\rm d}\delta_0}{{\rm d}q} = 1.76(74)\ {\rm rad}   
\end{dmath}
from the difference with the nearest energy to the kaon mass, and 
\begin{dmath}
\frac{{\rm d}\delta_0}{{\rm d}q} = 1.33(17)\ {\rm rad}   
\end{dmath}
from the next-to-nearest.


We can also obtain the derivative from the dispersive prediction of Colangelo {\it et al}~\cite{Colangelo:2001df}. The derivative with respect to $s=E_{\pi\pi}^2$, computed at our lattice $\pi\pi$ energy using Eqs. 17.1-17.3 of Ref.~\cite{Colangelo:2001df} with $m_\pi=135$ MeV, is found to be
\begin{dmath}
\frac{{\rm d}\delta_0}{{\rm d}s} = 3.36(3)\times 10^{-6}\ {\rm rad}\ {\rm MeV}^{-2}\,,   
\end{dmath}
where the error is the statistical error arising from the uncertainty in the lattice spacing and measured lattice $\pi\pi$ energy. Note that this result is obtained at the physical pion mass, which is 5\% smaller than our lattice value. In order to estimate the impact of the difference in pion masses on this derivative we use NLO chiral perturbation theory~\cite{Bijnens:1997vq,Colangelo:2001df} (ChPT) to estimate the derivative with respect to energy at $k=0.1$ GeV, at which ChPT is expected to be reliable. Assuming that the slope with respect to $\sqrt{s}$ is roughly constant (which is well motivated by the dispersion theory result, cf. Fig. 7 of Ref.~\cite{Colangelo:2001df}) we estimate the change in $\frac{{\rm d}\delta_0}{{\rm d}s}$ evaluated at our lattice $\pi\pi$ energy as 1.2\%. This value is small relative to the final systematic error we assign to  the derivative in Sec.~\ref{sec-llfactorsys} and can therefore be neglected here. Finally, applying ${\rm d}s/{\rm d}q = 4.18(5)\times 10^5$ MeV$^2$, where again the errors are statistical, we obtain
\begin{dmath}
\frac{{\rm d}\delta_0}{{\rm d}q} = 1.405(5)\ {\rm rad}\,. \label{eq-delta-deriv-colangelo}  
\end{dmath}

The near-linearity of the dispersive prediction suggests that a linear ansatz,
\begin{dmath}
\frac{{\rm d}\delta_0}{{\rm d}E_{\pi\pi}} \approx \frac{\delta_0}{E_{\pi\pi}-2m_\pi}
\end{dmath}
may also be appropriate. With this ansatz we find
\begin{dmath}
\frac{{\rm d}\delta_0}{{\rm d}q} = 1.259(36)\ {\rm rad}\,.\label{eq-deltaderiv-linq} 
\end{dmath}

Given that the derivative of the phase shift is a subleading contribution and that the above values are all in reasonable agreement, we expect that the Lellouch-L\"uscher factor can be obtained reliably.  The variation in these results will be taken into account in our systematic error in Sec.~\ref{sec-llfactorsys}.

In our 2015 work~\cite{Bai:2015nea} we also considered a linear ansatz in $q$, 
\begin{dmath}
\frac{{\rm d}\delta_0}{{\rm d}q} \approx \frac{\delta_0}{q}
\end{dmath}
for which we obtain
\begin{dmath}
\frac{{\rm d}\delta_0}{{\rm d}q} = 0.790(22)\ {\rm rad}\,.  
\end{dmath}
This value is not as well motivated as the ansatz in Eq.~\eqref{eq-deltaderiv-linq} and is in disagreement with all four of the above results. Given the good agreement between our measured phase-shifts and the above estimates of the derivative with the dispersive predictions, we will not include this result in our systematic error estimate.

\subsection{Optimal $\pmb{\pi\pi}$ operator}
\label{sec-optimal-pipi-op}

For use later in this document we define here an optimal operator that maximally projects onto the $\pi\pi$ ground state relative to the first-excited state.

Under the excellent assumption that the backwards-propagating component of the time dependence is small in the fit window, the two-point functions can be described as a sum of exponentials:
\begin{dmath}
C^{\pi\pi}_{\alpha\beta}(t) = \sum_i A_\alpha^i A_\beta^i e^{-E_i t}\,,
\end{dmath}
where again Greek indices denote operators and Roman indices states. We wish to define an optimized operator that projects onto the ground state:
\begin{dmath}
{\cal O}_{\rm opt} = \sum_\alpha {\cal O}_\alpha r_\alpha\,,
\end{dmath}
for which
\begin{dmath}
C^{\pi\pi}_{\rm opt}(t) = \langle 0|{\cal O}_{\rm opt}^\dagger(t) {\cal O}_{\rm opt}(0)|0\rangle \approx [A_{\rm opt}^0]^2  e^{-E_0 t}\,,\label{eq-Ooptpipicorr}
\end{dmath}
where the approximate equality indicates that additional exponential terms resulting from excited-state contamination, although suppressed, still exist for an optimal operator composed of a finite number of $\pi\pi$ operators. Expanding the Green's function,
\begin{dmath}
\langle 0|{\cal O}_{\rm opt}^\dagger(t) {\cal O}_{\rm opt}(0)|0\rangle = \sum_{\alpha\beta} r_\alpha \langle 0|{\cal O}_\alpha^\dagger(t) {\cal O}_\beta(0) |0\rangle  r_\beta \\
= \sum_i \sum_{\alpha\beta} r_\alpha A_\alpha^i A_\beta^i r_\beta e^{-E_i t} = \sum_i \left[\sum_\alpha  A_\alpha^i r_\alpha \right]^2  e^{-E_i t}\,.
\end{dmath}
Without loss of generality we can fix $A_{\rm opt}^0=1$, which alongside Eq.~\eqref{eq-Ooptpipicorr} is sufficient to define $r_i$:
\begin{dmath}
\sum_\alpha A_\alpha^i r_\alpha  = \delta_{i,0}\,.
\end{dmath}
If the number of states is equal to the number of operators this can be interpreted as a matrix equation,
\begin{dmath}
{\bf A}\vec r = \hat 0\,,
\end{dmath}
where the row index of $\bf A$ is the state index $i$ and the column index the operator index $\alpha$. Here $\hat 0$ is a unit vector in the 0-direction, and as such
\begin{dmath}
\vec r = {\bf A}^{-1}\hat 0\,.
\end{dmath}
which gives
\begin{dmath}
r_\alpha = [{\bf A}^{-1}]_{\alpha, 0}
\end{dmath}
i.e. $\vec r$ is the first column of the inverse matrix.

\begin{figure}[tb]
\centering
\includegraphics[width=0.48\textwidth]{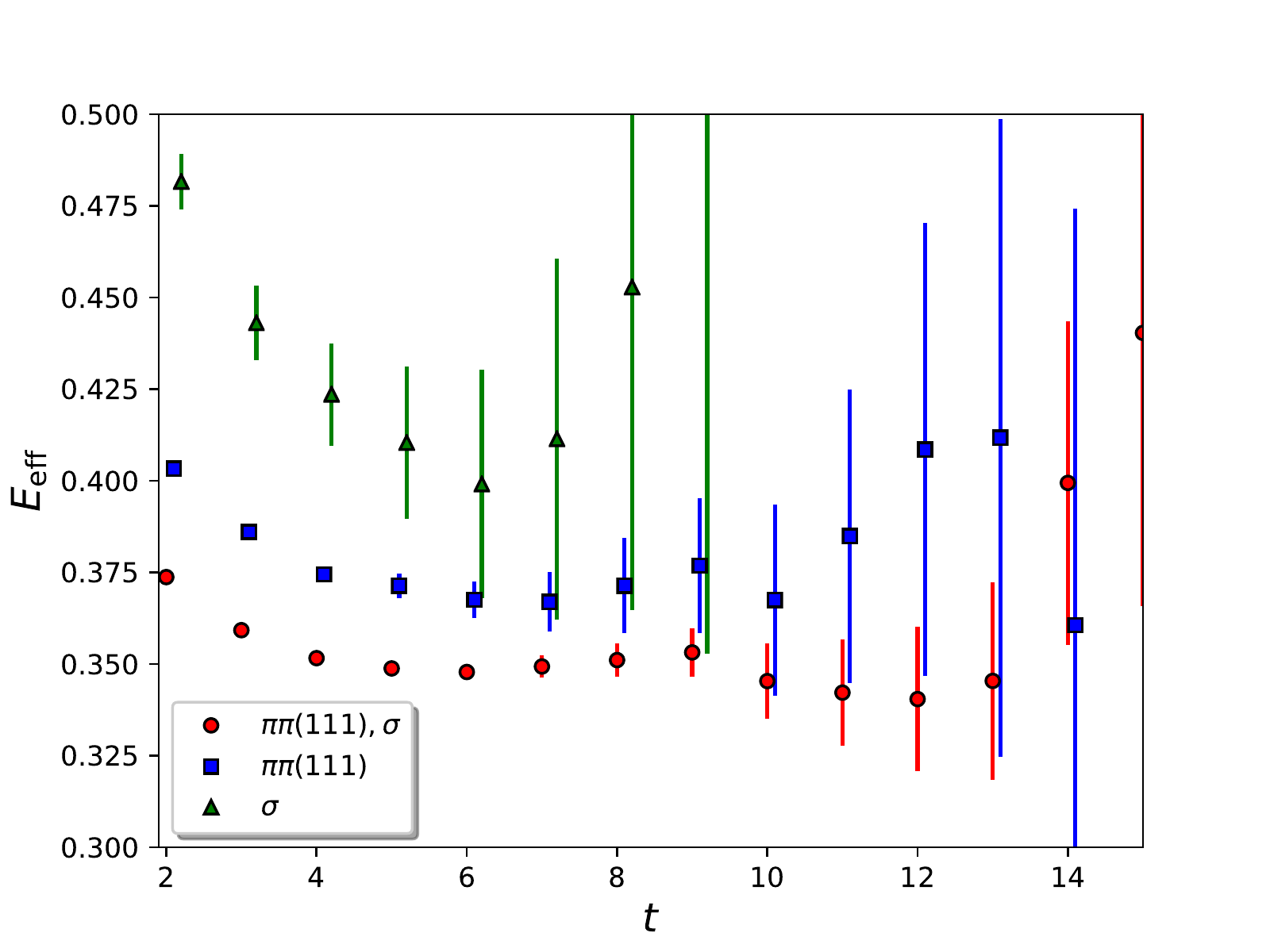}
\caption{A comparison of the effective ground-state energy obtained from the optimal operator ({\it i.e.} the optimal combination of the $\sigma$ and $\pi\pi(111)$ operators, and labeled ``$\pi\pi(111),\,\sigma$'' here) with the energies obtained from the $\sigma$ and $\pi\pi(111)$ operators separately. \label{fit-pipiopteffenergy}}
\end{figure}

As our $\pi\pi$ fits include only two states, we drop the noisier $\pi\pi(311)$ operator in order to form a square matrix of correlation functions. We then obtain
\begin{equation}
\vec r\,^T = (5.24(18)\times 10^{-7}, -2.86(17)\times 10^{-4}) 
\end{equation}
where the elements are the coefficients of the $\pi\pi(111)$ and $\sigma$ operators, respectively. In Fig.~\ref{fit-pipiopteffenergy} we compare the effective energy obtained with the optimal operator to that of the $\pi\pi(111)$ and $\sigma$ operators alone. We observe a marked reduction in the ground-state energy and a noticeable improvement in the length of the plateau region resulting from the removal of excited-state contamination, as well as a significant improvement in the statistical error. This optimal operator will also be used in our matrix element fits in the following section.

\section{Results from three-point correlation functions for $\pmb{\Delta I=1/2,~K\to\pi\pi}$ decays}
\label{sec:ThreePointResults}
In this section we detail the measurement and fitting of the $K\to\pi\pi$ three-point Green's functions, from which the unrenormalized matrix elements $\langle (\pi\pi)_{I=0}|Q_i|K^0\rangle$ are obtained.

\subsection{Overview of measurements}

\begin{figure}[tb]
\centering
\begin{subfigure}[t]{0.35\textwidth}
\includegraphics[width=\textwidth]{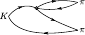}
\caption{{\it type1}}
\end{subfigure}
\hspace{2cm}
\begin{subfigure}[t]{0.35\textwidth}
\includegraphics[width=\textwidth]{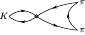}
\caption{{\it type2}}
\end{subfigure}
\begin{subfigure}[t]{0.35\textwidth}
\includegraphics[width=\textwidth]{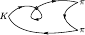}
\caption{{\it type3}}
\end{subfigure}
\hspace{2cm}
\begin{subfigure}[t]{0.35\textwidth}
\includegraphics[width=\textwidth]{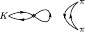}
\caption{{\it type4}}
\end{subfigure}
\caption{The four classes of $K\to\pi\pi$ Wick contractions. \label{fig-type1234}  }
\end{figure}

On the lattice we measure the following three-point functions,
\begin{dmath}
C_i(t,t_{\rm sep}^{K\to\snk})  = \langle 0 | {\cal O}_{\rm snk}^\dagger(t_{\rm sep}^{K\to\snk}) Q_i(t) {\cal O}_{\tilde K^0}(0) |0 \rangle\,, \label{eq-3ptfuncdef}
\end{dmath}
where $t$ denotes the time separation between the kaon and four-quark operators, and $t_{\rm sep}^{K\to\snk}$ the time separation between the kaon and the $\pi\pi$ ``sink'' operator, ${\cal O}_{\rm snk}$. As described in Ref.~\cite{Blum:2011pu}, the Wick contractions of these functions fall into four categories based on their topology, as illustrated in Fig.~\ref{fig-type1234}. 

Note that here and below we take care to differentiate between the G-parity kaon state $\tilde K^0$, which is a G-parity even eigenstate of the finite-volume Hamiltonian, and the physical kaon $K^0$ that is not an eigenstate of the system. The matrix elements of the physical kaon are related to those of the G-parity kaon by a constant multiplicative factor of $\sqrt{2}$ that serves as the analogue of the Lellouch-L\"uscher finite-volume correction as described in Sec. VI.B. of Ref.~\cite{Christ:2019sah}.

In order to maximize statistics we translate the three-point function over multiple kaon timeslices and average the resulting measurements. As the statistical error is dominated by the {\it type3} and {\it type4} diagrams these are measured with kaon sources on every timeslice, $0\leq t_K < L_T$. The far more precise {\it type1} and {\it type2} contributions are measured every eighth timeslice in order to reduce the computational cost. For the remainder of this section we will assume all correlation functions to have been averaged over the kaon timeslice where appropriate. 

We compute each diagram with 5 different time separations between the kaon and the $\pi\pi$ sink operators, $t_\sep^{K\to\snk}\in\{10,12,14,16,18\}$, with the $\Delta S=1$ four-quark operator inserted on all intervening timeslices.  Note these five time separations specify the time between the kaon operator and the closest single-pion factor in the $\pi\pi$ operator for those cases when the $\pi\pi$ operator is a product of single-pion operators evaluated on different time slices. (This convention of specifying the minimum time separation from those $\pi\pi$ operators which are non-local in the time is followed throughout this paper.) As these $\pi\pi$ operators comprise back-to-back moving pions with zero total momentum, we must measure each diagram for all possible orientations of the pion momenta in order to project onto the rotationally symmetric state. 

The {\it type3} and {\it type4} diagrams both contain a light or strange quark loop beginning and ending at the operator insertion point that results in a quadratic divergence regulated by the lattice cutoff.  This divergence is removed by defining the subtracted operators~\cite{Blum:2011pu,Bernard:1985wf},
\begin{dmath}
Q_i \to Q_i - \alpha_i \bar s\gamma^5 d\,. \label{eq-suboptdef}
\end{dmath}
We will henceforth denote the unsubtracted operator with a hat notation, $\hat Q_i$. The coefficients $\alpha_i$ in Eq.~\eqref{eq-suboptdef} are defined by imposing the condition,
\begin{dmath}
\langle 0 | \Big\{\hat Q_i(t) - \alpha_i(t) [\bar s\gamma^5 d](t) \Big\} {\cal O}_{\tilde K^0}(0)|0\rangle = 0\,,\label{eq-subtractedopcond}
\end{dmath}
where we have allowed $\alpha_i$ to vary with time as this was found to offer a minor statistical improvement. Although the matrix element of this pseudoscalar operator vanishes by the equations of motion for energy-conserving kinematics and is therefore not absolutely necessary for our calculation, the subtraction reduces the systematic error resulting from the small difference between our $\pi\pi$ and kaon energies while simultaneously reducing the statistical error and suppressing excited-state contamination. 

Due to having vacuum quantum numbers, the $I=0$ $\pi\pi$ operators project also onto the vacuum state and this off-shell matrix element dominates the signal unless an explicit vacuum subtraction is performed,
\begin{dmath}
C_i(t,t_{\rm sep}^{K\to\snk}) \to C_i(t,t_{\rm sep}^{K\to\snk}) - \langle 0 | {\cal O}_{\rm snk}^\dagger(t_{\rm sep}^{K\to\snk})|0\rangle\langle 0| Q_i(t)  {\cal O}_{\tilde K^0}(0)|0 \rangle\,.\label{eq-vacsub}
\end{dmath}
However, due to our definition of the subtraction coefficient $\alpha_i$ in Eq.~\eqref{eq-subtractedopcond}, the vacuum matrix elements appearing in the right-hand side vanish making this subtraction unnecessary. In practice this cancellation is not exact in our numerical analysis for the following reason: While the $\pi\pi$ ``bubble'' $\langle 0|{\cal O}_{\rm snk}^\dagger|0\rangle$ is formally time-translationally invariant we observed a minor statistical advantage in evaluating this quantity with the $\pi\pi$ operator on the same timeslice as it appears in the full disconnected Green's function that is being subtracted, such that it is maximally correlated. Therefore, for the right-most term in Eq.~\eqref{eq-vacsub} we compute
\begin{dmath}
\frac{1}{n_{t_K}}\sum_{t_K\in\{t_K\}} \langle 0| {\cal O}_{\rm snk}^\dagger(t_K+t_{\rm sep}^{K\to\snk}) |0\rangle \langle 0| \left\{ \hat Q_i(t+t_K) -\alpha_i(t)[\bar s\gamma^5 d](t+t_K)\right\} {\cal O}_{\tilde K^0}(t_K) |0 \rangle\,,
\end{dmath}
where $t_K$ is the kaon timeslice and $\{t_K\}$ the set of timeslices upon which measurements were performed, i.e. with the product of the $K\to$vacuum matrix element and the $\pi\pi$ bubble performed under the average over the kaon source timeslice rather than after. As suggested by the above, the coefficients $\alpha_i(t)$ are computed separately from the $t_K$-averaged matrix elements and therefore the cancellation between the two terms in brackets is exact only up to the degree to which the time translation symmetry is realized at finite statistics. Due to our large statistics we found the difference in the fitted $Q_6$ matrix element obtained with and without the vacuum subtraction to be at the 0.1\% level.

We perform measurements with all three two-pion operators described in Sec.~\ref{sec-pipi2pt}. For the $K\to\pi\pi$ matrix elements of the four-quark operators, the full set of Wick contractions for the $\pi\pi(111)$ and $\pi\pi(311)$ sink operators can be found in Appendix B.1 and B.2 of Ref.~\cite{ZhangThesis}, and those of the $\sigma$ operator in Appendix~\ref{sec:appendix-sigmacon} of this document. The Wick contractions for the $K\to\pi\pi$ matrix elements of the pseudoscalar operator (with all three sink operators) as well as the $K\to$vacuum matrix elements of this and the four-quark operators are provided in Appendix~\ref{sec:appendix-Pop} of this document.

\begin{figure}[tb]
\centering
\includegraphics[width=0.45\textwidth]{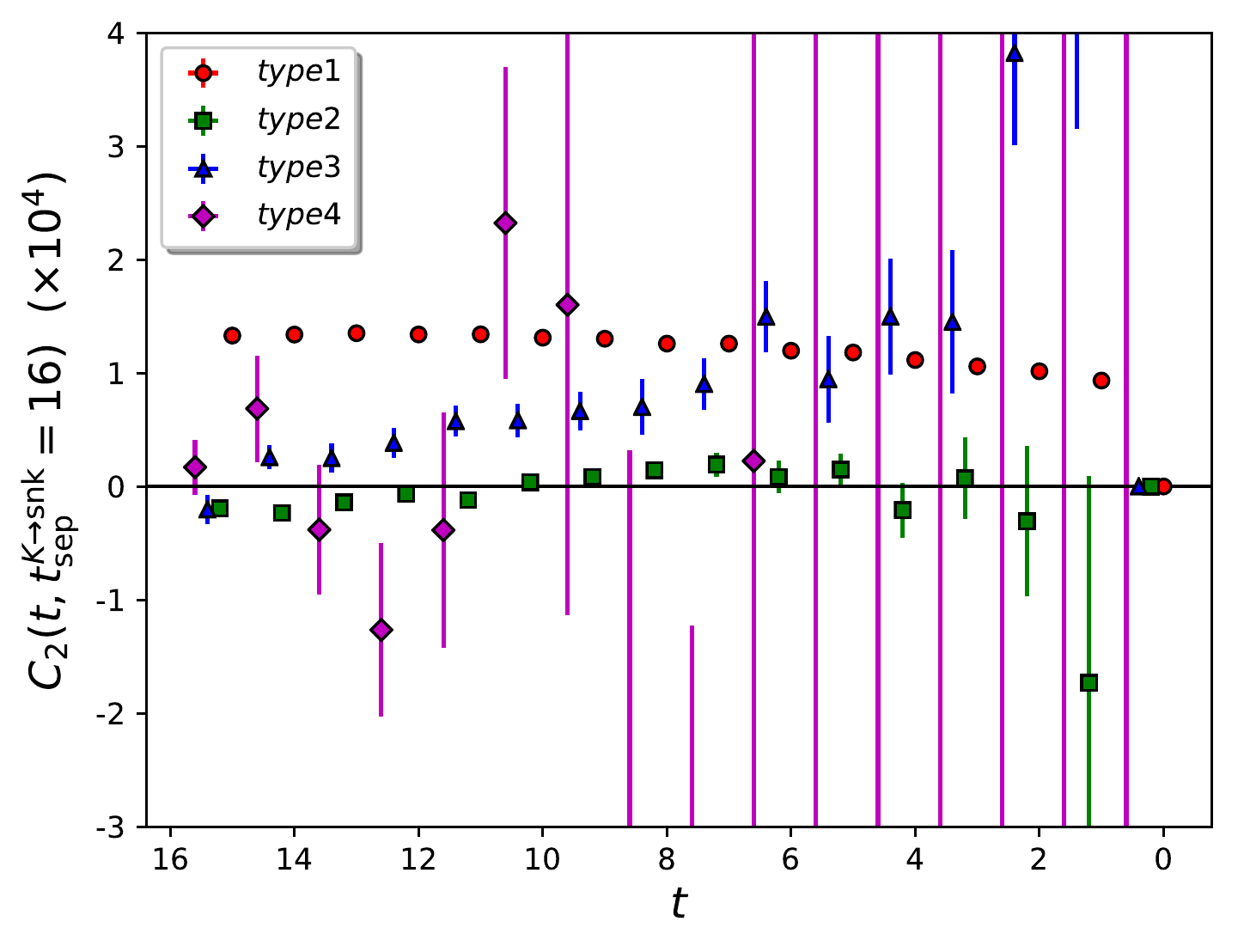}
\includegraphics[width=0.45\textwidth]{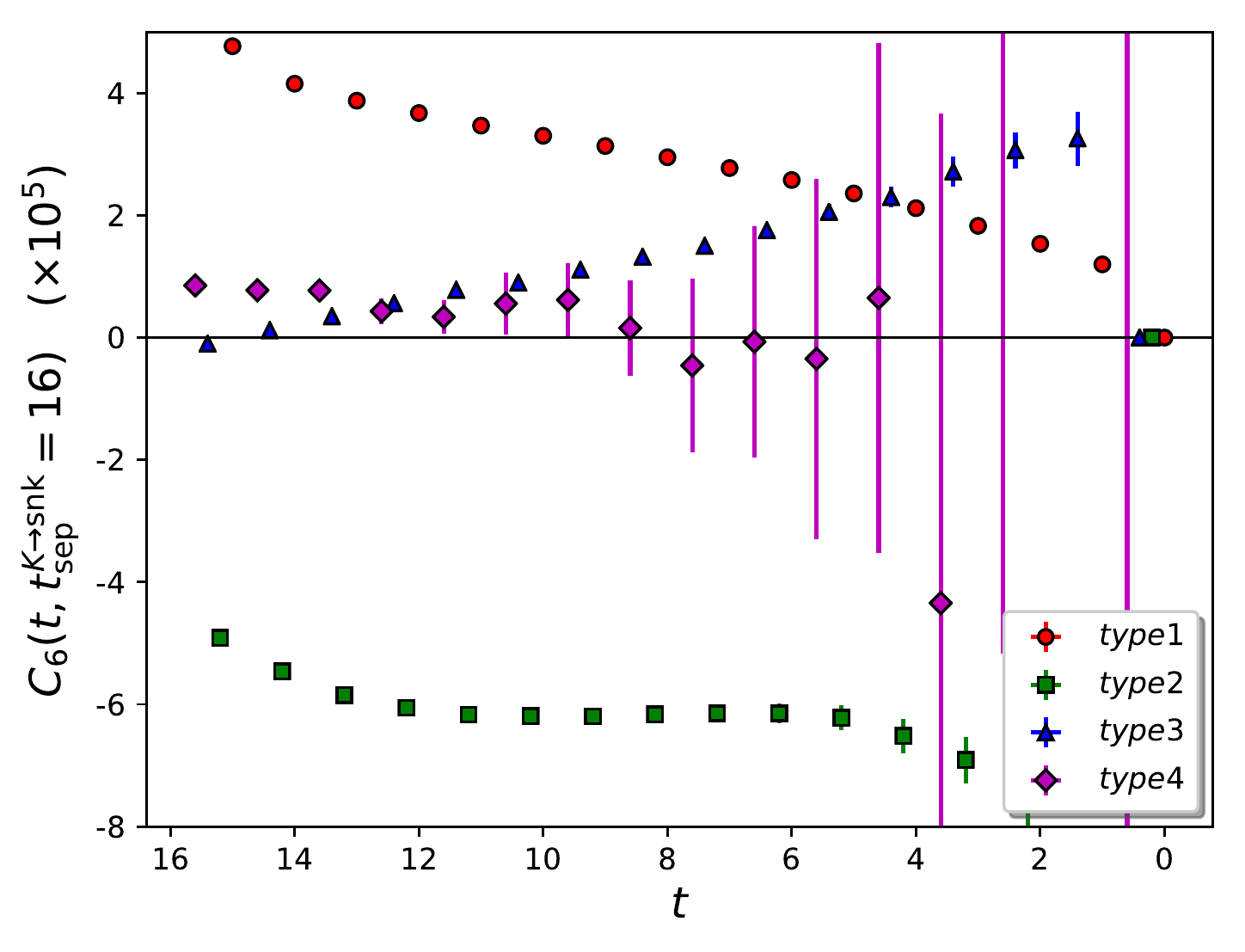}
\caption{The contributions of the four Wick contraction topologies ${\it type1}$-${\it type4}$ to the $C_2$ (left) and $C_6$ (right) three-point functions with the $\pi\pi(111)$ sink operator, plotted as a function of the time separation between the kaon and the four-quark operator, $t$, at fixed $t_{\rm sep}^{K\to\snk}=16$. For clarity we plot with an inverted x-axis such that the $\pi\pi$ sink operator is on the left-hand side. These correlation functions include the subtraction of the pseudoscalar operator. \label{fig-Q2Q6typecontribs}  }
\end{figure}

In Fig.~\ref{fig-Q2Q6typecontribs} we plot the contributions of the four classes of Wick contraction illustrated in Fig.~\ref{fig-type1234} to the three-point functions of the (subtracted) $Q_2$ and $Q_6$ operators with the $\pi\pi(111)$ sink operator. As the individual topologies are not separately interpretable as Green's functions of the QCD path integral, their time dependence is not necessarily described by the propagation of physical eigenstates of the QCD Hamiltonian. As such we cannot combine our data sets with different $t_{\rm sep}^{K\to\snk}$ when generating such plots, and instead plot with a single, fixed $t_{\rm sep}^{K\to\snk}=16$. Despite the inability to interpret the time dependence physically, we can look at the relative contributions of each topology within the central region of the plot in which the behavior of the combined data is dominated by the kaon and $\pi\pi$ ground-states, i.e. the region in which we perform our fits below. Our final choices of cut incorporate data from this set in the range $6 \leq t \leq 11$ (cf. Sec.~\ref{sec-finalfitresults}). In this window we observe that for both the $C_2$ and $C_6$ correlation functions, the contribution of the noisy, ${\it type4}$ disconnected diagrams is largely consistent with zero, albeit with much larger errors for the former. $C_2$ appears dominated by the ${\it type1}$ and ${\it type3}$ diagrams, which both contribute with the same sign, with a negligible contribution from the ${\it type2}$ diagrams. The contribution of the ${\it type1}$ and ${\it type3}$ diagrams appears to behave similarly for the $C_6$ three-point function, however here we observe a strong cancellation between those and the ${\it type2}$ diagrams.

\subsection{Determination of $\pmb{\alpha_i}$}
\label{sec-alphai-comput}

The subtraction coefficients $\alpha_i$ are computed via Eq.~\eqref{eq-subtractedopcond} as the following ratio of two-point functions,
\begin{dmath}
\alpha_i(t) = \frac{\langle 0 |  \hat Q_i(t) {\cal O}_{\tilde K^0}(0)|0 \rangle}{  \langle 0 | [\bar s\gamma^5 d](t) {\cal O}_{\tilde K^0}(0) | 0\rangle }\,,
\label{eq:alpha-determine}
\end{dmath}
where the average of the correlation functions over the kaon source timeslice is implicit as above.


The Wick contractions for the $\langle 0|\hat Q_i(t) {\cal O}_{\tilde K^0}(0)|0\rangle$ two-point functions are identical to the components of the {\it type4} $K\to\pi\pi$ diagrams that are connected to the kaon. While these connected components are formally independent of the sink two-pion operator, in practice these quantities were computed using code that was organized differently for the $\pi\pi$ and $\sigma$ operators.  As described in Appendix~\ref{sec:appendix-Pop} of this paper and Appendix B.2 of Ref.~\cite{ZhangThesis}, the factors entering the {\it type4} diagrams that determine the $\alpha_i$ were constructed from two separate bases of functions of the quark propagators, one for the $\sigma$ and the other for the $\pi\pi(\ldots)$ operators, where for each basis $\gamma^5$ hermiticity was used in a different way. While $\gamma^5$ hermiticity is an exact relation, the fact that we are using a stochastic approximation for the high modes of the all-to-all propagator allows small differences to arise between the values of the $\alpha_i$ computed in these two bases. We therefore have separate results for the $\alpha_i$ from the $\pi\pi$ and $\sigma$ three-point functions calculations.

\begin{figure}[tbp]
\centering
\includegraphics[width=0.32\textwidth]{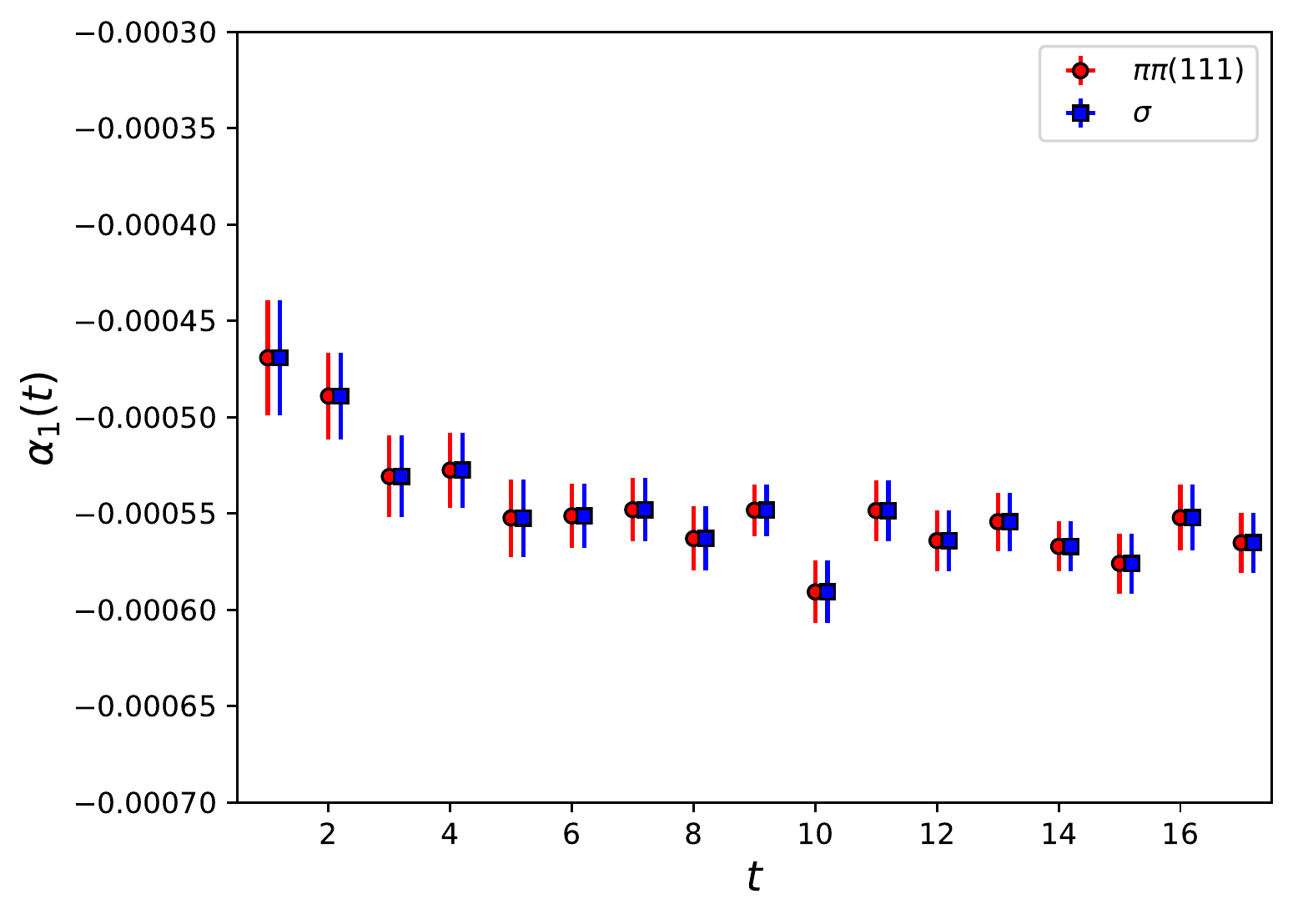}
\includegraphics[width=0.32\textwidth]{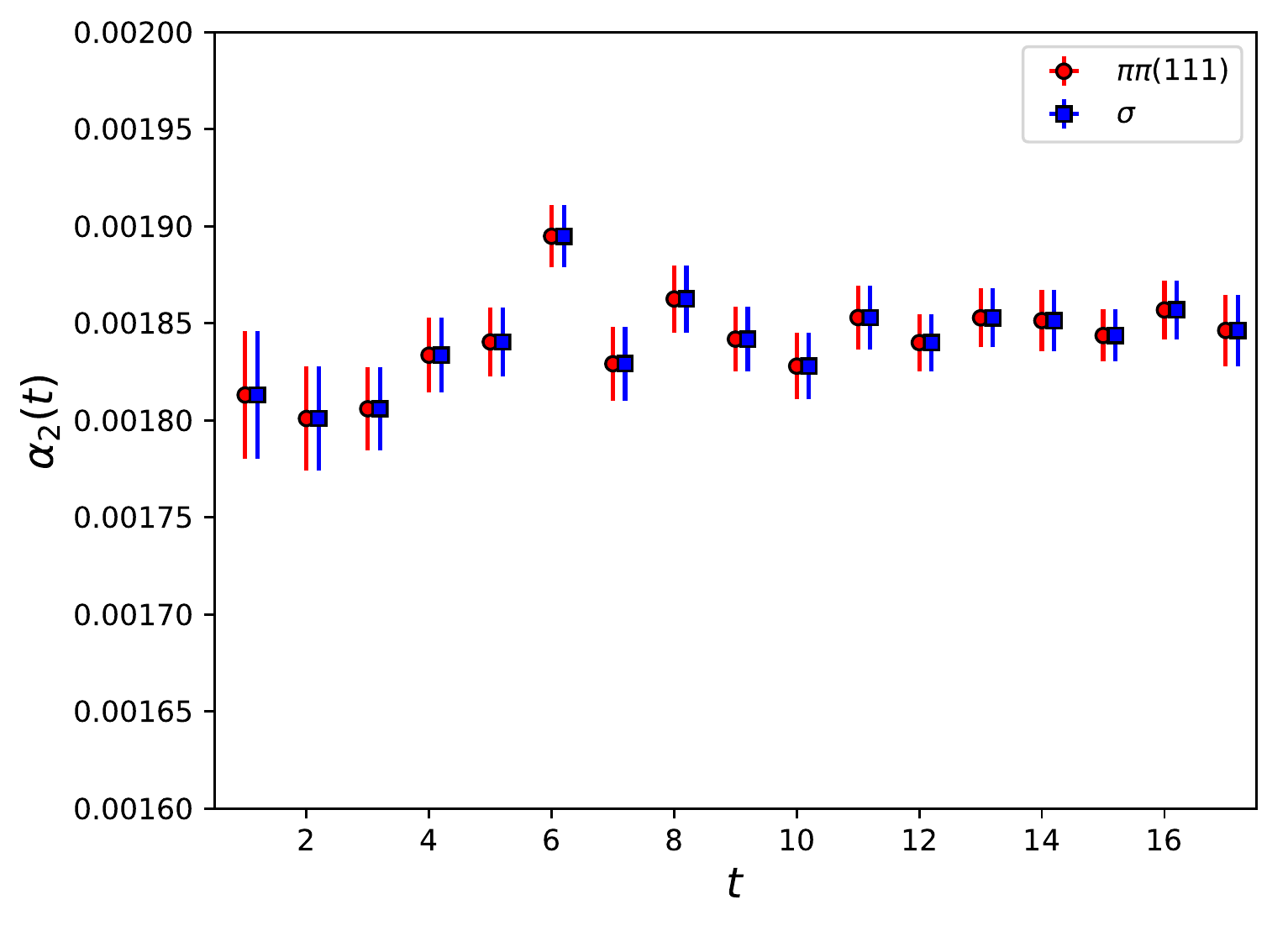}
\includegraphics[width=0.32\textwidth]{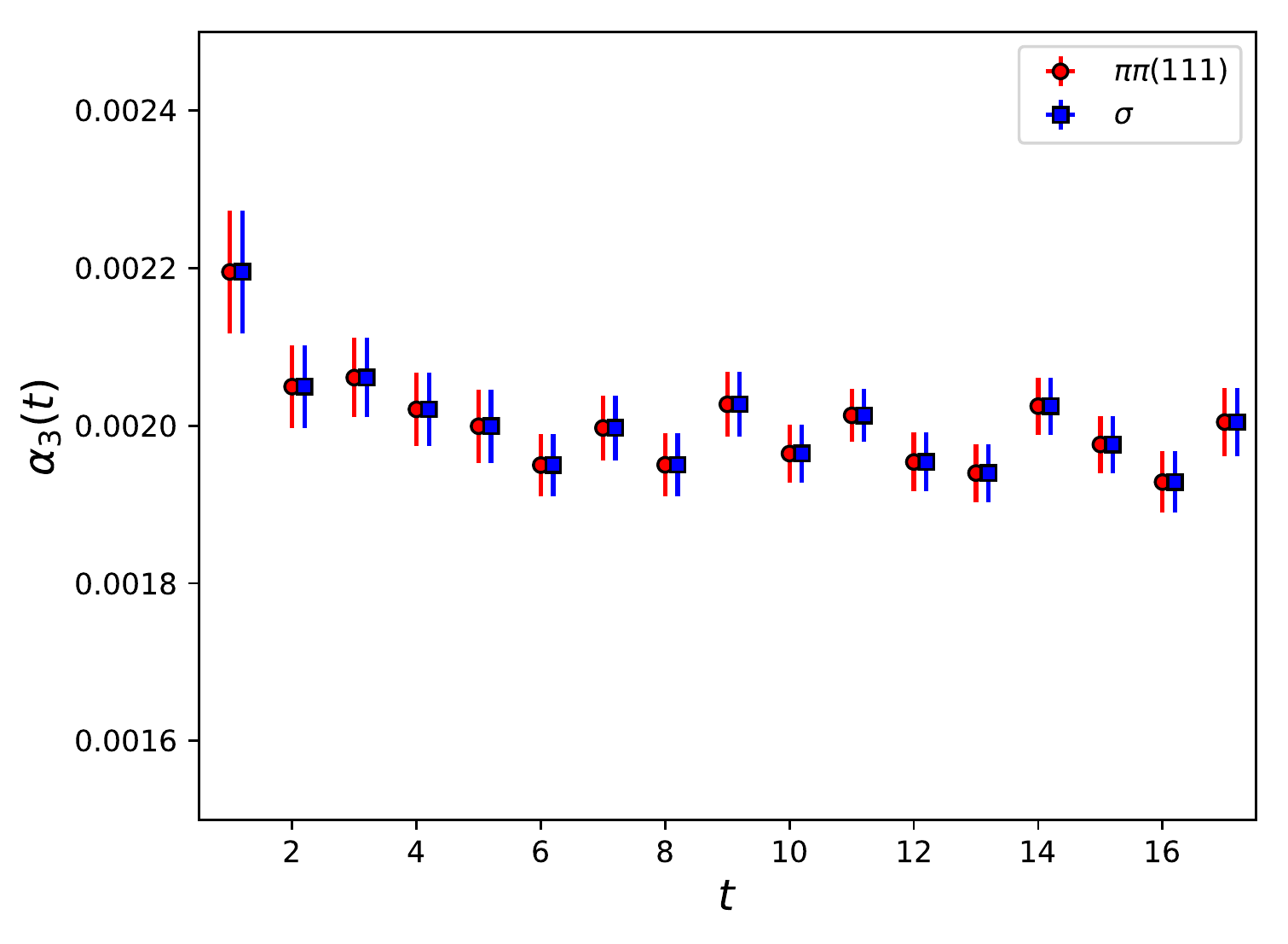}
\\
\includegraphics[width=0.32\textwidth]{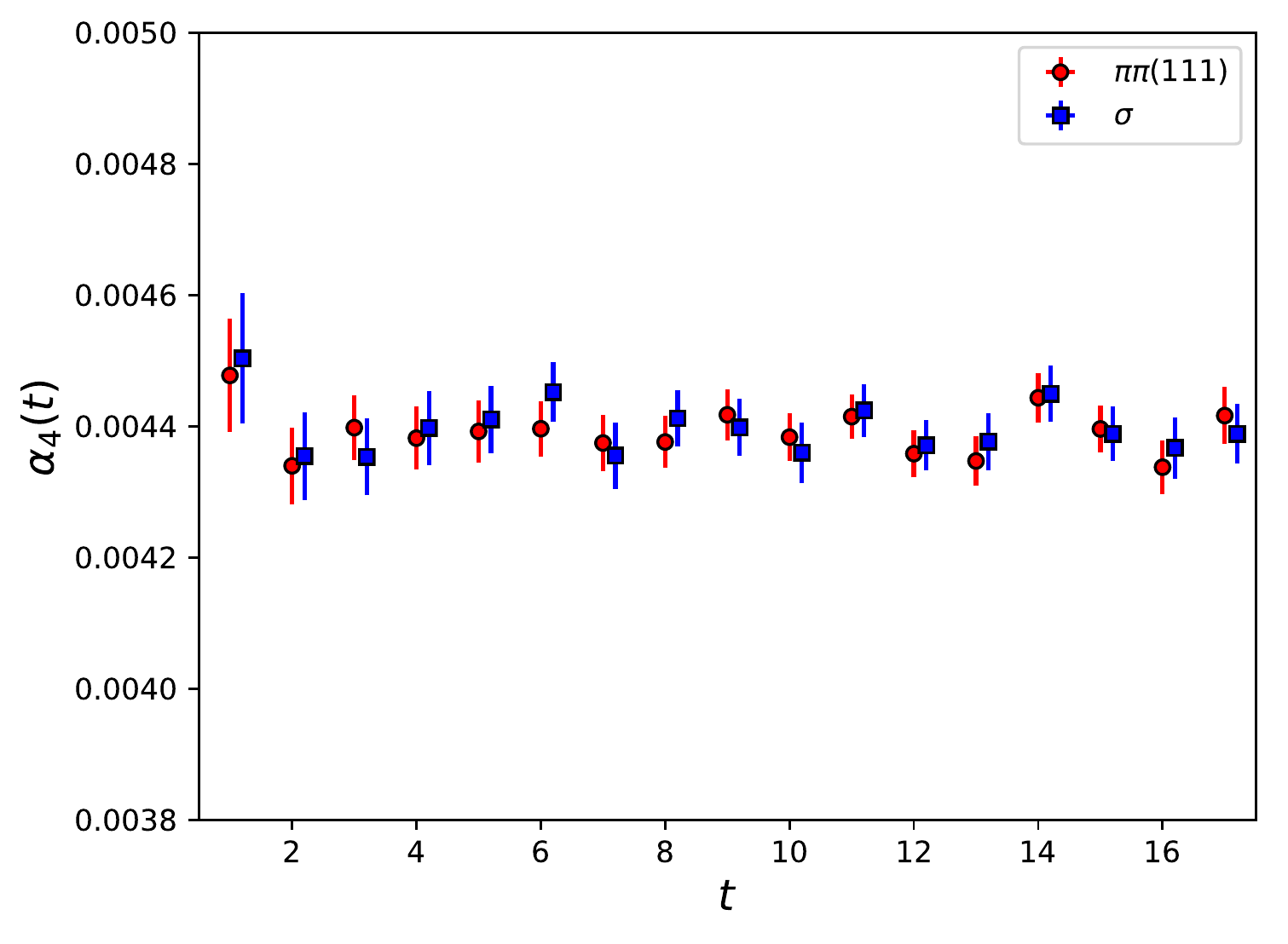}
\includegraphics[width=0.32\textwidth]{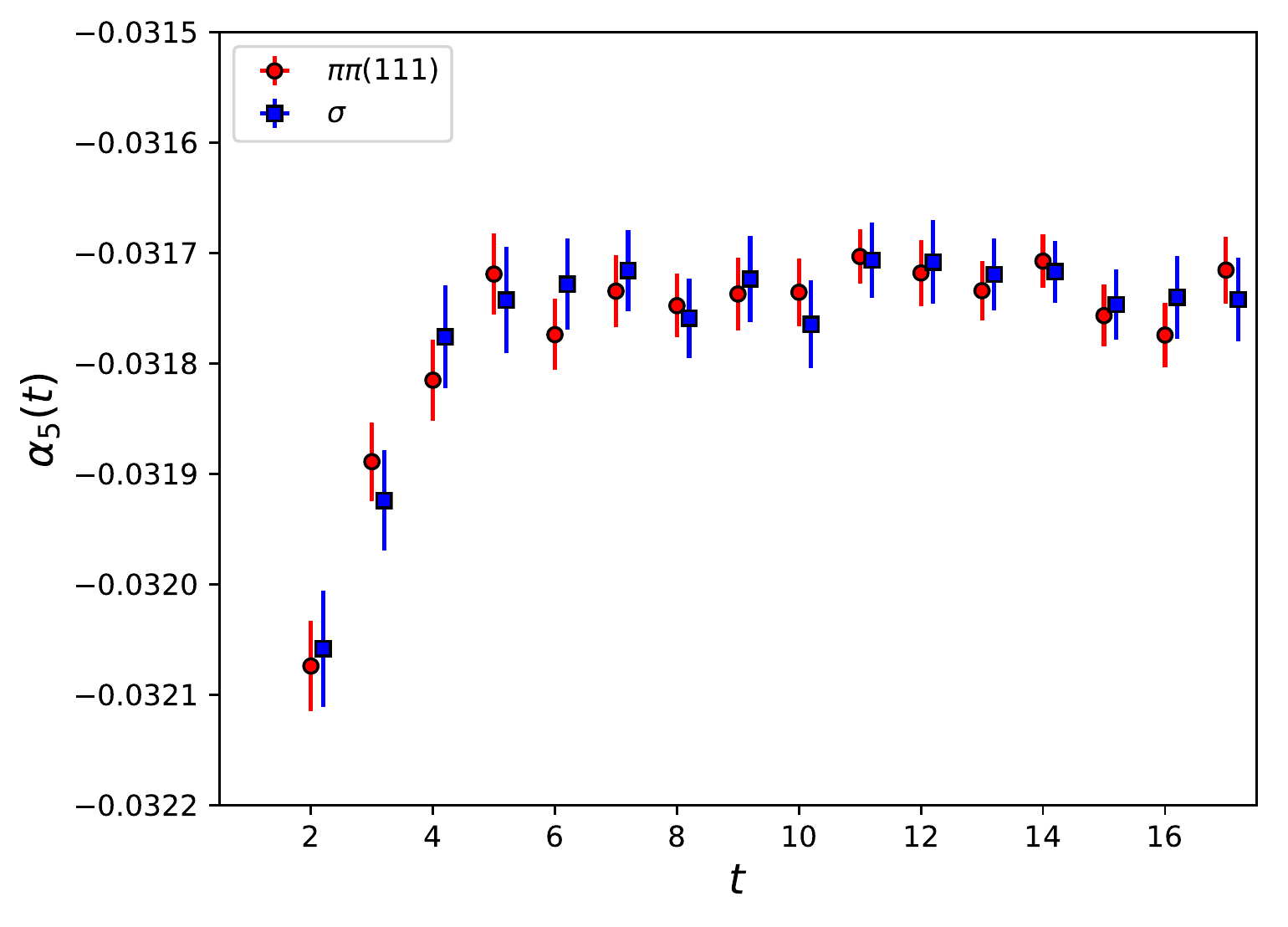}
\includegraphics[width=0.32\textwidth]{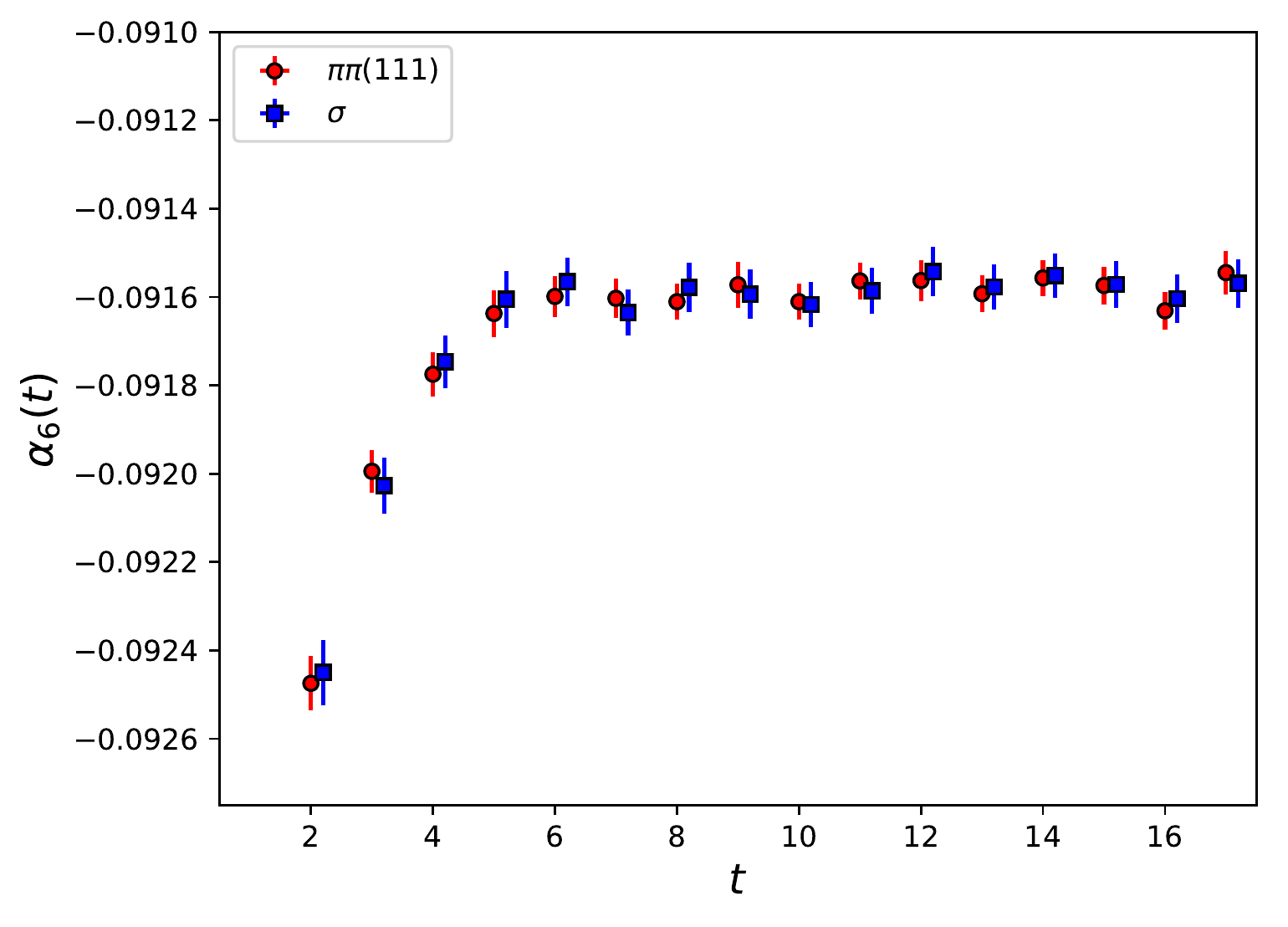}
\\
\includegraphics[width=0.32\textwidth]{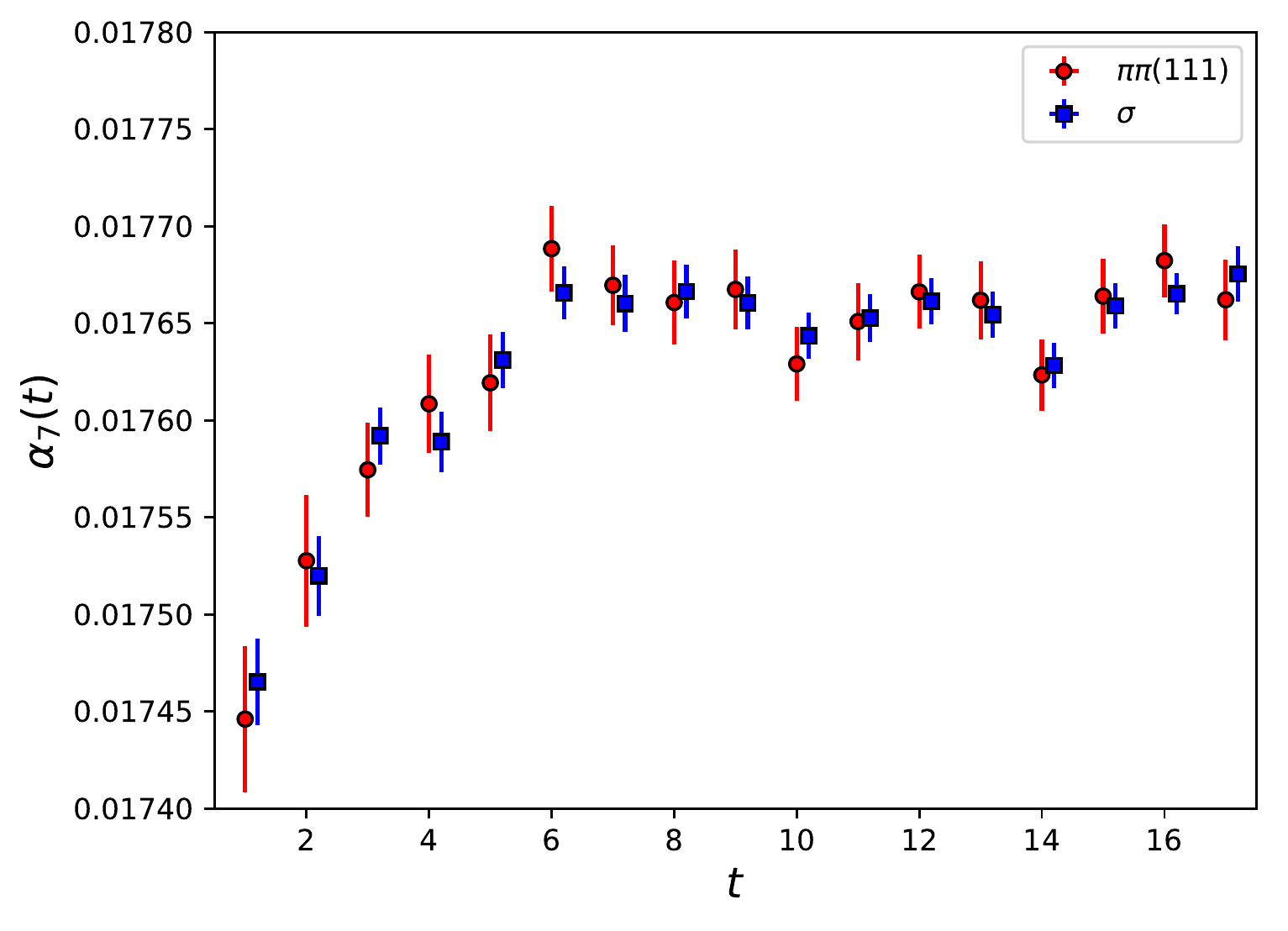}
\includegraphics[width=0.32\textwidth]{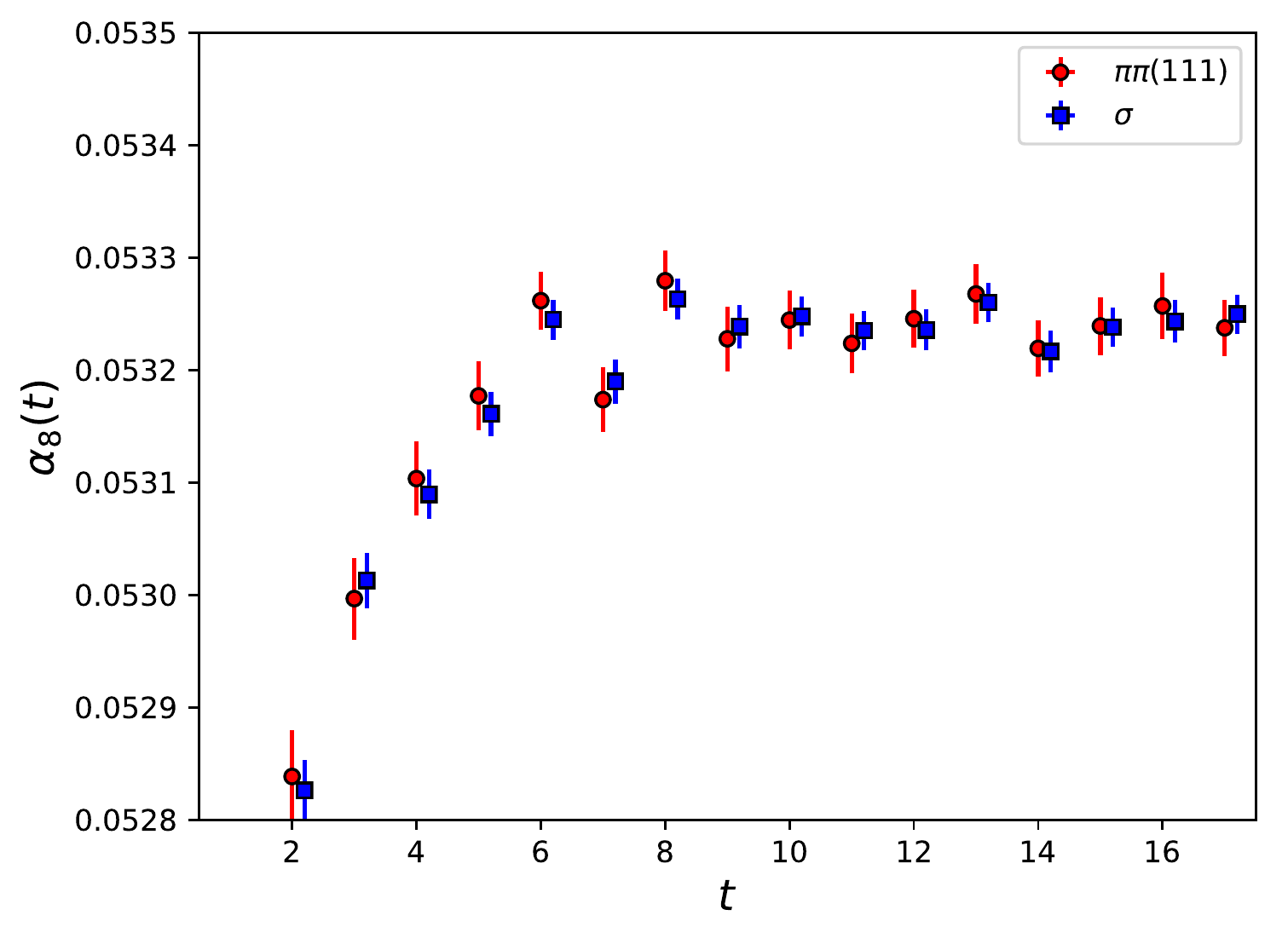}
\includegraphics[width=0.32\textwidth]{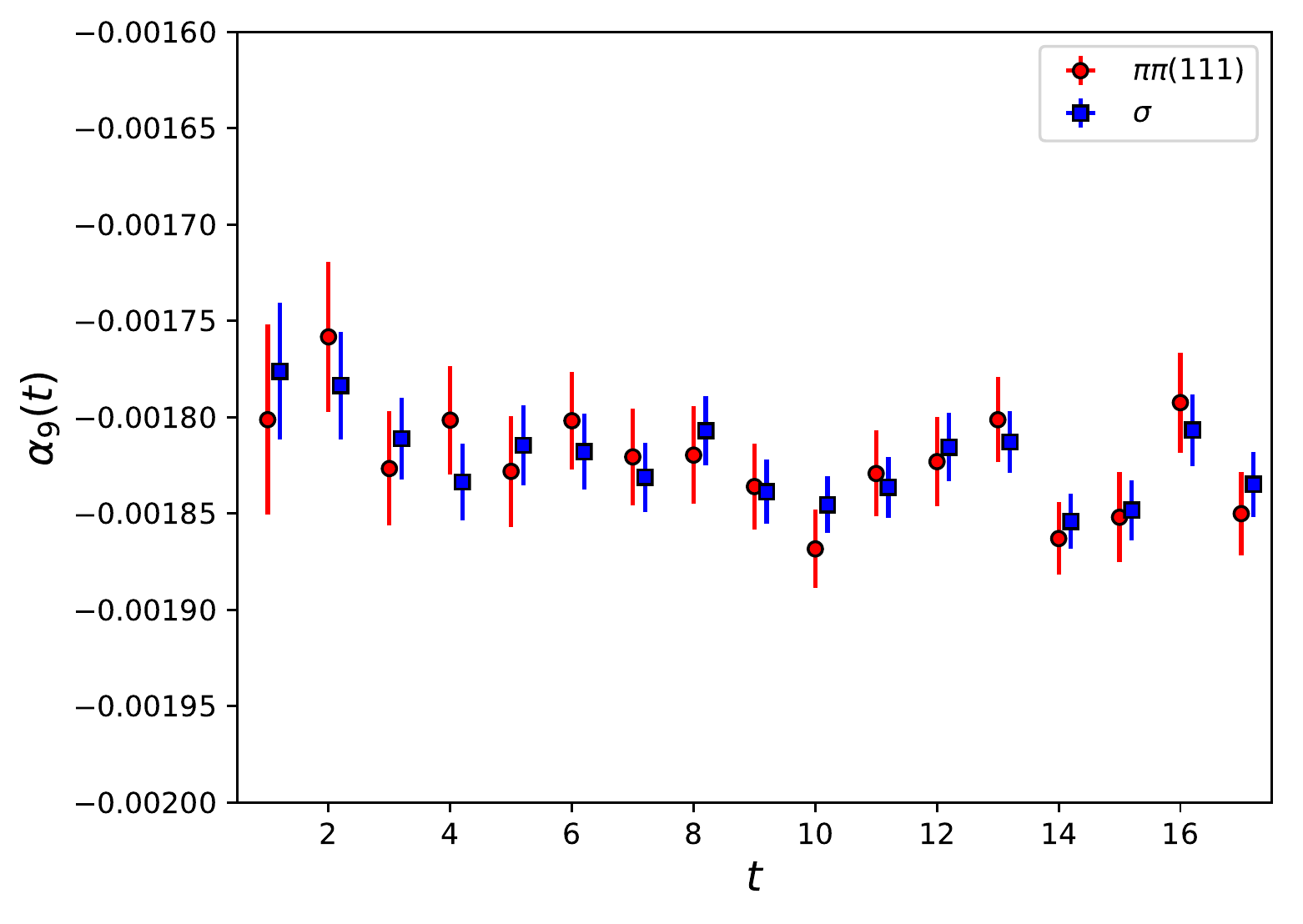}
\\
\includegraphics[width=0.32\textwidth]{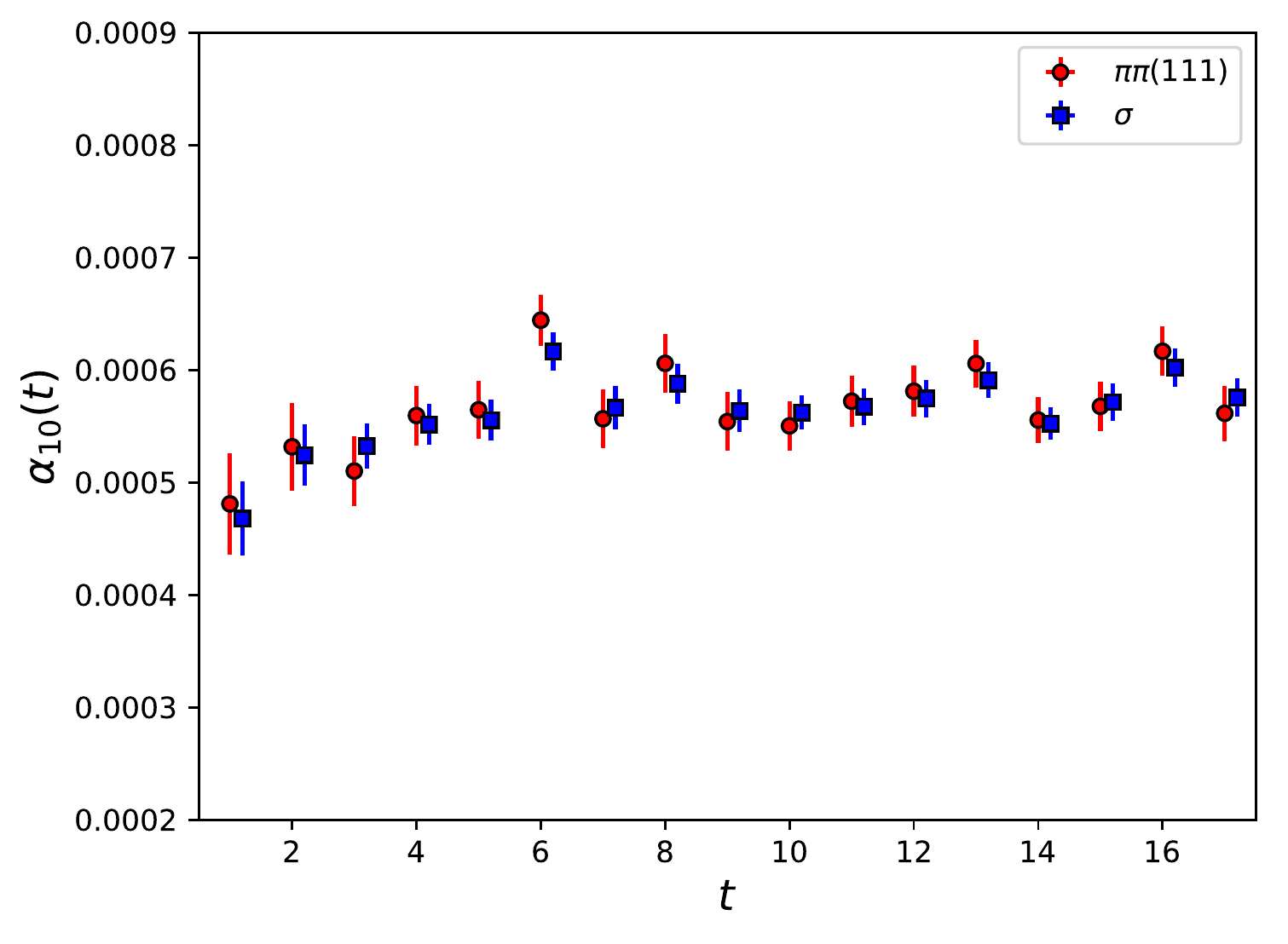}
\caption{The pseudoscalar subtraction coefficient $\alpha_i$ as a function of time for each of the ten operators in the following order: $\hat Q_1$-$\hat Q_3$ on the first line, $\hat Q_4$-$\hat Q_6$ on the second, $\hat Q_7$-$\hat Q_9$ on the third and $\hat Q_{10}$ on the fourth. Red circles denote data obtained in the basis of correlation functions used for the $\pi\pi(111)$ operator, and blue squares for the $\sigma$ sink operator. \label{fit-alpha} }
\end{figure}

In Fig.~\ref{fit-alpha} we plot the time dependence of the $\alpha_i$ for all ten operators. We observe excellent agreement between the results obtained from the two different bases of contractions as expected. For a number of operators we find statistically significant but relatively small excited-state contamination for small $t$ that in all cases appears to die away by $t=6$. While the effects of this contamination are unlikely to significantly affect our final results, the cuts that we later apply to our fits nevertheless exclude data with $t<6$.

\subsection{$ \langle \pmb{\pi\pi} | \pmb{\bar s\gamma^5 d} | \pmb{\tilde K^0}\rangle$  matrix elements}
\label{sec-pipiPK-matelem-results}

The $K\to\pi\pi$ matrix elements of the pseudoscalar operator $\bar s\gamma^5 d$ are required to perform the subtraction of the divergent loop contribution. In this section we independently analyze these matrix elements in order to understand their time dependence and the corresponding effect of the subtraction on the amount of excited state contamination in the final $K\to\pi\pi$ result. 

In the limit of large time separation between the source/sink operators and the four-quark operator, only the lowest-energy $\pi\pi$ and kaon states are present.  Since the pseudoscalar matrix elements vanish by the equations of motion when the decay conserves energy and the kaon and $\pi\pi$ ground-state energies in our calculation differ by only 2\%, we expect the subtraction to result in only a negligible shift in the central value but a marked improvement in the statistical errors in this limit. However at finite time separations, the contributions of the excited states may take a long time to die away due to the increasing magnitude of the corresponding matrix elements between initial and final states of different energies. It is this concern that prompts us to study this system more carefully.

The lattice three-point function
\begin{dmath}
C_P(t, t_{\rm sep}^{K\to{\rm snk}}) =\langle 0|{\cal O}_{\rm snk}^\dagger(t_{\rm sep}^{K\to{\rm snk}})\ [\bar s\gamma^5 d](t)\ {\cal O}_{\tilde K^0}(0)|0\rangle\label{eq-KPpipi-corr}
\end{dmath}
for a generic sink $\pi\pi$ operator,  ${\cal O}_{\rm snk}$, has the following time dependence:
\begin{dmath}
C_P(t, t_{\rm sep}^{K\to{\rm snk}}) = \sum_{ij} A^i_{\rm in} A^j_{\rm out} M_P^{ij} \exp\left(-E^i_{\rm in}t\right)\exp\left(-E^j_{\rm out}(t_{\rm sep}^{K\to{\rm snk}} - t)\right)\,,
\end{dmath}
where the subscript `in' refers to the incoming kaonic state, `out' to the outgoing two-pion state, and $M_P^{ij}$ is the matrix element for the term involving in and out states $i$ and $j$, respectively. It is convenient to define an ``effective matrix element'' by dividing out the ground-state time dependence and operator amplitudes,
\begin{dmath}
M_P^{\rm eff, snk}(t', t_{\rm sep}^{K\to{\rm snk}}) =M_P^{00} + \sum_{i,j\neq 0} A^{\prime\,i}_{\rm in} A^{\prime\,j}_{\rm out} M_P^{ij} \exp\left(-\Delta E^i_{\rm in}(t_{\rm sep}^{K\to{\rm snk}} - t')\right)\exp\left(-\Delta E^j_{\rm out}t'\right)\,, \label{eq-opPK-norm-tdep}
\end{dmath}
where 
\begin{dmath}
t' = t_{\rm sep}^{K\to{\rm snk}} - t
\end{dmath}
is the separation between the four-quark operator and the sink and
\begin{dgroup}
\begin{dmath}
A^{\prime\,i}_{\rm in/out} = A^i_{\rm in/out}/A^{0}_{\rm in/out}\,,
\end{dmath}
\begin{dmath}
\Delta E^i_{\rm in/out} = E^i_{\rm in/out} - E^0_{\rm in/out}\,.
\end{dmath}
\end{dgroup}
Note that $M_P^{\rm eff, snk}$ is dependent on the sink operator through the terms involving the excited states, in which a ratio of ground and excited state amplitudes appears.

\begin{figure}[t]
\includegraphics[width=0.48\textwidth]{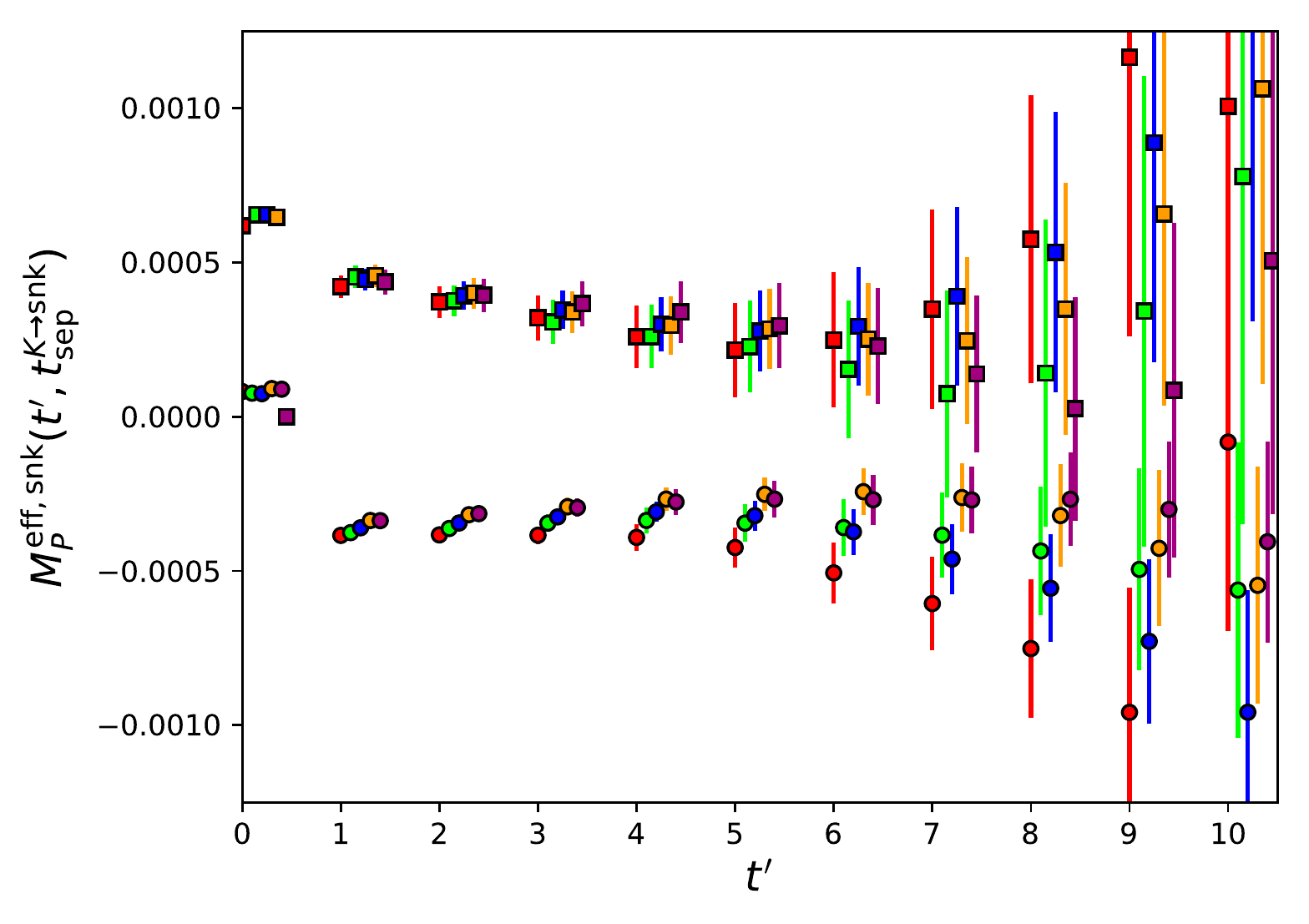}
\includegraphics[width=0.48\textwidth]{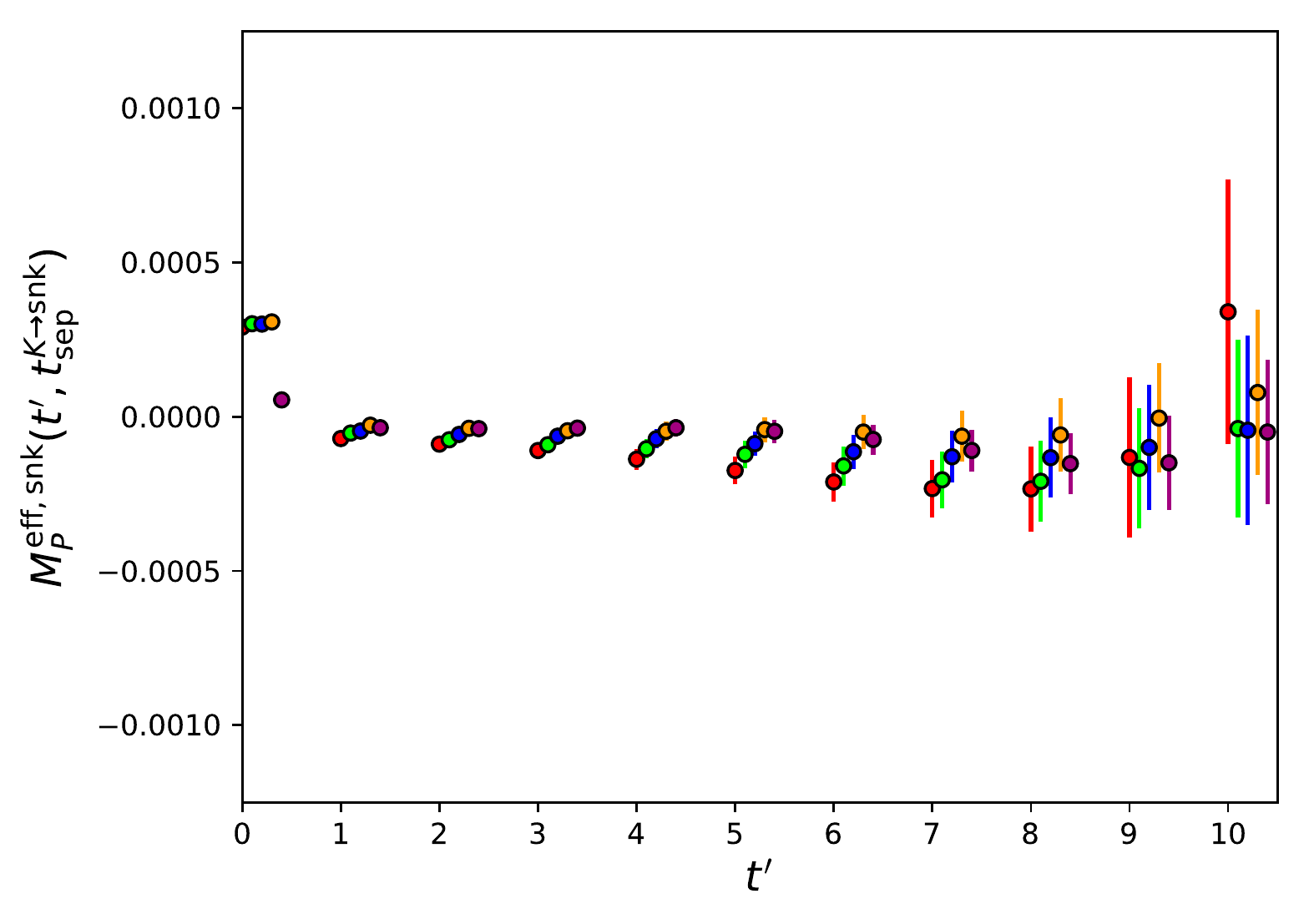}
\caption{The effective pseudoscalar matrix element $M_P^{\rm eff,snk}$ as a function of the time separation between the four-quark operator and the sink, $t'$. In the left pane we show the data for the $\pi\pi(111)$ operator (circles) and the $\sigma$ operator (squares) separately, and in the right pane we show the same for the optimal operator. Colored data correspond to the different $t_{\rm sep}^{K\to{\rm snk}}$ as follows: red (10), green (12), blue (14), orange (16) and mauve (18). The data for each of these different separations are staggered in order such that $t_{\rm sep}^{K\to{\rm snk}}=10$ is the left-most point of each cluster and $t_{\rm sep}^{K\to{\rm snk}}=18$ the right-most. \label{fig-opPK-for-top-snk} }
\end{figure}

We measure the correlation function Eq.~\eqref{eq-KPpipi-corr} for each of our three two-pion operators. Note that a vacuum subtraction is also required here and is performed in the same way as for the four-quark operators. In Fig.~\ref{fig-opPK-for-top-snk} we plot $M_P^{\rm eff, snk}$ for the $\pi\pi(111)$ and $\sigma$ operators for each of the five values of $t_{\rm sep}^{K\to{\rm snk}}$. The corresponding data for the $\pi\pi(311)$ operator is much noisier and has therefore been excluded. The form of this plot can be explained as follows: As $E^0_{\rm in} \approx E^0_{\rm out}$ we expect $M_P^{00}$ to be small. If we then assume that the dominant excited state contributions come from the term involving the excited kaon state and ground $\pi\pi$ state ($i=1,j=0$) and the term with the ground kaon state and the first excited $\pi\pi$ state ($i=0,j=1$), then we expect the data to behave as
\begin{dmath}
M_P^{\rm eff, snk}(t', t_{\rm sep}^{K\to{\rm snk}}) \approx A^{\prime\,1}_{\rm in} M_P^{10} \exp\left(-\Delta E^1_{\rm in}t_{\rm sep}^{K\to{\rm snk}}\right)\exp\left(+\Delta E^1_{\rm in}t'\right)
+ A^{\prime\,1}_{\rm out} M_P^{01} \exp\left(-\Delta E^1_{\rm out}t'\right)\,.
\end{dmath}
This ansatz then implies an exponentially falling contribution from the excited pion state and an exponentially growing piece from the excited kaon state, giving rise to a bowl-like shape assuming that $A^{\prime\,1}_{\rm in}$ and $A^{\prime\,1}_{\rm out}$ have the same sign, which appears to the the case here. Furthermore, the exponentially-growing piece in $t'$ is expected to be larger for smaller $t_{\rm sep}^{K\to{\rm snk}}$, and indeed we observe that the turnover point at which the exponentially-growing term begins to dominate occurs sooner for smaller $t_{\rm sep}^{K\to{\rm snk}}$. 

While the effective matrix elements of both sink operators initially trend towards zero, for the more precise $\pi\pi(111)$ data it seems that none of the five data sets are statistically consistent with zero at their maxima, suggesting we do not reach the limit of ground-state dominance. This is not necessarily an issue for our calculation given that the subtraction will heavily suppress these contributions in our final result, and furthermore the inclusion of multiple sink operators will improve our ability to extract the $\pi\pi$ ground-state matrix element. In order to disentangle these two effects it is convenient to examine the three-point function for the optimized sink operator discussed in Sec.~\ref{sec-optimal-pipi-op}. The time dependence of $M_P^{\rm eff, snk}$ for this operator is also shown in Fig.~\ref{fig-opPK-for-top-snk}. By definition this operator heavily suppresses $A^{\prime\,j}_{\rm out}$ for $j > 0$, and indeed we find the data to be much flatter in the low-$t'$ region and also considerably closer to zero. The exponential growth and $t_{\rm sep}^{K\to{\rm snk}}$ dependence that enters due to the excited kaon term is expected to be largely unaffected by this transformation, however it seems that in several cases the plateaus extend much further into the large-$t'$ region than previously. It is likely that is due to an accidental cancellation owing to the fact that $A^{\prime\,1}_{\rm out}$ is positive for the $\pi\pi(111)$ operator and negative for the $\sigma$ operator (cf. Tab.~\ref{tab-pipi2ptfit-both}) and hence the exponentially-growing terms for these operators have opposite signs.

We conclude by discussing the expected size of the excited-state contamination in the matrix elements of the subtracted four-quark operators arising from the pseudoscalar operator. In the $K\to\pi\pi$ calculation, this dimension-3 operator is introduced to remove what in the continuum limit would be a quadratic divergence resulting from the self-contraction between two of the four quark operators appearing in those operators $\hat Q_i$ with a component transforming in the $(8,1)$ or $(8,8)$ representations of $SU(3)_L\times SU(3)_R$.  In our lattice calculation these terms behave as $1/a^2$ when expressed in physical units.  To leading order in $a$ this $1/a^2$ coefficient does not depend on the external states and is therefore removed from our $\langle 0|(\pi\pi)\hat Q_i \tilde K^0|0\rangle$ amplitude by the subtraction defined above, even though the coefficients $\alpha_i$ are determined from the $\langle 0|\hat Q_i \tilde K^0|0\rangle$ matrix element in Eq.~\eqref{eq:alpha-determine}.  Because of the chiral structure of the $(8,1)$ and $(8,8)$ operators, these coefficients have the structure: $\alpha_i {\sim}\frac{m_s-m_d}{a^2} + \ldots$~\cite{Blum:2001xb}, where the ellipsis represents terms which are not power-divergent. 

Thus, the $\bar s\gamma^5 d$ subtraction removes the leading $1/a^2$ term in the matrix element of $\hat Q_i$, leaving behind a finite piece of size ${\sim}(m_s-m_d)\Lambda_{\rm QCD}^2\bar s\gamma^5 d$.  This remainder is not physical and depends on the condition chosen to define the $\alpha_i$.  However, it will contribute to our final result if $E_{\pi\pi} \ne m_K$.  For the ground-state component ($i=0$,$j=0$) this term is thus heavily suppressed by the factor $(E^0_{\pi\pi}-m_K)$. However for the excited states we expect this piece to be on the order of the physical contribution from the dimension-6 four-quark operator.  
 As such it may result in a modest enhancement of the excited state matrix elements.  Providing we are able to demonstrate that we have the excited $\pi\pi$ and kaon states under control through appropriate cuts on our fitting ranges, this should pose no obstacle to our calculation.

\subsection{Description of fitting strategy}

For a lattice of sufficiently large time extent that around-the-world terms in which states propagate through the lattice temporal boundary can be neglected, and assuming that the four-quark operator is sufficiently separated from the kaon source that the kaon ground state is dominant, the three-point Green's functions $C_i$ of the weak effective operators defined in Eq.~\eqref{eq-3ptfuncdef} have the general form,
\begin{equation}
C_i(t,t_{\rm sep}^{K\to\snk})  = \sum_j \frac{1}{\sqrt{2}} A_K A^j_\snk e^{-m_K t} M_i^j e^{-E_j(t_{\rm sep}^{K\to\snk} -t)}\,, \label{eq-fitform}
\end{equation}
where $M_i^j=\langle(\pi\pi)^j|Q_i|K^0\rangle$ is the matrix element of the four-quark operator $Q_i$ with the $\pi\pi$ state $j$, with $M_i^0$ corresponding to the physical $K\to\pi\pi$ matrix elements required to compute $A_0$. The factor of $1/\sqrt{2}$ relates the matrix element involving the kaon G-parity eigenstate to that of the physical kaon~\cite{Christ:2019sah}. Here $A_K$ is the amplitude of the G-parity kaon operator, $A^j_\snk$ are the amplitudes of the sink operator with the state $j$, and $E_j$ is the energy of that state. These parameters are fixed to those obtained from the two-point function fits in Sec.~\ref{sec:TwoPointResults}: $A_K$ and $m_K$ to the results given in Tab.~\ref{tab-pik2ptfit}, and $A^j_\snk$ and $E_j$ to the results obtained from the three-operator, two-state $\pi\pi$ fits given in the second column of Tab.~\ref{tab-pipi2ptfit-both}.

We perform simultaneous correlated fits over multiple sink operators to the form Eq.~\eqref{eq-fitform} in order to determine the matrix elements $M_i^j$, allowing for one or more states $j$.  Independent one-state fits are also performed to the optimized sink operator defined in Sec.~\ref{sec-optimal-pipi-op}. The fits are performed to each weak effective operator separately, in the 10-operator basis (the relationship between these 10 linearly-dependent operators serves as a useful cross-check of the fit results) using the strategy outlined in Sec.~\ref{sec-fitstrategy}. We apply a cut $t_{\rm min}$ on the separation $t$ between the kaon and the four-quark operator in order to isolate the ground-state kaon, and also a cut $t'_{\rm min}$ on the separation $t'=t_{\rm sep}^{K\to\snk} -t$ between the four-quark and sink operators. These cuts, the number of sink operators, and the number of excited $\pi\pi$ states included in the fit are varied in order to study systematic effects. 

For use below we again define an ``effective matrix element'' in which the ground-state $\pi\pi$ and kaon amplitudes and time dependence are multiplied out,
\begin{dmath}
M_i^{\rm eff, snk}(t') = C_i(t,t_{\rm sep}^{K\to\snk})  \left( \frac{1}{\sqrt{2}} A_K A^0_\snk e^{-m_K t}e^{-E_0(t_{\rm sep}^{K\to\snk} -t)}\right)^{-1}
=  M_i^0 + \sum_j \frac{A^j_\snk}{A^0_{\rm snk}} M_i^j e^{-(E_j-E_0)t'}\,. \label{eq-Meffkpipi}
\end{dmath}
These effective matrix elements converge exponentially to the ground-state matrix element at large $t'$. Note that, unlike in Sec.~\ref{sec-pipiPK-matelem-results}, we are assuming that a cut, $t_{\rm min}$, on the separation between the kaon and four-quark operators has been applied that is sufficient to isolate the contribution of the kaon ground state. As a result, these effective matrix elements can be assumed to be independent of $t_{\rm sep}^{K\to\snk}$ and a weighted average of our five datasets of different $t_{\rm sep}^{K\to\snk}$ can be applied to improve the statistical resolution of the data presented in our plots.

\subsection{Fit results}
\label{sec-fitresults}

\begin{figure}[tbp]
\includegraphics[width=0.48\textwidth, trim={0 0 0 0.3cm}, clip]{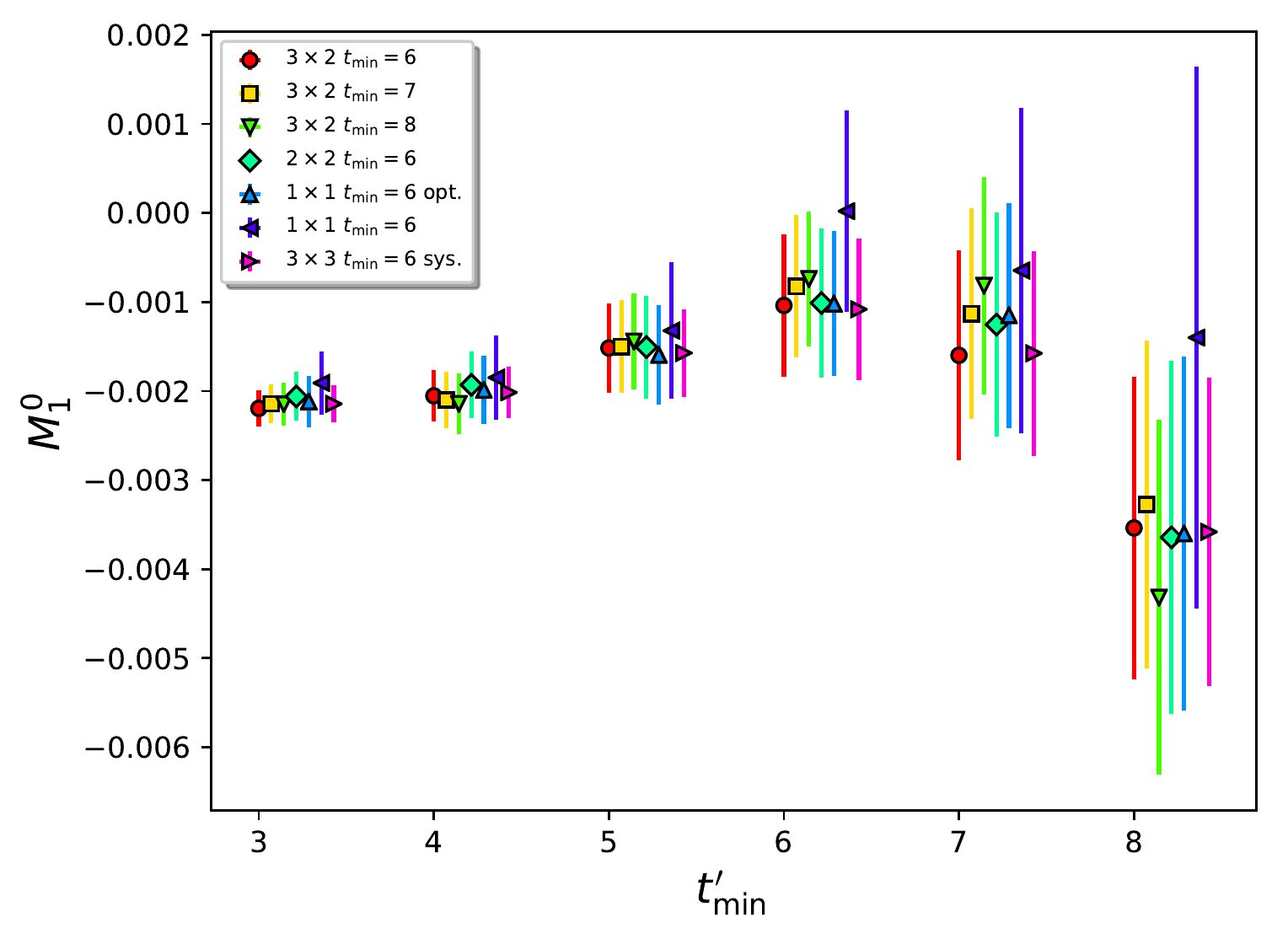}
\includegraphics[width=0.48\textwidth, trim={0 0 0 0.3cm}, clip]{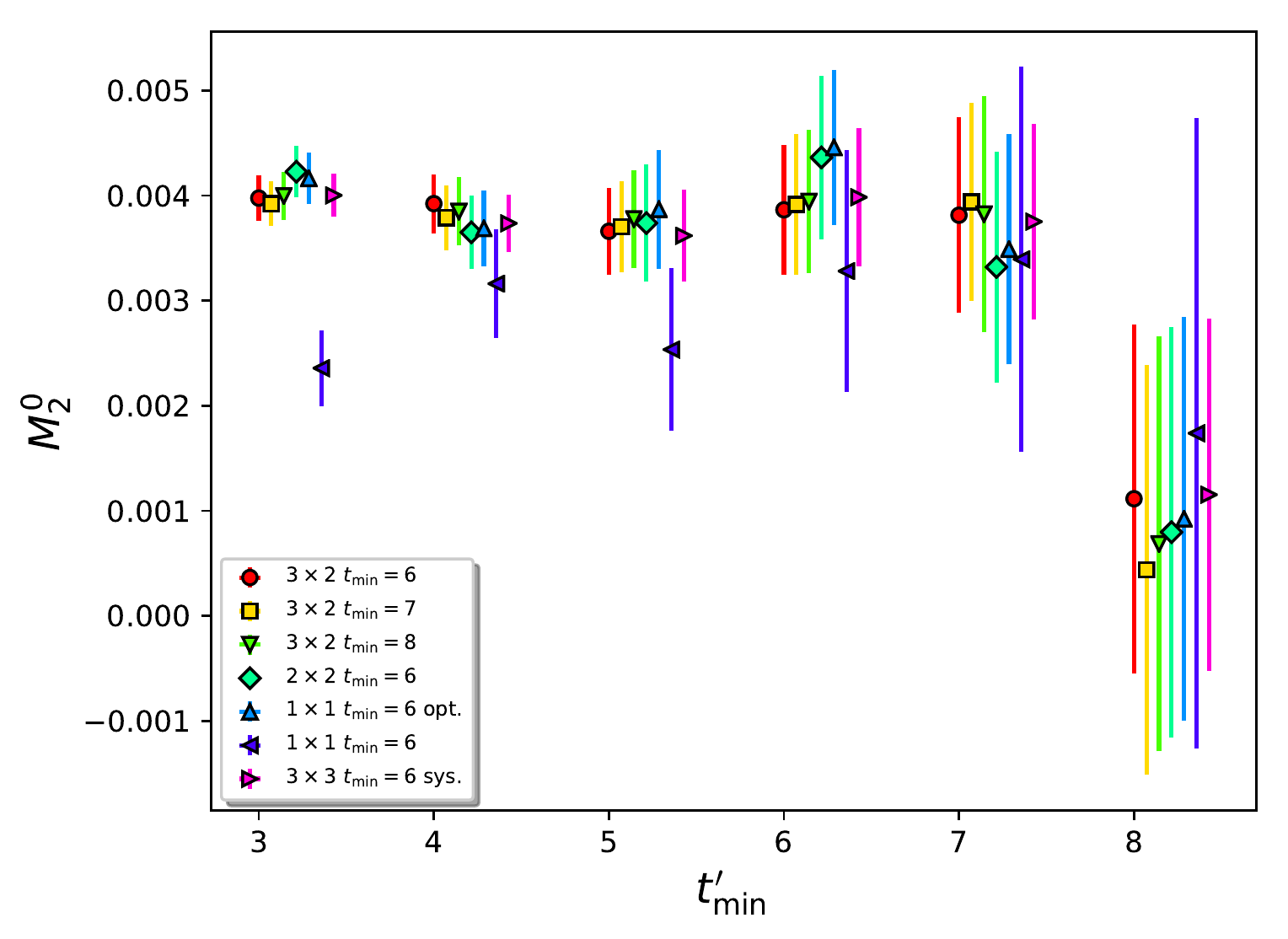} \\
\includegraphics[width=0.48\textwidth, trim={0 0 0 0.3cm}, clip]{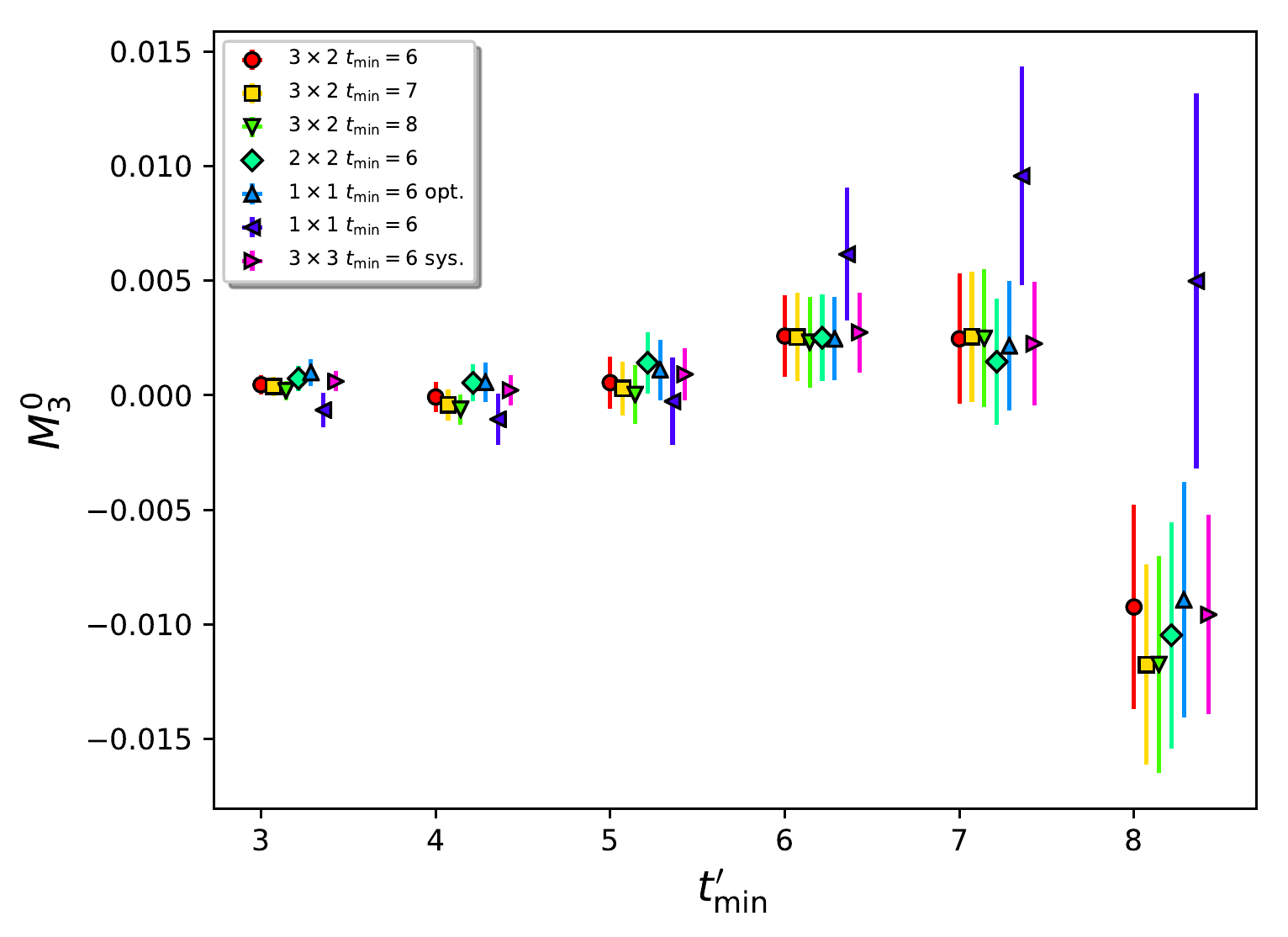}
\includegraphics[width=0.48\textwidth, trim={0 0 0 0.3cm}, clip]{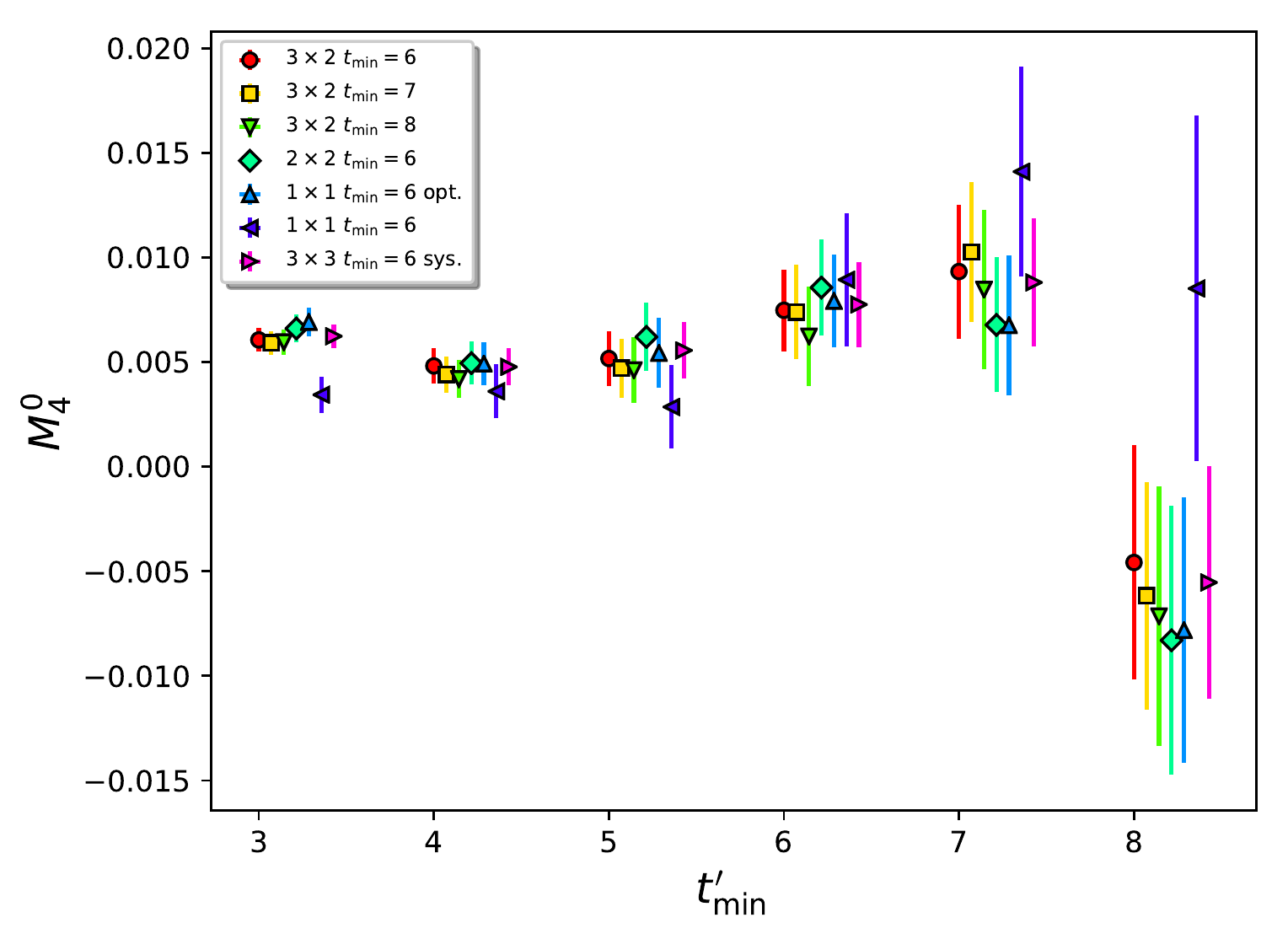} \\
\includegraphics[width=0.48\textwidth, trim={0 0 0 0.3cm}, clip]{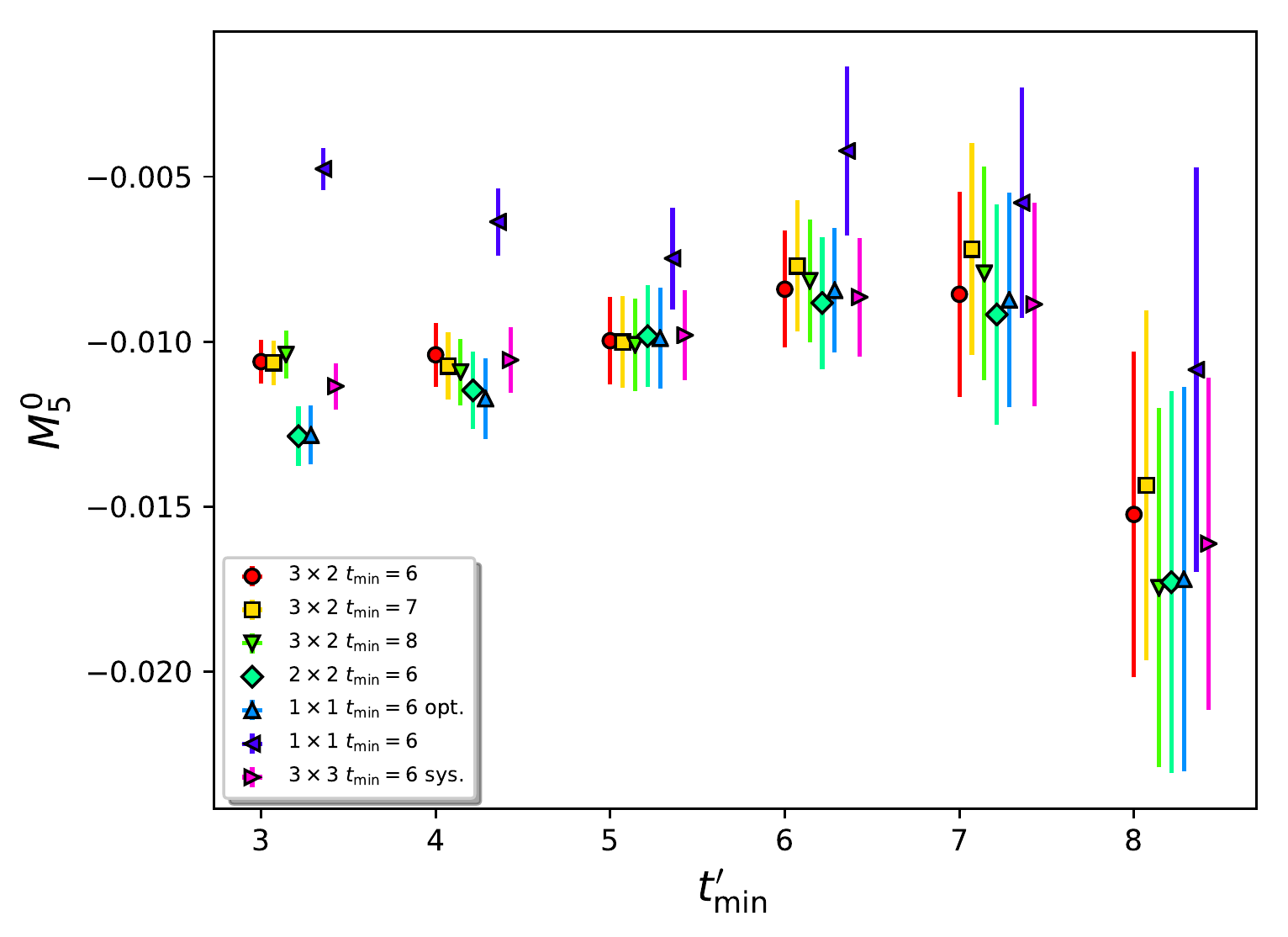}
\includegraphics[width=0.48\textwidth, trim={0 0 0 0.3cm}, clip]{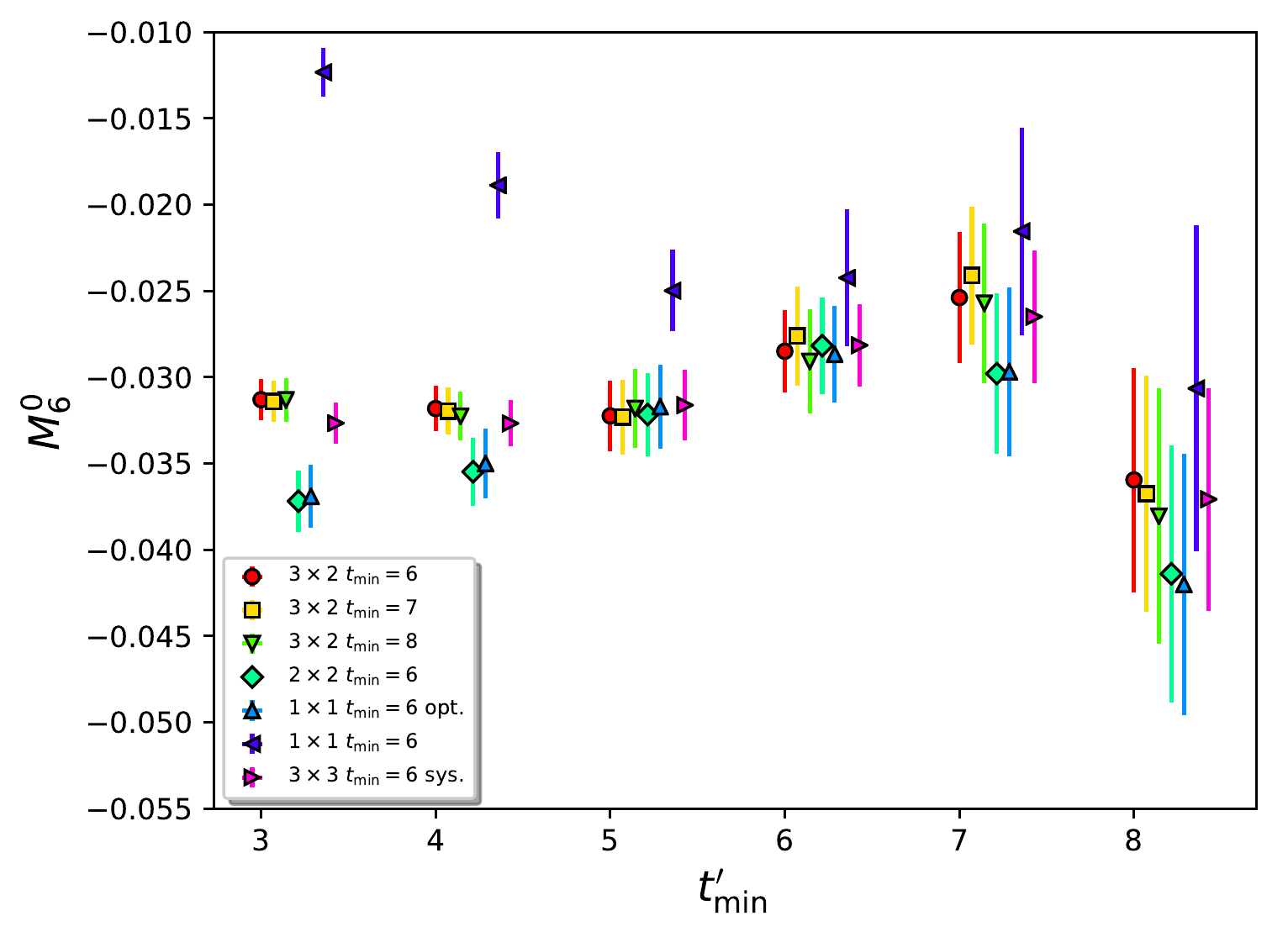}
\caption{Fit results in lattice units for the $K\to\pi\pi$ ground-state matrix elements $M_1^0-M_6^0$ as a function of $t'_{\rm min}$, the minimum time separation between the four-quark and sink operators that is included in the fit. The results have been shifted along the x-axis for clarity in order of their appearance in the legend. The legends are given in the format \#ops $\times$ \#states followed by the cut $t_{\rm min}$ on the time separation between the kaon and the four-quark operators. Here ``opt.'' is the fit to the optimal operator and ``sys.'' is used to estimate the systematic error resulting from a third state. \label{fig-matelemQ1toQ6} }
\end{figure}

\begin{figure}[tbp]
\includegraphics[width=0.48\textwidth, trim={0 0 0 0.3cm}, clip]{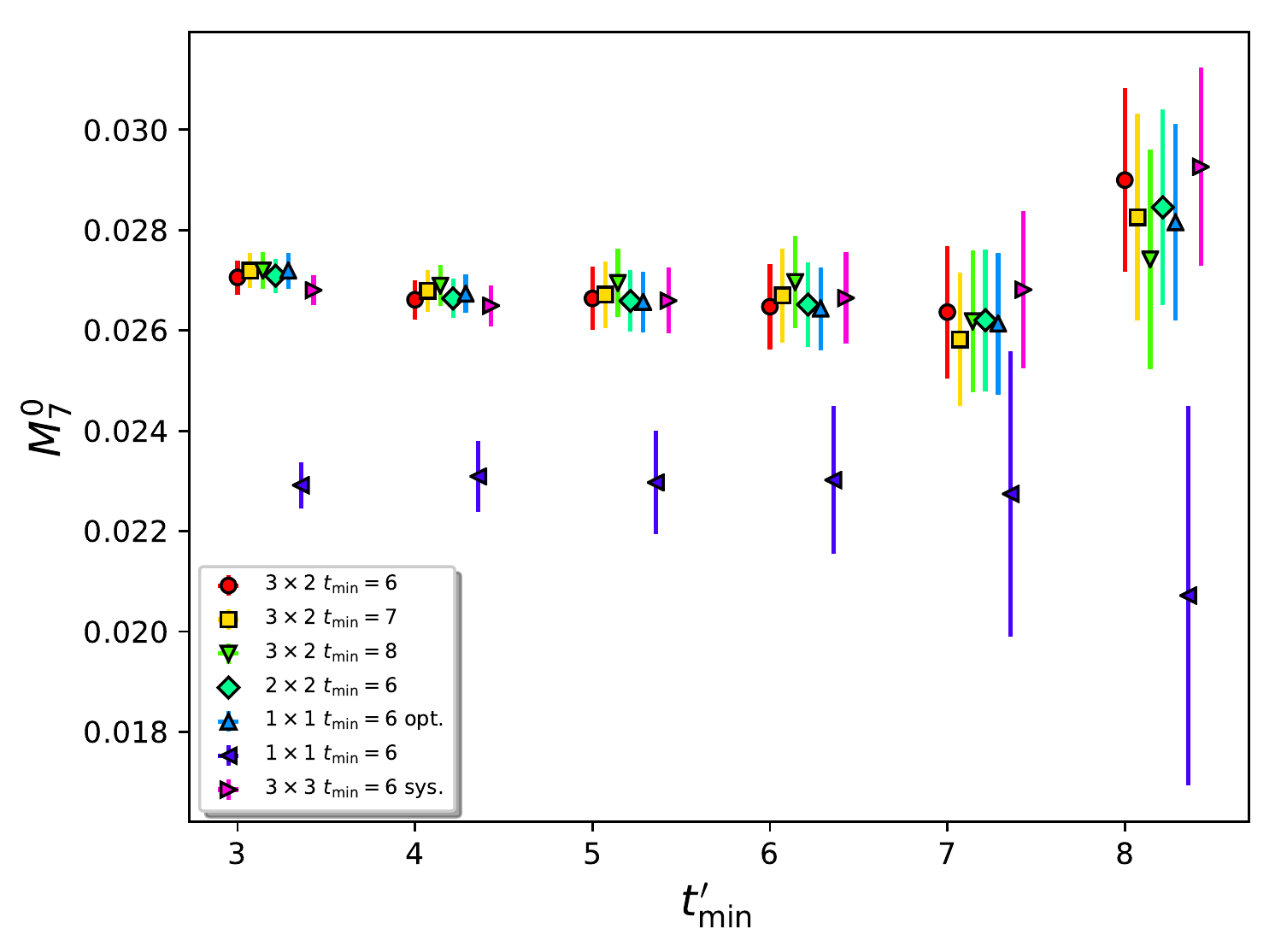}
\includegraphics[width=0.48\textwidth, trim={0 0 0 0.3cm}, clip]{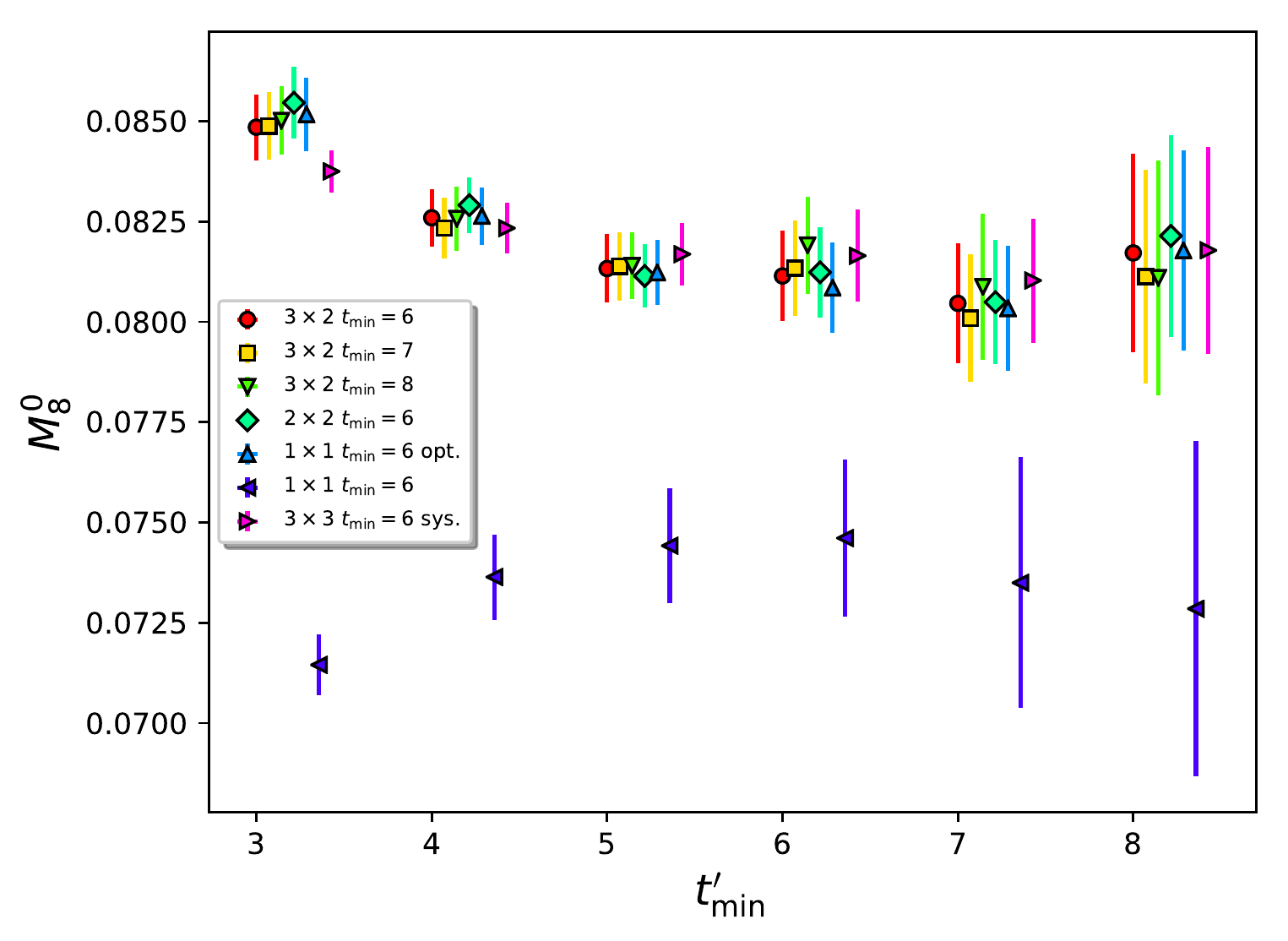} \\
\includegraphics[width=0.48\textwidth, trim={0 0 0 0.3cm}, clip]{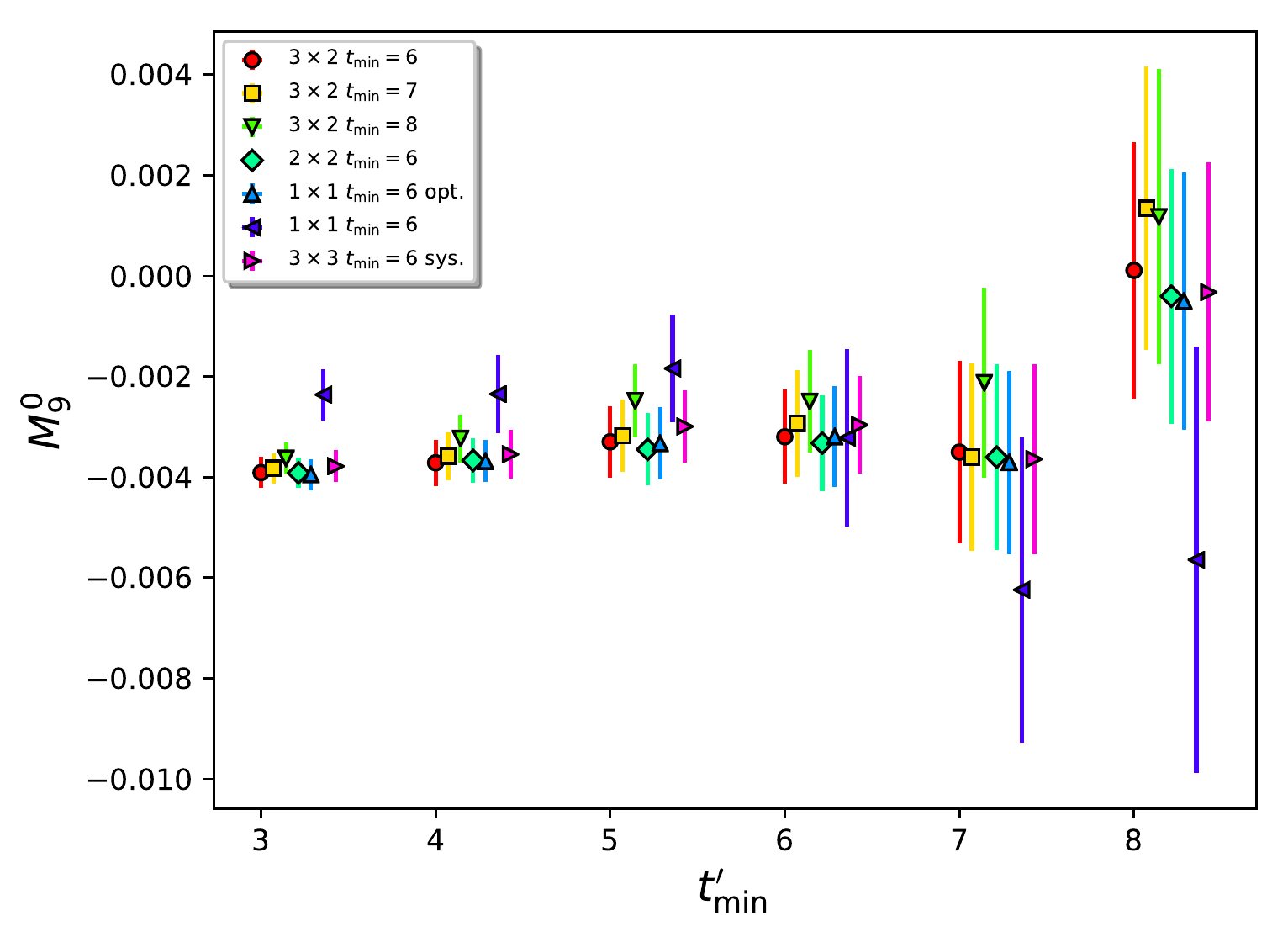}
\includegraphics[width=0.48\textwidth, trim={0 0 0 0.3cm}, clip]{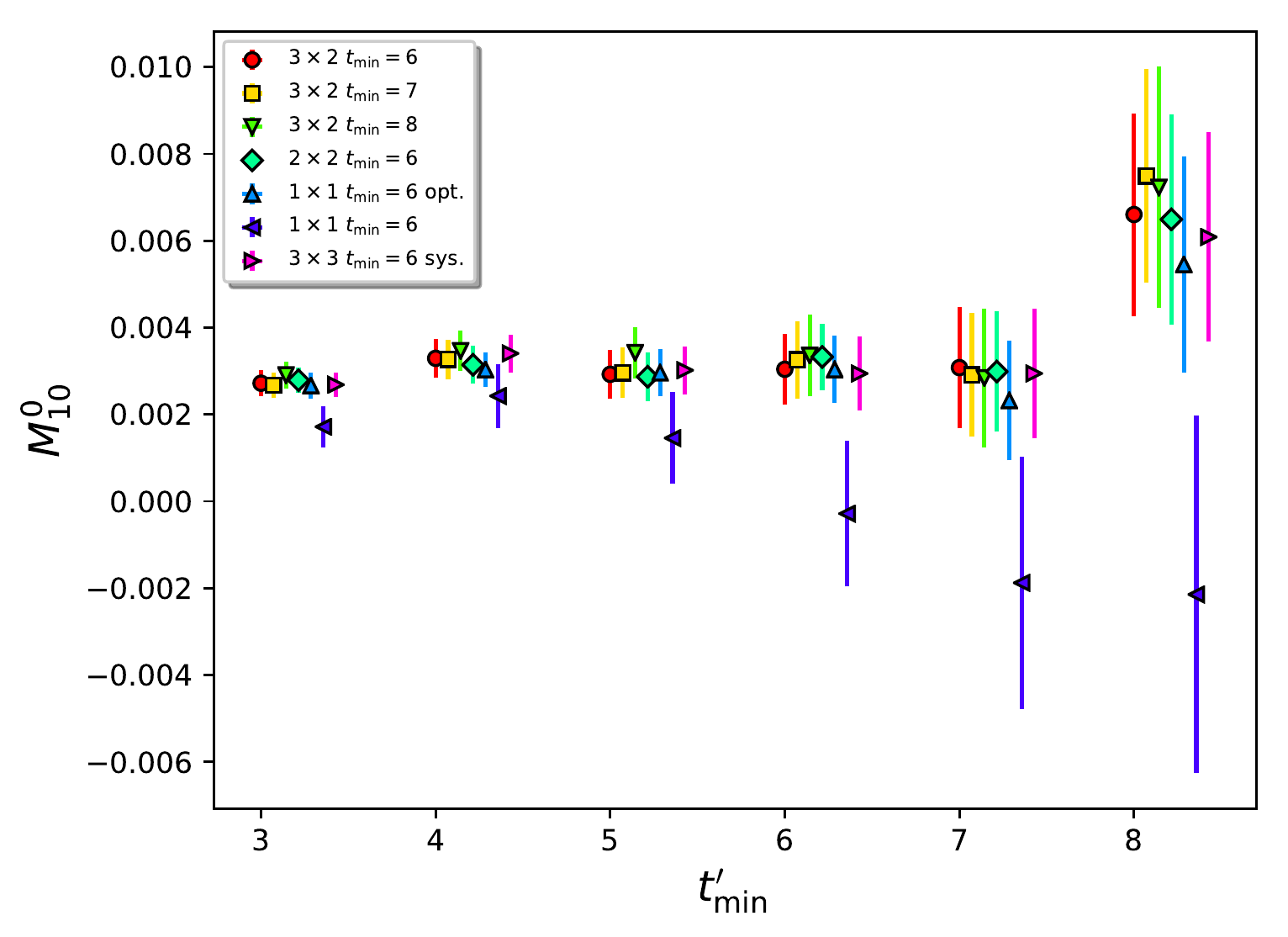}
\caption{The extension of Fig.~\ref{fig-matelemQ1toQ6} to the ground-state matrix elements $M_7^0-M_{10}^0$.
\label{fig-matelemQ7toQ10} }
\end{figure}

In this section we examine the results of fitting various subsets of our data, with the goal of finding an optimal fit window in which systematic errors arising from both excited $\pi\pi$ and kaon states are minimized.

In Figs.~\ref{fig-matelemQ1toQ6} and~\ref{fig-matelemQ7toQ10} we plot the fitted ground-state matrix elements $M_i^0$ as a function of $t'_{\rm min}$ for various choices of $t_{\rm min}$, the number of sink operators and the number of states. The three-operator fits are performed using the $\pi\pi(311)$, $\pi\pi(111)$ and $\sigma$ sink operators; for the two-operator fits we drop the noisier $\pi\pi(311)$ data; and for the one-operator fits we further drop the $\sigma$ data. The one-operator, one-state fits are equivalent to those performed in our 2015 work, albeit with more statistics and more reliable $\pi\pi$ energies and amplitudes.

The discussion below will be focused on these figures. We will first discuss general features addressing the quality of the data and the reliability of the fits, and will then concentrate on searching for evidence of systematic effects (or lack thereof) arising from kaon and $\pi\pi$ excited states. Based on those conclusions we will then present our final fit results.

\subsubsection{Discussion of data and fit reliability}

\begin{figure}[tb]
\includegraphics[width=0.48\textwidth, trim={0 0 0 0.3cm}, clip]{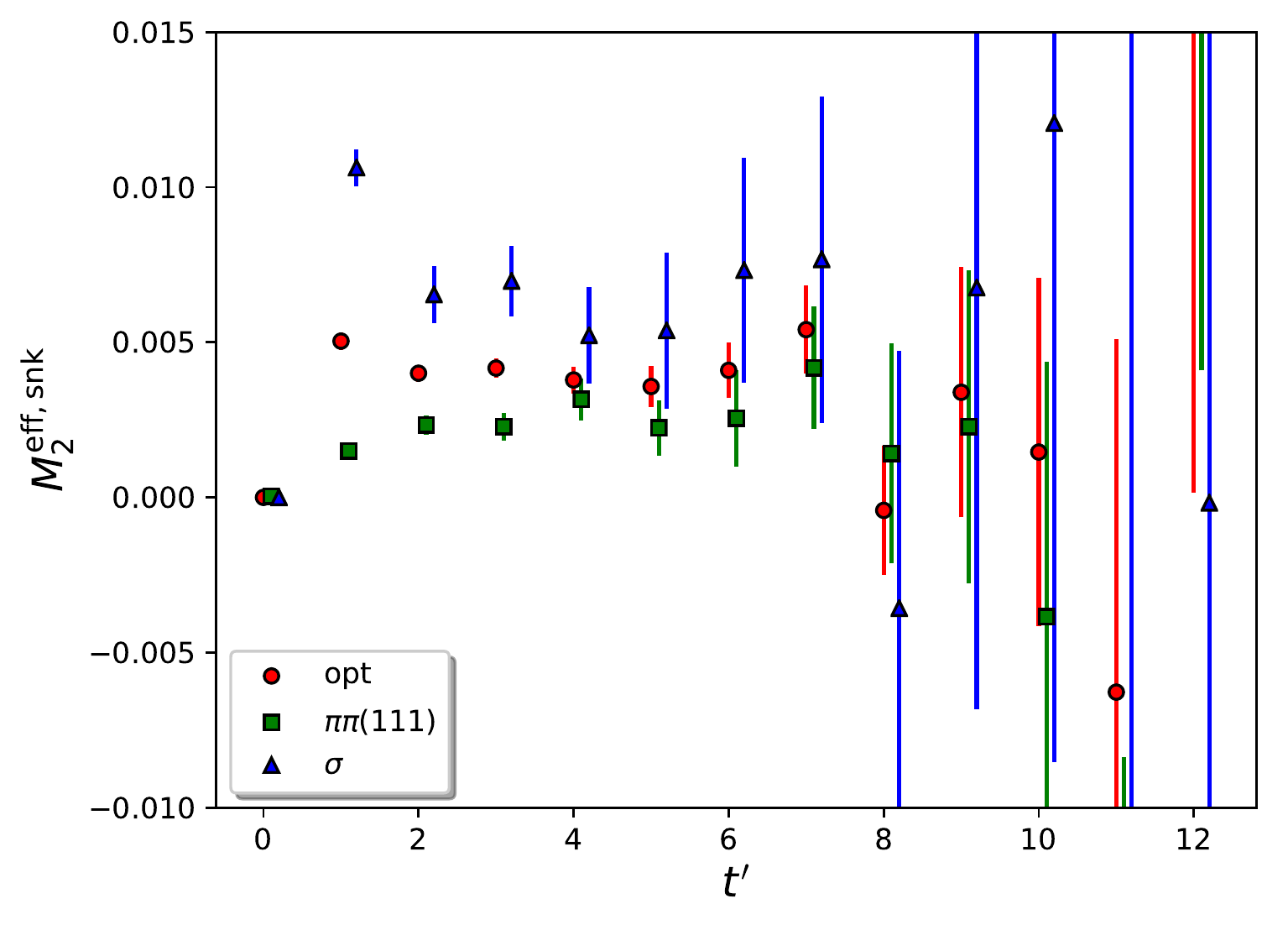}
\includegraphics[width=0.48\textwidth, trim={0 0 0 0.3cm}, clip]{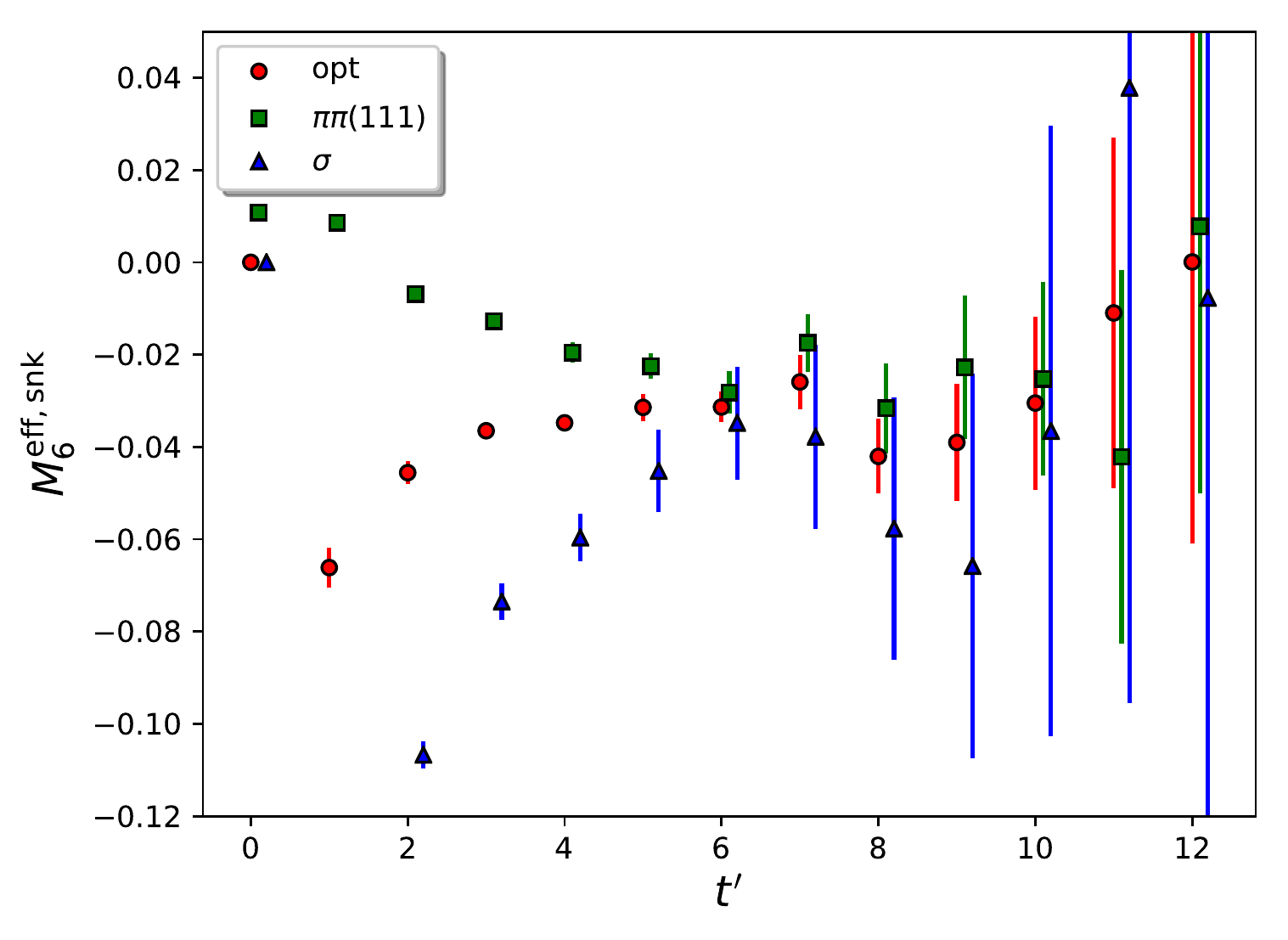}
\caption{The effective matrix elements $M^{\rm eff, snk}_2$ (left) and $M^{\rm eff, snk}_6$ (right) for the $\pi\pi(111)$ and $\sigma$ sink operators and the two-operator two-state optimal sink operator (labeled ``opt.'' here), plotted as a function of $t'$. The error-weighted average has been applied to the five different $K\to{\rm sink}$ separations subject to a cut of $t_{\rm min}=6$. \label{fig-Q2Q6compareoptop} }
\end{figure}

We will first comment on the fits to the optimal operator, labeled ``opt.'' in the figures. This approach is outwardly advantageous in that the fits are performed to a single state and the covariance matrix is considerably smaller. In Fig.~\ref{fig-Q2Q6compareoptop} we compare the $t'$ dependence of the $M_2^{\rm eff,snk}$ and $M_6^{\rm eff,snk}$ effective matrix elements of this optimal operator to that of the $\pi\pi(111)$ and $\sigma$ operators alone, where we note a marked improvement in the quality of the plateau. This behavior, which is also accounted for implicitly in the multi-state fits, demonstrates the power of the multi-operator technique for isolating the ground state. In Figs.~\ref{fig-matelemQ1toQ6} and~\ref{fig-matelemQ7toQ10} we observe that the fit results for the optimal operator agree very well with the multi-state fit results in all cases. While this approach does not appear to offer any statistical advantage, the strong agreement suggests that our complex multi-state correlated fits are under good control.

\begin{table}[tb]
\centering
\begin{tabular}{c|c||c|c}
\hline\hline
%
%
$i$ & P-value & $i$ & P-value \\
\hline
1 & 0.314 & 6 & 0.446 \\
2 & 0.737 & 7 & 0.843 \\
3 & 0.02  & 8 & 0.88 \\
4 & 0.123 & 9 & 0.581 \\
5 & 0.421 & 10 & 0.545
\end{tabular}
\caption{The p-values assessing how well the data with $t'\geq 7$ is described by the model for the $C_i$ correlation functions obtained by fitting to 3 operators and 2 states with $t'_{\rm min}=5$ and $t_{\rm min}=6$.\label{tab-tmin7pvaluetest} }
\end{table}

In Figs.~\ref{fig-matelemQ1toQ6} and~\ref{fig-matelemQ7toQ10} we observe for several ground-state matrix elements a trend in the fit results up to an extremum at $t'_{\rm min}=7$, followed by a statistically significant correction at the level of 1-2$\sigma$ for the fits with $t'_{\rm min}=8$. In this and Sec.~\ref{sec-syserr-excstate} we present substantial evidence that the systematic errors resulting from excited kaon and $\pi\pi$ states are minimal, which makes it unlikely that this rise is associated with excited state contamination. Certainly if it were due to excited $\pi\pi$ states we would expect an improvement as more sink operators are added, but there is little evidence of such, and likewise if excited kaon states were the cause we would expect an improvement as we increase the $t_{\rm min}$ cut, whereas no significant change is observed. The most likely explanation is a statistical fluctuation in our correlated data set, and indeed in Fig.~\ref{fig-Q2Q6compareoptop} we see evidence of such a fluctation peaking at $t'=7$ which is likely driving this phenomenon. 

Given the above, an interesting question we can ask is whether the models we obtain from our fits with $t'_{\rm min}=5$, which in all cases lie within the plateau region before this rise, are a good description of the subset of data with $t'\geq 7$, or in other words how likely it is that these data are consistent with this model allowing only for statistical fluctuations. In Tab.~\ref{tab-tmin7pvaluetest} we list the p-values for these data using the model obtained by fitting to 3 sink operators and 2 states with $t'_{\rm min}=5$ and $t_{\rm min}=6$, computed using the technique discussed in Sec.~\ref{sec-fitstrategy} (with no free parameters). We observe excellent p-values in all cases bar $M^0_3$, and to a lesser extent $M^0_4$. The lower p-values for these operators are common for all of the multi-operator fits and are likely associated with the statistical fluctuations described above which are more apparent for these matrix elements (cf. Fig.~\ref{fig-matelemQ1toQ6}). We expect that such unusual statistical fluctuations will be found when so many different operators and fitting ranges are examined.  Of most importance in a calculation of Im$(A_0)$ is $M^0_6$, for which we find that the model obtained with $t'_{\rm min}=5$ is an excellent description of the data with $t'\geq 7$. The p-value is in fact little different from the value $p=0.525$ obtained by fitting to these data directly, suggesting that the models are equally good descriptions despite the tension in the ground-state matrix elements.

\begin{figure}[tb]
\centering
\includegraphics[width=0.48\textwidth]{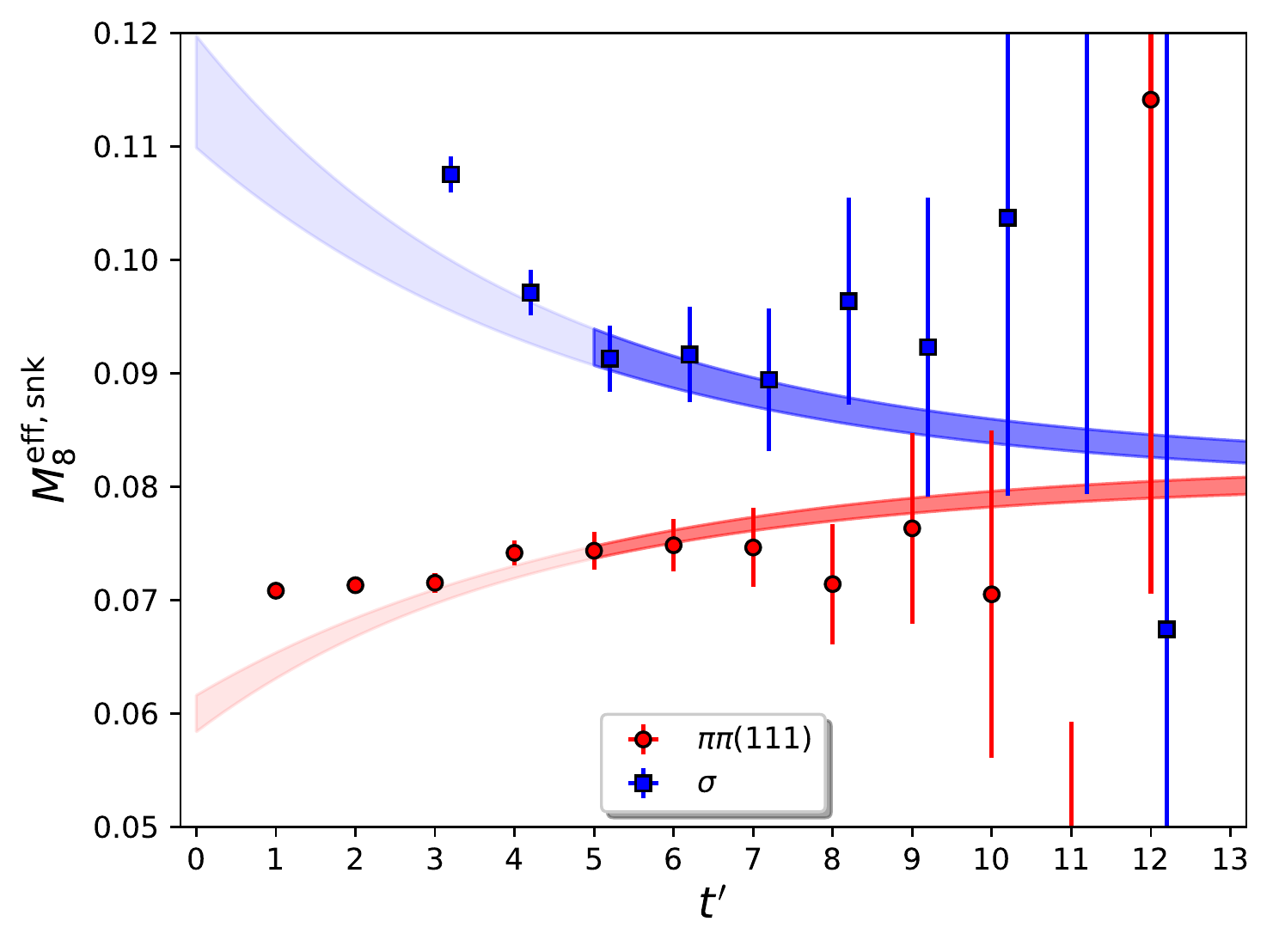}
\caption{The $M_8^{\rm eff, snk}$ effective matrix element for the $\pi\pi(111)$ (red circles) and $\sigma$ (blue squares) sink operators overlaid by curves showing the function $M^{\rm eff,snk}_8(t')$ predicted using the 
parameters obtained by fitting the data with $t_{\rm min}=6$ and $t'_{\rm min}=5$. The lighter part of the band is the portion of the curve outside of the fit region. An error-weighted average over $t_{\rm sep}^{K\to{\rm snk}}$ has been performed to the data. Recall that the effective matrix elements are defined (Eq.~\ref{eq-Meffkpipi}) such that the result converges to the ground-state matrix element at large $t'$.\label{fig-Q8_3x2_tmin6_tpmin5} }
\end{figure}

For $M_7^0$ and $M_8^0$ (and to a lesser extent, $M_{10}^0$) we observe a discrepancy between the one-operator and multi-operator results at the 1-2$\sigma$ level that persists even to large $t'_{\rm min}$. Given the very clear plateaus in the multi-state fit results, this disagreement is likely due again to statistical effects in these correlated data. This is evidenced for example in Fig.~\ref{fig-Q8_3x2_tmin6_tpmin5} in which we overlay the $M_8^{\rm eff, snk}$ effective matrix element for the $\pi\pi(111)$ and $\sigma$ sink operators by the multi-operator fit curve. We observe that the fit curve for the $\pi\pi(111)$ operator is completely compatible with the data but favors a value that is consistently within the upper half of the error bar, suggesting that the apparent flatness of the $\pi\pi(111)$ effective matrix element represents a false plateau, and the fact that the multi-operator method is capable of resolving the behavior is a testament to its power.

\subsubsection{Excited kaon state effects}

We now address excited kaon state effects. Because the data rapidly becomes noisier as we move the four-quark operator closer to the kaon operator and thus further away from the $\pi\pi$ operator, such effects are not expected to be significant. The simplest test is to vary the cut on the time separation between the kaon operator and the four-quark operator, $t_{\rm min}$. The first three points from the left of each cluster in Figs.~\ref{fig-matelemQ1toQ6} and~\ref{fig-matelemQ7toQ10} show the result of varying $t_{\rm min}$ between 6 and 8 at fixed $t'_{\rm min}$. As expected we observe no statistically significant dependence on this cut.


\begin{figure}[tb]
\includegraphics[width=0.48\textwidth, trim={0 0 0 0.3cm}, clip]{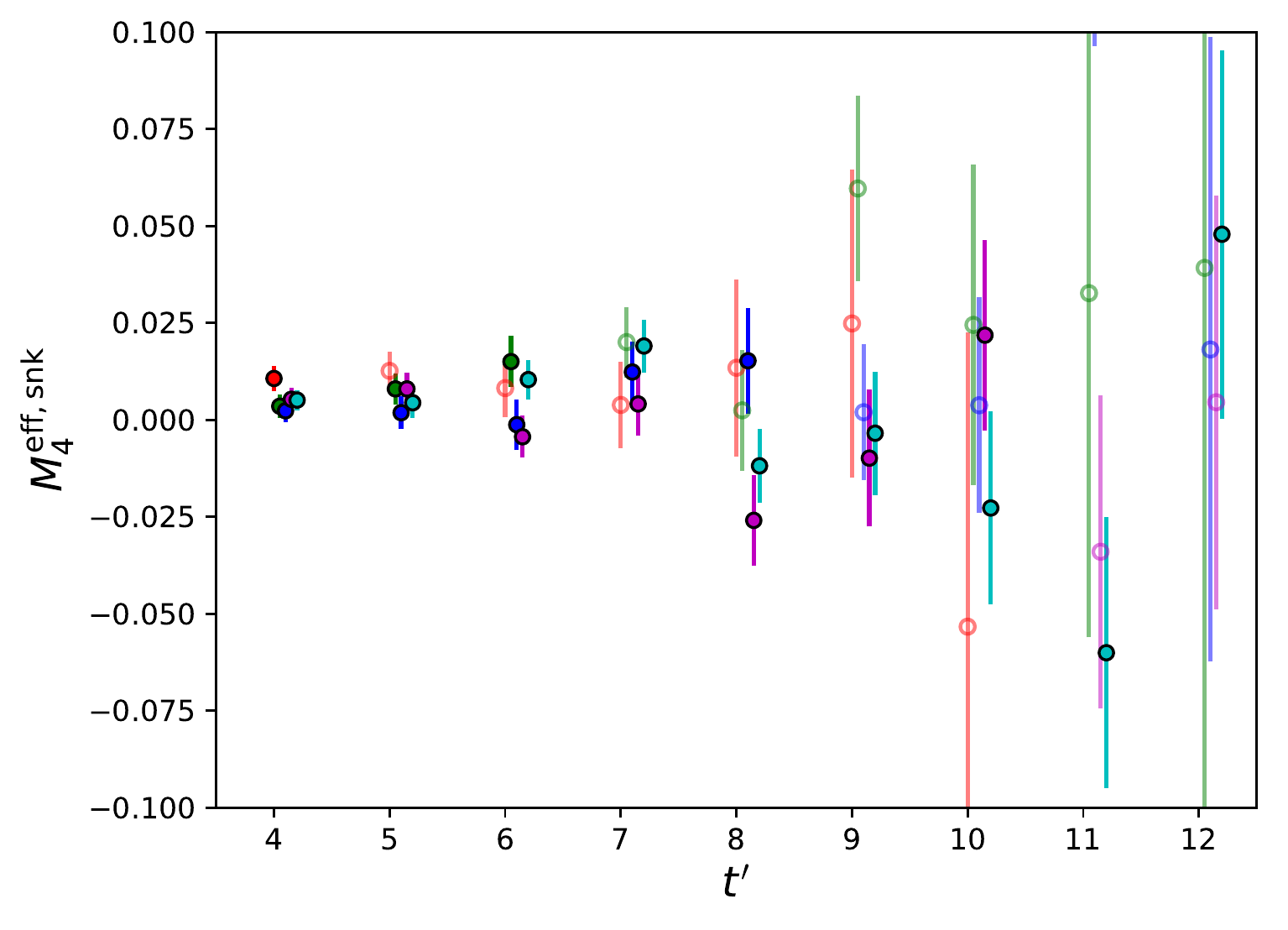}
\includegraphics[width=0.48\textwidth, trim={0 0 0 0.3cm}, clip]{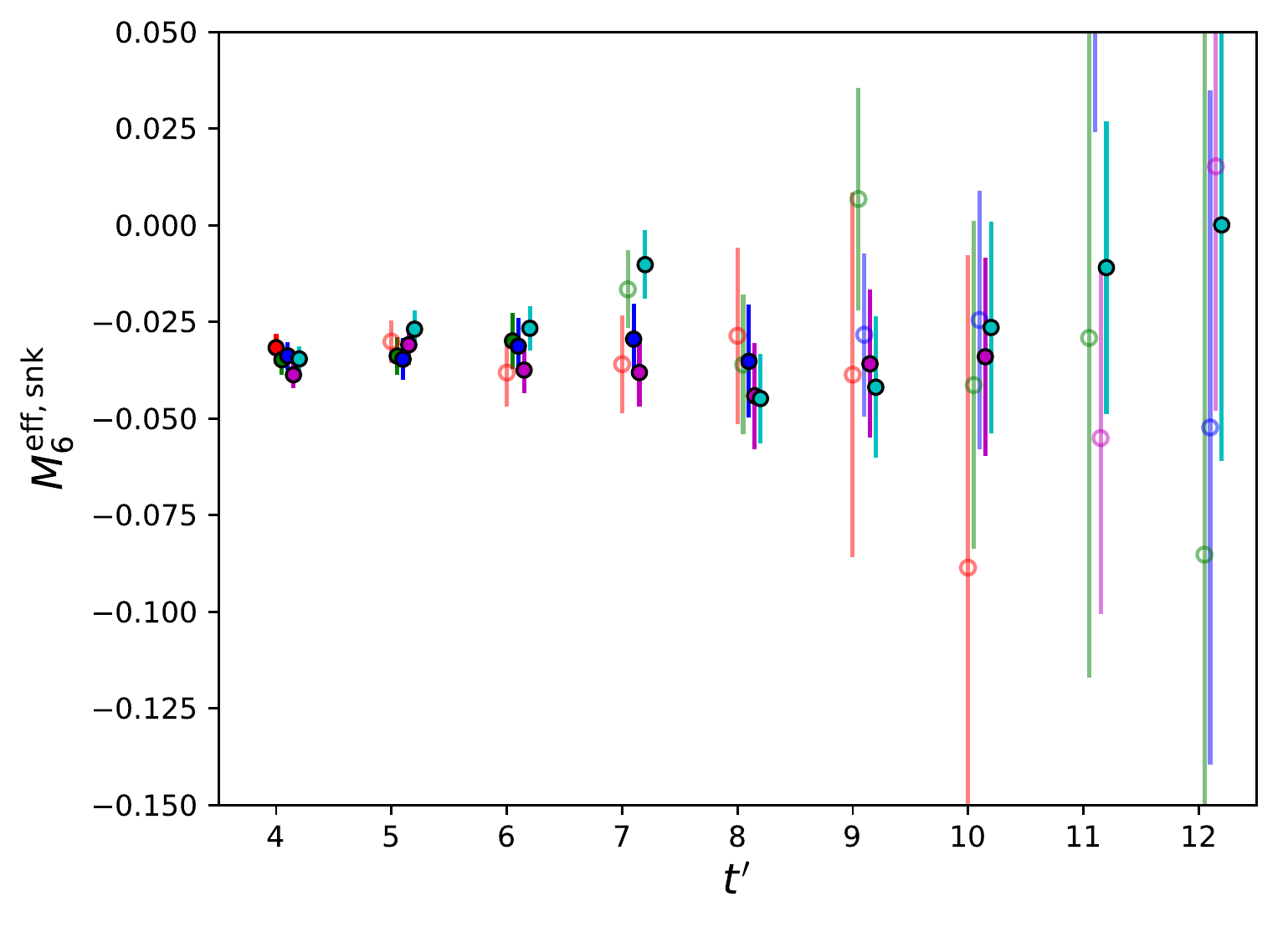}
\caption{The $t'$ dependence of the $M_4^{\rm eff,snk}$ (left) and $M_6^{\rm eff, snk}$ (right) effective matrix elements with the optimal sink operator. Each cluster of points, separated for clarity, shows the data for the five different $K\to{\rm snk}$ separations: 10,12,14, 16, and 18, from left to right in that order. The darker, filled points are those that lie within our cut of $t_{\rm min}=6$, and the lighter, hollow points are those excluded. \label{fig-Q4Q6opttprimedep} }
\end{figure}

We can also test for excited kaon effects by examining the data near the kaon operator in more detail, alongside looking for trends in the five different $K\to{\rm sink}$ separations at fixed $t'$. The optimal operator proves convenient for examining this behavior as it neatly combines the two dominant sink operators and should be flat within the fit window. In Fig.~\ref{fig-Q4Q6opttprimedep} we plot the data for the $M^{\rm eff,snk}_4$ and $M^{\rm eff,snk}_6$ effective matrix elements with a distinction drawn between data included and excluded by a cut on the kaon to four-quark operator time separation of $t_{\rm min}=6$. We find no apparent evidence of excited kaon state contamination even for data excluded by the cut, nor do we observe any trends of the data in the $K\to{\rm sink}$ separation. 

We therefore conclude that excited kaon effects in our results are negligible.

\subsubsection{Excited $\pi\pi$ state effects}

The dominant fit systematic error is expected to be due to excited $\pi\pi$ states. Fortunately, given that we can change both the number of operators and the number of states alongside varying the fit window within a region where our data is most precise, there are a number of tests we can perform to probe this source of error. 

We begin by comparing the multi-operator fits to the one-operator ($\pi\pi(111)$) fit, the latter being equivalent to the procedure used for our 2015 work. In the majority of cases we see little evidence of excited state contamination in the one-operator data, as evidenced by its agreement with the multi-operator fits as well as the strong consistency between the fits as we vary the fit window. However for the $M_5^0$ and $M_6^0$ matrix elements we observe strong evidence of excited-state contamination in these fits at smaller $t'_{\rm min}$. Fig.~\ref{fig-matelemQ1toQ6} clearly demonstrates how these effects are suppressed as we add more operators: Initially the one-operator results converge with the 3-operator results at $t'_{\rm min}=5$ and 6, respectively, at which point the excited states appear to be sufficiently suppressed. Introducing a second operator and state we eliminate part of this contamination and the convergence appears earlier, at $t'_{\rm min}=4$ and 5, respectively. Finally, in adding the third operator we find results that are essentially flat from $t'_{\rm min}=3$. This suggests that the 5\% excited-state systematic error on our 2015 result which used $t'_{\rm min}=4$ was significantly underestimated for these matrix elements.

In general we observe excellent agreement between two and three-operator fits with two-states. Unfortunately, as mentioned above, the $\pi\pi(311)$ data are considerably noisier than those of the other operators, and the associated $\pi\pi$ energy and amplitudes are less-well known, and as such these data contribute relatively little to the fit. Nevertheless we do observe that for the $M_5^0$ and $M_6^0$ matrix elements, the introduction of the third operator results in values that for low $t'_{\rm min}$ (3 or 4) are in considerably better agreement with the results for larger $t'_{\rm min}$, suggesting that in the regime in which these data are less noisy (i.e. closer to the $\pi\pi$ operator) the third operator is acting to remove some residual excited-state contamination. We conclude that it is beneficial to include the third operator.

In order to study the possibility of residual contamination from a third state we perform three-operator, three-state fits to the matrix elements using the $\pi\pi$ two-point function fit parameters given in the third column of Tab.~\ref{tab-pipi2ptfit-both} and the same fit ranges for $t$ and $t'$ used in the three-operator, two-state fits. The results for the ground-state matrix elements are also included in Figs.~\ref{fig-matelemQ1toQ6} and~\ref{fig-matelemQ7toQ10} with the label ``sys.''. We find that including this third state has very little impact and the results agree very well with the three-operator, two-state fits. This again suggests that we have the $\pi\pi$ excited-state systematic error under control.

\begin{figure}[tbp]
\centering
\includegraphics[width=0.48\textwidth]{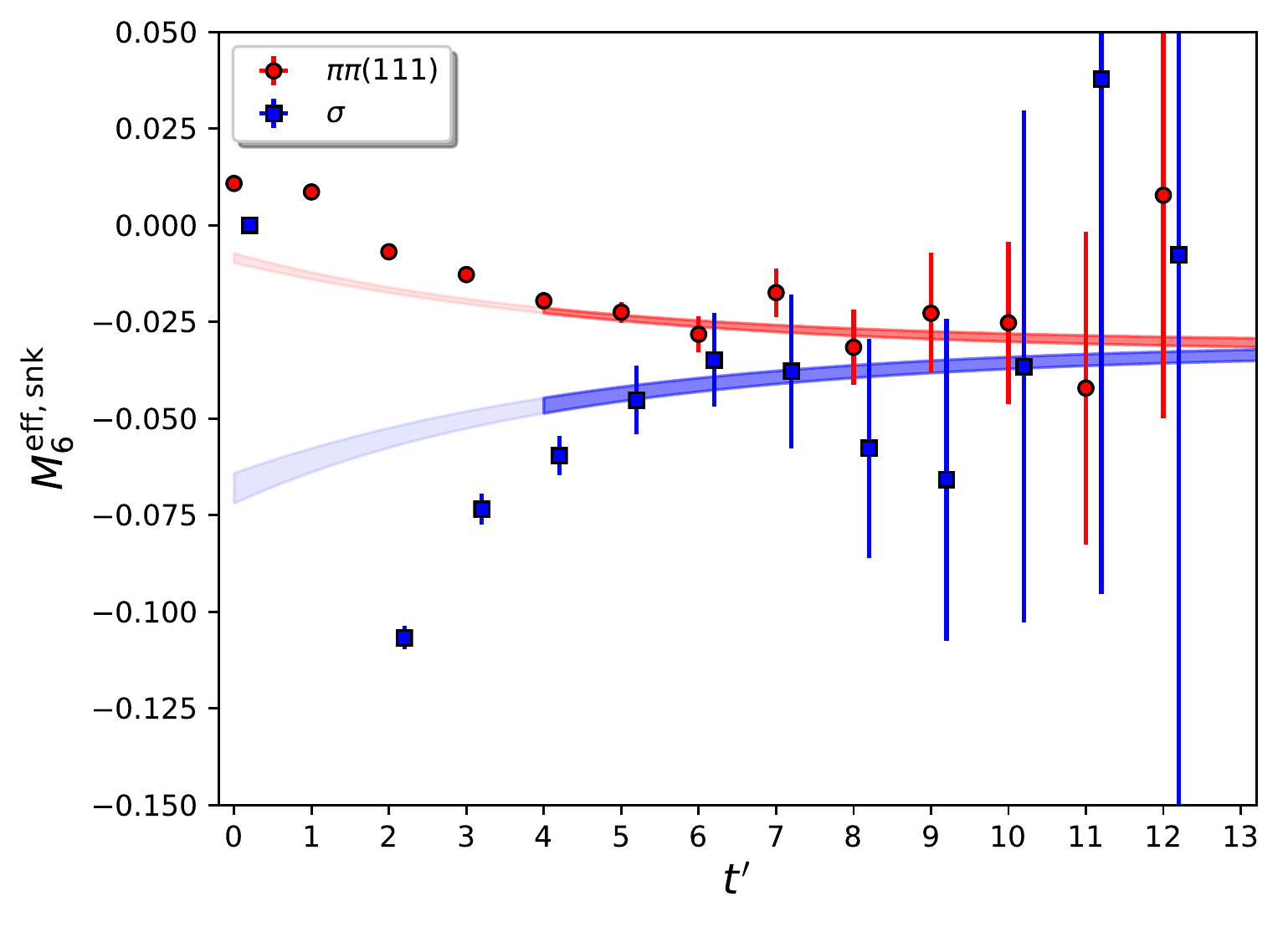}
\includegraphics[width=0.48\textwidth]{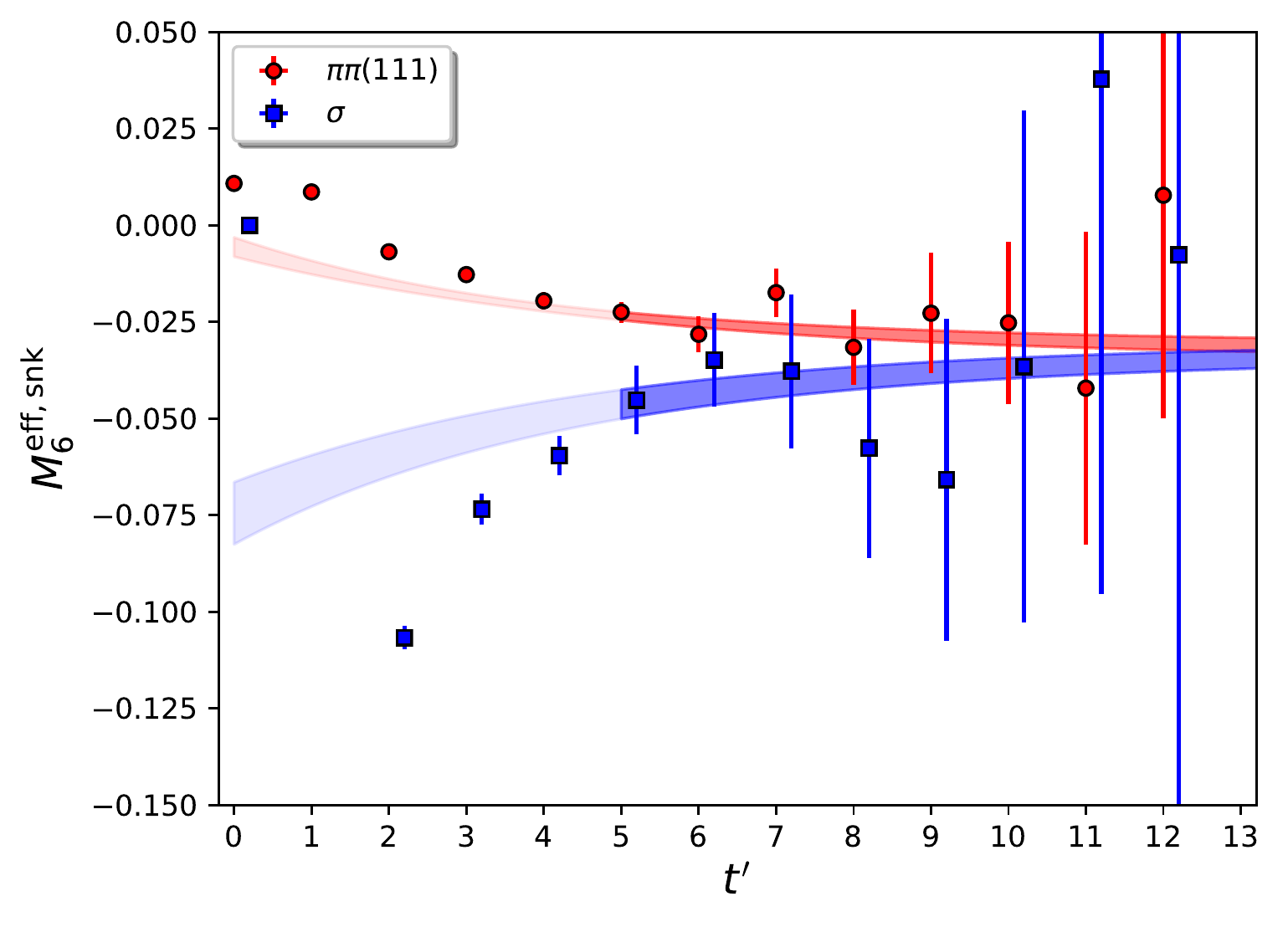}\\
\includegraphics[width=0.48\textwidth]{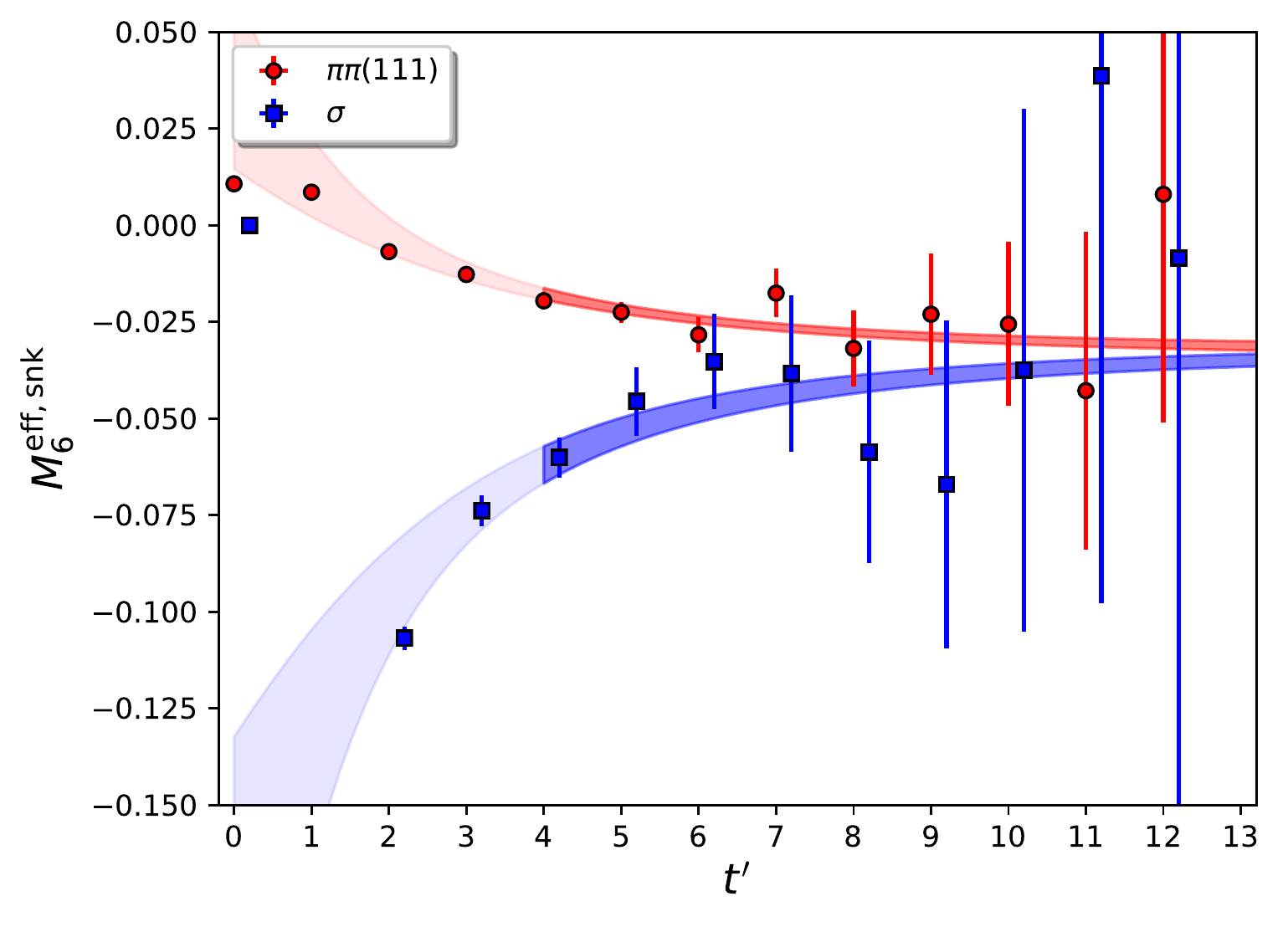}
\includegraphics[width=0.48\textwidth]{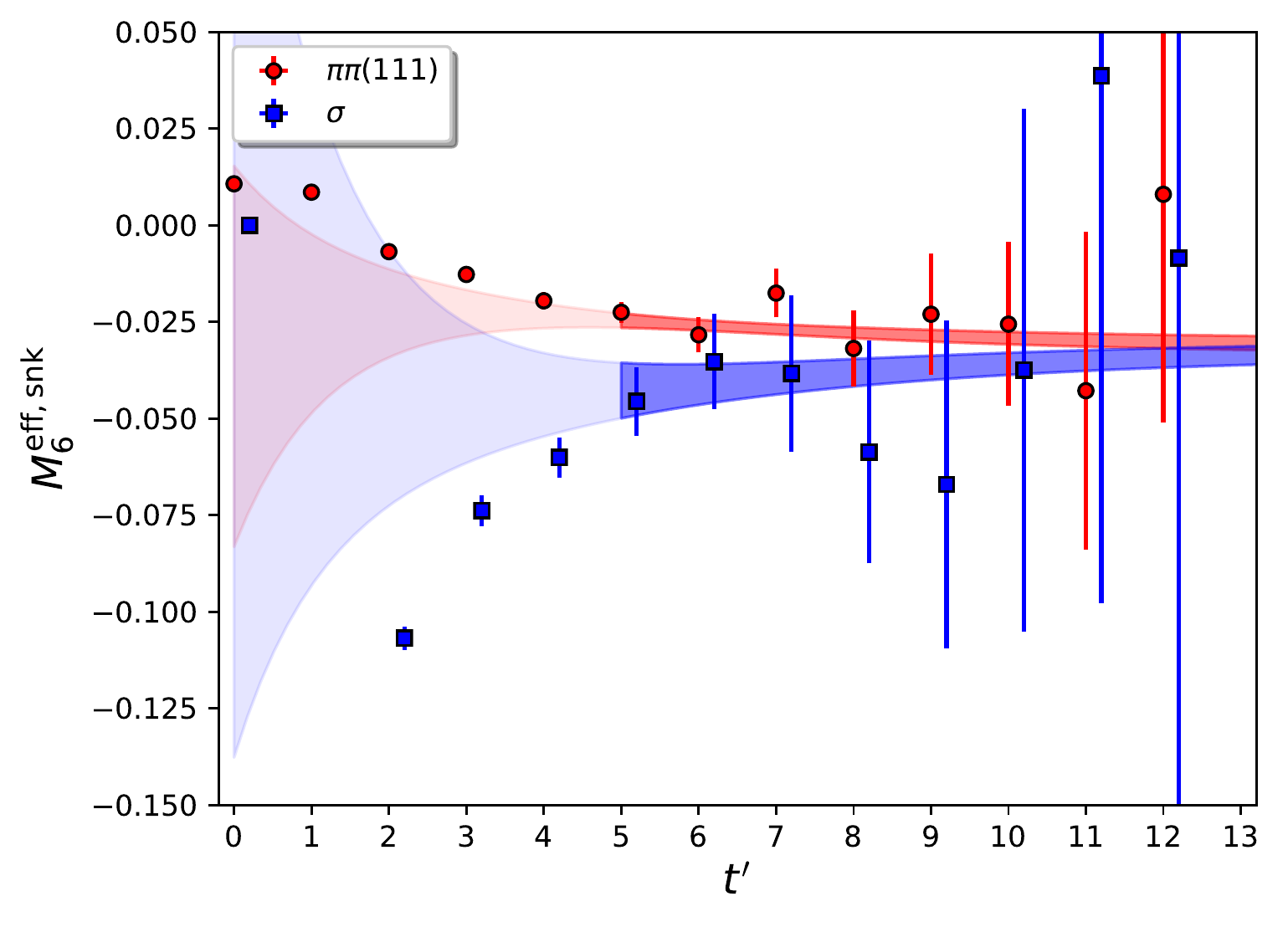}
\caption{The effective matrix element $M_6^{\rm eff, snk}$ for the $\pi\pi(111)$ (red circles) and $\sigma$ (blue squares) sink operators, overlaid by the fit curves. The lighter part of the band is the portion of the curve outside of the fit region. The upper panels are for the 2-state fits and the lower panels are for the 3-state fits. In each case the left panel is for $t'_{\rm min}=4$ and the right panel $t'_{\rm min}=5$. All fits are performed with 3 operators and use $t_{\rm min}=6$. \label{fig-Q6_3x2_3x3_tmin6_tpmin4_5-comp} }
\end{figure}

A further test for excited-state contamination is to study the agreement of the fit curves with the data outside of the fit region. To this end in Fig.~\ref{fig-Q6_3x2_3x3_tmin6_tpmin4_5-comp} we plot the $\pi\pi(111)$ and $\sigma$ operator data for the $M_6^{\rm eff,snk}$ effective matrix element overlaid by the fit curves for the 3-operator, 2-state fits, and for the 3-operator, 3-state fits described above, using $t'_{\rm min}=4$ and 5. The fitted ground-state matrix elements in these cases are all in complete agreement to within a fraction of their statistical errors. We observe that the 3-operator, 2-state fit curve with $t'_{\rm min}=5$ describes well the $\pi\pi(111)$ data at $t'=4$ but shows a tension for the $\sigma$ data at this timeslice. Fitting with $t'_{\rm min}=4$ does not resolve this tension, suggesting the effects of a third state are visible in the $\sigma$ operator data at $t'=4$. This is consistent with the pattern of couplings of the operators to the states in Tab.~\ref{tab-pipi2ptfit-both} which show a significant reduction in the couplings to higher states for the $\pi\pi(111)$ operator but almost equal-sized couplings of the $\sigma$ operator to all three states. The 3-operator, 3-state fit with $t'_{\rm min}=5$ does not appear to well resolve the contribution of the third state, which is consistent with our observation that this state is no longer visible in the $\pi\pi$ two-point data from this timeslice. However with $t'_{\rm min}=4$ we are able to resolve the effect of this state, and observe excellent agreement of the model with the data even down to very low times. It should be noted however that the third-state energy of $E_2=0.94(10)$ (in lattice units) obtained by our fits is somewhat larger than the value of $E_2=0.692$ predicted by dispersion theory suggesting that the effects of even higher excited states may be playing a role here. Nevertheless the strong agreement between the ground-state matrix elements for all of these fits suggest that the residual effects of the higher excited states on the 3-operator, 2-state fits are negligible. 

For our final result we choose to focus upon the three-operator, two-state fits. While the majority of the corresponding curves in Figs.~\ref{fig-matelemQ1toQ6} and~\ref{fig-matelemQ7toQ10} are essentially flat from $t'_{\rm min}=3$, we opt for a conservative and uniform cut of $t'_{\rm min}=5$ at which we can strongly claim an absence of significant excited-state effects. In the Sec.~\ref{sec-syserr-excstate} we will consider means by which we can assign a systematic error to this result.

\subsubsection{Final fit results}
\label{sec-finalfitresults}

\begin{table}[tb]
\centering
\begin{tabular}{c|c||c|c}
\hline\hline
Param & Value & Param & Value\\
\hline
$M_{1}^0$ & $-0.00152(50)$ & $M_{2}^0$ & $ 0.00366(41)$\\
$M_{1}^1$ & $ 0.0015(22)$ & $M_{2}^1$ & $-0.0050(21)$\\
p-value & 0.488 & p-value & 0.743\\
\hline
$M_{3}^0$ & $ 0.0005(11)$ & $M_{4}^0$ & $ 0.0052(13)$\\
$M_{3}^1$ & $ 0.0018(52)$ & $M_{4}^1$ & $-0.0045(59)$\\
p-value & 0.036 & p-value & 0.139\\
\hline
$M_{5}^0$ & $-0.0100(13)$ & $M_{6}^0$ & $-0.0322(20)$\\
$M_{5}^1$ & $ 0.0182(49)$ & $M_{6}^1$ & $ 0.0563(81)$\\
p-value & 0.458 & p-value & 0.159\\
\hline
$M_{7}^0$ & $ 0.02664(63)$ & $M_{8}^0$ & $ 0.08133(85)$\\
$M_{7}^1$ & $-0.0158(26)$ & $M_{8}^1$ & $-0.0464(45)$\\
p-value & 0.913 & p-value & 0.676\\
\hline
$M_{9}^0$ & $-0.00330(71)$ & $M_{10}^0$ & $ 0.00292(57)$\\
$M_{9}^1$ & $ 0.0051(30)$ & $M_{10}^1$ & $-0.0048(27)$\\
p-value & 0.327 & p-value & 0.56\\
\hline
\end{tabular}
\caption{Final $K\to\pi\pi$ matrix element results in lattice units obtained from a three-operator, two-state fit with $t_{\rm min}=6$ and $t'_{\rm min}=5$. Here $M_i^j$ refers to the matrix element of the $Q_i$ operator with $\pi\pi$ state $j$. \label{tab-fitresultfinal} }
\end{table}

As discussed above we choose the 3-operator, 2-state fit with $t'_{\rm min}=5$ for our final result. As we observe no significant dependence on the cut on the separation between the kaon and four-quark operators we will choose $t_{\rm min}=6$. In Tab.~\ref{tab-fitresultfinal} we present the full set of p-values and parameters for these fits. We obtain acceptable p-values in the majority of cases, with the notable exception of the $Q_3$ four-quark operator for which $p=4\%$. We find that this p-value is not improved by increasing $t'_{\rm min}$, and also that the p-value of the one-operator, one-state fit with the same fit range -- with which our chosen value is in excellent agreement -- has a p-value of 15\%. The low probability is therefore unlikely to be associated with any systematic effect and can be attributed to low-probability statistical effects. 

We conclude this section with a comparison of the statistical errors of the matrix elements $M_2^0$ and $M_6^0$ to those determined in our 2015 analysis. Previously we obtained
\begin{dmath}\begin{array}{rl}
M_{2}^0 &= 0.00424(116) \\
M_{6}^0 &= -0.0189(47)\,.
\end{array}\end{dmath}
Comparing these values to those in Tab.~\ref{tab-fitresultfinal} we find that the errors have reduced by factors of 2.8 and 2.4 for $M_2^0$ and $M_6^0$, respectively. Comparing the 3-operator, 2-state fits to the 1-operator, 1-state fits in Fig.~\ref{fig-matelemQ1toQ6} we observe that the larger improvement for $M_2^0$ can be explained by the additional operators, however for $M_6^0$ these two approaches have similar errors. The fact that the error on $M_6^0$ has improved considerably more than the factor of 1.9 expected by the increase in statistics can therefore be attributed to the improved precision of the $\pi\pi$ two-point function fits observed in Sec.~\ref{sec-pipi2pt}.


\section{Non-perturbative renormalization of lattice matrix elements}
\label{sec:NPR}
The Wilson coefficients are conventionally computed in the $\msbar$ (NDR) renormalization scheme, and therefore we are required to renormalize our lattice matrix elements also in this scheme. This is achieved by performing an intermediate conversion to a non-perturbative {\it regularization invariant momentum scheme with symmetric kinematics} (RI-SMOM). As the name suggests, these schemes can be treated both non-perturbatively on the lattice (provided the renormalization scale is sufficiently small compared to the Nyquist frequency $\pi/a$) and in continuum perturbation theory (providing the renormalization scale is sufficiently high that perturbation theory is approximately valid at the order to which we are working).  Thus, we can use continuum perturbation theory to match our RI-SMOM matrix elements to $\msbar$, avoiding the need for lattice perturbation theory. The matching factors have been computed to one-loop in Ref.~\cite{Lehner:2011fz}.

In our 2015 calculation we computed the renormalization matrix at a somewhat low renormalization scale of $\mu=1.529$ GeV in order to avoid large cutoff effects on our coarse, $a^{-1}=1.38$ GeV ensemble. Due to this low scale, the systematic error associated with the perturbative RI to $\msbar$ matching was our dominant error, with an estimated size of 15\%. In this paper we utilize the step-scaling procedure~\cite{Arthur:2010ht, Durr:2010vn, Durr:2010aw, Arthur:2011cn} (summarized below) in order to circumvent the limit imposed by the lattice cutoff and increase the renormalization scale to 4.0 GeV at which the error arising from the use of one-loop perturbation theory is expected to be significantly smaller. A separate step-scaling calculation to 2.29 GeV was performed in Ref.~\cite{McGlynnThesis} and we will utilize those results to study the scale dependence of the perturbative and discretization errors in our operator normalization.


\subsection{Summary of approach}
\label{sec-renorm-summary}

Due to operator mixing, the renormalization factors take the form of a matrix. This is most conveniently expressed in the seven-operator chiral basis in which the operators are linearly independent and transform in specific representations of the $SU(3)_L\otimes SU(3)_R$ chiral symmetry group, an accurate symmetry of our DWF formulation even at short distances. In this basis the renormalization matrix is block diagonal, with a $1\times 1$ matrix associated with the $Q_1'$ operator that transforms in the $(27,1)$ representation, a $4\times 4$ matrix for the $(8,1)$ operators $Q_{2}'$, $Q_{3}'$, $Q_{5}'$ and $Q_{6}'$ , and a $2\times 2$ matrix for the $(8,8)$ operators $Q_{7}'$ and $Q_{8}'$. 

In the RI-SMOM scheme the renormalized operators are generally defined thus,
\begin{dmath}
{\cal O}_i^{{\rm RI}} = Z^{{\rm RI}\leftarrow{\rm lat}}_{ij}{\cal O}_j^{{\rm lat}} 
\end{dmath}
where Einstein's summation conventions are implied and the label ``RI'' is used as short-hand for the RI-SMOM scheme. The renormalization factors are defined via
\begin{dmath}
Z_q^{-2} [P_m]^{\beta\alpha\delta\gamma}[\Gamma_{im}^{\rm RI}]^{\alpha\beta\gamma\delta}(p_1,p_2) = F_{im}\,,
\end{dmath}
where the index $m$ is not summed over. Here $\alpha-\delta$ are combined spin and color indices, $Z_q$ is the quark field renormalization, $q$ is a four-momentum that defines the renormalization scale and $P_m$ are ``projection matrices'' described below. The quantities $F_{im}$ on the right-hand side are found by evaluating the left-hand side of the equation at tree level. $\Gamma^{\rm RI}_{im}$ are computed as
\begin{dmath}
[\Gamma_{im}^{\rm RI}]^{\alpha\beta\gamma\delta}(p_1,p_2) = \left\langle E_m\sum_x e^{2iqx}{\cal O}_i^{{\rm RI}}(x)\right\rangle_{\rm amp.}^{\alpha\beta\gamma\delta} \label{eq-ampvertexRI}
\end{dmath}
where the sum is performed over the full four-dimensional lattice volume and $q=p_1-p_2$. Here $E_m$ are a set of seven four-quark operators that each create the four quark lines that connect to the weak effective operator, 
\begin{equation}\begin{array}{l}
\displaystyle E_1 = E_2 = E_4 = E_5 = s(-p_1) \bar d(p_2) u(-p_1) \bar u(p_2)\\
\displaystyle E_3 = E_6 = E_7 = s(-p_1) \bar d(p_2)\sum_{q=u,d,s} q(-p_1) \bar q(p_2)\,,
\end{array}\end{equation}
where the momentum arguments indicate the incoming momenta and the quark momenta satisfy {\it symmetric kinematics}: $p_1^2 = p_2^2 = (p_1 - p_2)^2 = q^2 \equiv \mu^2$. The subscript ``amp.'' in Eq.~\eqref{eq-ampvertexRI} implies that the external propagators are amputated by applying the ensemble-averaged inverse propagator, such that the resulting Green's function has a rank-4 tensor structure in the spin-color indices. 

These Green's functions are not gauge-invariant, hence the procedure must be performed using gauge-fixed configurations, for which we employ Landau gauge-fixing.  The use of momentum-space Green's functions introduces contact terms that prevent the use of the equations of motion so that additional operators, beyond those needed to determine on-shell matrix elements, must be introduced if all possible operator mixings are to be included, as is required if the RI-SMOM scheme is to have a continuum limit.  These are discussed below.  

Note that the Wick contractions of Eq.~\eqref{eq-ampvertexRI} result in disconnected penguin-like diagrams that interact only by gluon exchange; these diagrams are evaluated using stochastic all-to-all propagators and are typically noisy, requiring multiple random hits and hundreds of configurations. The presence of disconnected diagrams also precludes the use of partially-twisted boundary conditions and therefore limits our choices of the renormalization momentum scale to the allowed lattice momenta.  

The quark field renormalization $Z_q$ is also computed in the RI-SMOM scheme via the amputated vertex function of the local vector current operator, $\bar q\gamma^\mu q$, from which we compute $Z_V/Z_q$ where $Z_V$ is the corresponding renormalization factor for the local vector current. The factor $Z_V$ is not unity as the local vector current is not conserved, however it can be computed independently from the ratio of hadronic matrix elements containing the local and conserved (five-dimensional) vector current allowing $Z_q$ to be obtained from the above.  Alternatively, $Z_q$ can also be computed from the local axial-vector current operator $\bar q\gamma^\mu\gamma^5 q$.  Again the ratio $Z_A/Z_q$ is determined from a three-point function evaluated in momentum space and, providing non-exceptional kinematics are used, is equivalent up to negligible systematic effects at large momentum~\cite{Aoki:2007xm}.  The quantity $Z_A$ is then determined by comparing the pion-to-vacuum matrix elements of the local and approximately conserved (five-dimensional) axial current.

The independent projection matrices $P_m$ contract the external spin and color indices, and are chosen with a tensor structure that reflects that of the operator with the same index. For the weak effective operators, we can choose both parity-even and parity-odd projectors, which project onto the parity-even and parity-odd components of the amputated Green's function, respectively, and which should both provide the same result due to chiral symmetry. In practice however we have found that the parity-odd choices are better protected against residual chiral symmetry breaking effects that induce non-zero mixings between the different $SU(3)_L\otimes SU(3)_R$ representations (cf. Sec. 4.5 of Ref.~\cite{Liu:2012}), and so we will use the parity-odd projectors exclusively. We consider two different projection schemes: the ``$\gamma^\mu$ scheme'', for which the parity-odd projectors have the spin structure,
\begin{dmath}
P_m^{\gamma^\mu} = \pm\gamma_\mu \otimes (\gamma^5\gamma^\mu) - (\gamma^5\gamma_\mu) \otimes \gamma_\mu\,,
\end{dmath}
and the ``$\slashed{q}$ scheme'' with spin structure
\begin{dmath}
P_m^{\slashed{q}} = \pm\slashed{q} \otimes (\gamma^5\slashed{q}) - (\gamma^5\slashed{q}) \otimes \slashed{q}\,.
\end{dmath}
For the full set of parity-odd and parity-even projectors we refer the reader to Sec. 3.3.2 of Ref.~\cite{ZhangThesis}. 

Similar choices of $\gamma^\mu$ and $\qslash$ projector exist also for the quark field renormalization. We will follow the convention of describing our RI-SMOM schemes with a label of the form ${\rm SMOM}(A,B)$ where the quantities $A$ and $B$ in parentheses describe the choices of projector for the four-quark operator and $Z_q$, respectively. In this work we consider only the $\SMOMgg$ and $\SMOMqq$ schemes as previous studies of the renormalization of the neutral kaon mixing parameter $B_K$ indicate that the non-perturbative running is better described by perturbation theory for these two choices than for the two mixed schemes~\cite{Aoki:2010pe}. We will compare our final results obtained using both intermediate schemes in order to estimate the systematic perturbative and discretization errors in computing the RI to $\msbar$ matching.

\subsection{Operator mixing}

The seven weak effective operators mix with several dimension-3 and dimension-4 bilinear operators. For the parity-odd components these are $S_1=\bar s\gamma^5 d$, $S_2=\bar s \stackrel{\rightarrow}{\slashed D}\gamma^5 d$ and $S_3=\bar s \stackrel{\leftarrow}{\slashed D} \gamma^5d$ , where the arrow indicates the direction of the discrete covariant derivative. These are accounted for by performing the renormalization with subtracted operators,
\begin{dmath}
Q_i^{\prime{\rm sub, lat}} = Q_i^\prime + \sum_{j=1}^{3} b_j S^{\rm lat}_j\,.
\end{dmath}
The subtraction coefficients $b_j$ are obtained by applying the following conditions,
\begin{dmath}
P_j^{\beta\alpha} \left\langle  s(-p_1) \bar d(p_2) {\cal O}_i^{\rm sub, lat}(q) \right\rangle_{\rm amp.}^{\alpha\beta} = 0
\end{dmath}
with symmetric kinematics at the scale $q^2$. The projection operators can be found in Sec. 7.2.6 of Ref.~\cite{McGlynnThesis}. In practice we find that the subtraction coefficients are small due to the suppression of the mixing by a factor of the quark mass as a result of chiral symmetry, and also the observation that the amputated vertex function Eq.~\eqref{eq-ampvertexRI} with a four-quark external state and a two-quark operator necessarily involves only disconnected diagrams that are small at large momentum scales due to the running of the QCD coupling. 

Mixing also occurs with the dimension-5 chromomagnetic penguin operator and a similar electric dipole operator, conventionally labeled $Q_{11}$ and $Q_{12}$, respectively~\cite{Bertolini:1994qk}. These operators do not vanish by the equations of motion and therefore contribute also to the on-shell matrix elements, but break chiral symmetry and as such are expected to be heavily suppressed~\cite{Bertolini:1994qk,Buras:2018evv}. It is therefore conventional to neglect their effects in, for example, the determination of the Wilson coefficients~\cite{Buchalla:1995vs}.   In our DWF calculation the dimension-1 mixing coefficients of these dimension-5 operators will be of order the input quark masses used in our RI-SMOM calculations or the DWF residual mass --- effects, when combined with the required gluon exchange, should be at or below the percent level.  Thus, in this work we neglect these operators.

In addition to the lower-dimension operators there is also mixing with both gauge-invariant and gauge-noninvariant dimension-6 two-quark operators. These operators enter at next-to-leading order and above, and are therefore naturally small provided we perform our renormalization at large energy scales. 

The gauge-noninvariant dimension-6 operators vanish due to gauge symmetry and in many cases also by the equations of motion, and therefore do not contribute to on-shell matrix elements~\cite{Martinelli:2001ak}. These operators enter the renormalization only at the two-loop level~\cite{Blum:2001xb} and above, and given that the RI$\to\msbar$ matching factors are at present only available to one loop, the systematic effect of disregarding these operators is likely to be much smaller than our dominant systematic errors. Nevertheless we are presently investigating position-space renormalization~\cite{Tomii:2018zix} which does not require gauge fixing and therefore does not suffer from such mixing, and as such we may be able to remove this systematic error in future work.

Of the gauge-invariant dimension-6 operators,
\begin{dmath}
G_1 = \bar s\left[ D_\mu \left[ D^\mu, D_\nu \right] \right]\gamma^\nu(1-\gamma^5)d
\end{dmath}
is the only operator that mixes at one loop~\cite{Buras:1992tc}, with all others entering at two-loops and above. In Ref.~\cite{McGlynnThesis} we have investigated the impact of including the $G_1$ operator in our RI-SMOM renormalization and have computed the subsequent effect on the $K\to\pi\pi$ amplitudes. This can be achieved without the need for measuring matrix elements of $G_1$ between kaon and $\pi\pi$ states by taking advantage of the equations of motion to rewrite those matrix elements for on-shell kinematics in terms of the matrix elements of the conventional four-quark operators, such that the entire effect of this operator is captured by changes in the values of the $(8,1)$ elements of the renormalization matrix. Note that at present the results including the $G_1$ operator have been computed only at the 2.29 GeV renormalization scale and not the 4.0 GeV scale used for our final result. However, as demonstrated in Ref.~\cite{McGlynnThesis} and also in Sec.~\ref{sec-G1syserrs}, the effects of including $G_1$ are at the few percent level as expected, implying that the resulting systematic error is small compared to our other errors.

\subsection{Step-scaling}

Step-scaling~\cite{Arthur:2010ht, Durr:2010vn, Durr:2010aw, Arthur:2011cn} allows for the circumvention of the upper limit on the renormalization scale imposed by the lattice spacing through independently computing the non-perturbative running of the renormalization matrix to a higher scale using a finer lattice. The multiplicative factor relating the RI-SMOM operators renormalized at two different scales can be obtained from the ratio
\begin{dmath}
\Lambda^{\rm RI}(\mu_2, \mu_1) = Z^{{\rm RI}\leftarrow{\rm lat}}(\mu_2)(Z^{{\rm RI}\leftarrow{\rm lat}}(\mu_1))^{-1}\,, \label{eq-stepscalingmatrix}
\end{dmath}
where $\mu_1$ is a renormalization scale that lies below the cutoff on the original coarser lattice while $\mu_2$ is a higher scale, likely inaccessible on the coarser lattice. The quantity $\Lambda^{\rm RI}(\mu_2, \mu_1)$ is computed on finer lattices for which $\mu_2$ also lies below the cutoff and can be applied thus,
\begin{dmath}
Z^{{\rm RI}\leftarrow{\rm lat}} (\mu_2)
            = \Lambda^{\rm RI}(\mu_2, \mu_1) Z^{{\rm RI}\leftarrow{\rm lat}} (\mu_1)
\end{dmath}
in order to raise the renormalization scale to $\mu_2$, giving the renormalization matrix $Z^{{\rm RI}\leftarrow{\rm lat}} (\mu_2)$ which non-perturbatively converts our course-lattice operators into an RI scheme defined at a scale $\mu_2$ potentially much larger than the inverse of our coarse lattice spacing. The step-scaling matrix computed via Eq.~\eqref{eq-stepscalingmatrix} has discretization errors arising from the (fine) lattice spacing, but, providing one is careful to fix the orientation of the momenta, the matrix has a well-defined continuum limit as first described in Ref.~\cite{Arthur:2010ht}. We will take advantage of step scaling to avoid having to match perturbatively to $\msbar$ directly at the lower energy scales allowed by our coarse, $a^{-1}=1.38$ GeV lattice.

\subsection{Details and results of lattice calculation}

We use the step-scaling procedure to obtain the renormalization matrix at a scale of $\mu_2=4.006$ GeV by matching between our $\beta=1.75$, $a^{-1}=1.378(7)$ GeV (32ID) ensemble and a second, finer ensemble with $\beta=2.37$ and $a^{-1}=3.148(17)$ whose properties are described in Ref.~\cite{Blum:2014tka} under the label ``32Ifine''. These ensembles have periodic spatial boundary conditions rather than G-parity boundary conditions, but as previously mentioned, boundary effects can be neglected for these high-energy Green's functions. Such quantities are also constructed to be insensitive to the quark mass scale, and therefore we can disregard the unphysically heavy 170 MeV and 370 MeV pion masses on the 32ID and 32Ifine ensembles, respectively. Note also that, although we do not take the continuum limit of the step-scaling matrix computed on the 32Ifine ensemble, the fine lattice spacing and the typically small size of discretization effects on such quantities~\cite{Boyle:2017skn} suggest the induced error is also negligible compared to our other errors. We remind the reader that these calculations do not include the $G_1$ operator, and its absence in our calculation is treated as a source of systematic error in Sec.~\ref{sec:syserrs}.

Due to the presence of disconnected diagrams in our calculation, the choices of quark momenta are restricted to the discrete values allowed by the finite-volume. The closest match between allowed momenta on the 32ID and 32Ifine ensembles that can be chosen as an intermediate scale is $\mu_1^{\rm 32ID}=1.531$ GeV and $\mu_1^{\rm 32Ifine}=1.514$ GeV, respectively. The fact that these scales differ by 1.1\% introduces a systematic error that, given the slow evolution of the QCD $\beta$-function, can be treated as negligible. 

We obtain the quark field renormalization for the 32Ifine ensemble via the vector current operator as described in Sec.~\ref{sec-renorm-summary}. For the 32ID ensemble we use the axial-vector operator as the corresponding renormalization factor, $Z_A$ has been measured to much higher precision than $Z_V$ (0.05\% versus 1.2\%, respectively)~\cite{Arthur:2012opa}. The measurements of $Z_A$ and $Z_V$ are treated as statistically independent from those of the amputated vertex functions and are incorporated into the calculation using the superjackknife technique.

On the 32ID ensemble we extend the calculation at $\mu_1^{\rm 32ID}=1.531$ GeV performed in our previous work and documented in Ref.~\cite{ZhangThesis} from 100 to 234 configurations, where for each configuration we have increased the number of stochastic sources used in the evaluation of the disconnected diagrams from 1 to 20, improving the statistical errors substantially. We measure the amputated Green's function Eq.~\eqref{eq-ampvertexRI} with quark momentum choices
\begin{dmath}
{p_1 = (0,4,4,0)\frac{2\pi}{L} }\,,\\
{p_2 = (4,4,0,0)\frac{2\pi}{L} }\,,
\end{dmath}
that satisfy symmetric kinematics $p_1^2 = p_2^2 = (p_1-p_2)^2 = (\mu_1^{\rm 32ID})^2$. Combined with the following measurements of the quark field renormalization coefficient in the $\gamma^\mu$ and $\slashed{q}$ schemes at $\mu_1^{\rm 32ID}$,
\begin{dmath}
{ Z_q^{\gamma^\mu}(\mu_1^{\rm 32ID}) = 0.7304(4) }\,,\\   
{ Z_q^{\slashed{q}}(\mu_1^{\rm 32ID}) = 0.8017(4) }\,,\\  
\end{dmath}
we obtain the renormalization matrices $Z_{ij}^{{\rm RI}\leftarrow{\rm lat}}$ for the $\SMOMgg$  and $\SMOMqq$ schemes given in Tab.~\ref{tab-Zmat1.53}.

For the measurement of the step-scaling matrix on the 32Ifine ensemble we likewise use
\begin{dmath}
{ p_1 = (1,1,2,0)\frac{2\pi}{L} }\,, \\
{p_2 = (0,1,1,4)\frac{2\pi}{L} }\,,
\end{dmath}
at the low scale $\mu_1^{\rm 32Ifine}=1.514$ GeV and
\begin{dmath}
{p_1 = (4,4,3,2)\frac{2\pi}{L} }\,,\\
{p_2 = (0,1,4,10)\frac{2\pi}{L} }\,,
\end{dmath}
at the high scale $\mu_2 = 4.006$ GeV. The corresponding values of $Z_q$ are
\begin{dmath}
{ Z_q^{\gamma^\mu}(\mu_1^{\rm 32Ifine}) = 0.8082(2) }\,,\\   
{ Z_q^{\slashed{q}}(\mu_1^{\rm 32Ifine}) = 0.8884(5)  }\,,\\ 
\end{dmath}
at $\mu_1^{\rm 32Ifine}=1.514$ GeV and
\begin{dmath}
{ Z_q^{\gamma^\mu}(\mu_2^{\rm 32Ifine}) =  0.80235(9) }\,,\\    
{ Z_q^{\slashed{q}}(\mu_2^{\rm 32Ifine}) =  0.83196(10) }\,,\\  
\end{dmath}
at $\mu_2 = 4.006$ GeV.

The results for the step-scaling matrix $\Lambda(4.006\ \mathrm{GeV},1.514\ \mathrm{GeV})_{ij}$ in both schemes are given in Tab.~\ref{tab-stepscale1.53_4.0}.  In Tab.~\ref{tab-Zmat4.00} we combine these step-scaling results with the 32ID $Z^{{\rm RI}\leftarrow{\rm lat}}$ results to produce the final renormalization matrices at 4.0 GeV, where the errors on the two independent ensembles have been propagated using the super-jackknife procedure.

As mentioned previously, we will also utilize step-scaled renormalization matrices computed at $\mu_2=2.29$ GeV both with and without the $G_1$ operator included. This calculation used an intermediate scale of $\mu=1.33$ GeV to match between the coarse and fine ensemble. Details of this calculation can be found in Ref.~\cite{McGlynnThesis}. In that work the statistical errors on $Z_V$ and $Z_A$ were not included in the results, and $Z_V$ was used rather than $Z_A$ in the determination of $Z_q$ on the 32ID ensemble. In order to match the procedure outlined above we have reanalyzed the data from that work, the results of which are presented in Tab.~\ref{tab-Zmat1.33} for $\mu=1.33$ GeV and Tab.~\ref{tab-Zmat2.29} for $\mu=2.29$ GeV. Note, at present only results in the $\SMOMgg$ scheme are available with $G_1$ included.

\begin{table}[tb]
\centering
\begin{tabular}{ccccccc}
\hline
$ 0.43216(43)$ & 0 & 0 & 0 & 0 & 0 & 0\\
0 & $ 0.4904(62)$ & $-0.0398(60)$ & $-0.0009(22)$ & $-0.0011(13)$ & 0 & 0\\
0 & $-0.0375(24)$ & $ 0.4937(25)$ & $-0.00242(93)$ & $ 0.00637(68)$ & 0 & 0\\
0 & $-0.011(19)$ & $-0.017(17)$ & $ 0.5138(63)$ & $-0.0968(38)$ & 0 & 0\\
0 & $ 0.0106(77)$ & $ 0.0304(80)$ & $-0.0328(28)$ & $ 0.3305(23)$ & 0 & 0\\
0 & 0 & 0 & 0 & 0 & $ 0.49839(49)$ & $-0.092841(93)$\\
0 & 0 & 0 & 0 & 0 & $-0.027045(31)$ & $ 0.30819(31)$\\
\hline
\end{tabular}
\newline
\vspace{1cm}
\newline
\begin{tabular}{ccccccc}
\hline
$ 0.46763(46)$ & 0 & 0 & 0 & 0 & 0 & 0\\
0 & $ 0.3670(66)$ & $-0.2593(71)$ & $-0.0025(25)$ & $-0.0005(15)$ & 0 & 0\\
0 & $ 0.1575(98)$ & $ 0.835(10)$ & $ 0.0019(38)$ & $-0.0006(23)$ & 0 & 0\\
0 & $-0.032(32)$ & $-0.016(30)$ & $ 0.519(11)$ & $-0.0952(63)$ & 0 & 0\\
0 & $-0.048(14)$ & $-0.077(17)$ & $-0.0578(46)$ & $ 0.3866(36)$ & 0 & 0\\
0 & 0 & 0 & 0 & 0 & $ 0.50244(50)$ & $-0.094095(95)$\\
0 & 0 & 0 & 0 & 0 & $-0.060488(73)$ & $ 0.37992(39)$\\
\hline
\end{tabular}
\caption{The elements of the $7\times 7$ $\SMOMgg$ (upper) and $\SMOMqq$ (lower) renormalization matrices $Z(1.531\mathrm{GeV})_{ij}^{{\rm RI}\leftarrow{\rm lat}}$ with renormalization scale $\mu=1.531$ GeV computed on the 32ID ensemble. \label{tab-Zmat1.53}  }
\end{table}

\begin{table}[tb]
\centering
\begin{tabular}{ccccccc}
\hline
$ 0.94514(24)$ & 0 & 0 & 0 & 0 & 0 & 0\\
0 & $ 0.976(49)$ & $-0.155(41)$ & $-0.022(19)$ & $ 0.023(15)$ & 0 & 0\\
0 & $-0.105(20)$ & $ 1.055(19)$ & $-0.0130(69)$ & $-0.0062(64)$ & 0 & 0\\
0 & $-0.10(15)$ & $-0.13(12)$ & $ 0.855(56)$ & $ 0.243(47)$ & 0 & 0\\
0 & $ 0.0010(750)$ & $-0.058(70)$ & $-0.031(27)$ & $ 1.728(24)$ & 0 & 0\\
0 & 0 & 0 & 0 & 0 & $ 0.96601(25)$ & $ 0.23304(65)$\\
0 & 0 & 0 & 0 & 0 & $ 0.00911(46)$ & $ 1.8170(26)$\\
\hline
\end{tabular}
\newline
\vspace{1cm}
\newline
\begin{tabular}{ccccccc}
\hline
$ 0.89837(24)$ & 0 & 0 & 0 & 0 & 0 & 0\\
0 & $ 1.110(77)$ & $ 0.099(62)$ & $-0.002(24)$ & $ 0.023(17)$ & 0 & 0\\
0 & $-0.486(49)$ & $ 0.532(41)$ & $-0.026(16)$ & $ 0.009(10)$ & 0 & 0\\
0 & $-0.19(28)$ & $-0.20(22)$ & $ 0.844(82)$ & $ 0.242(58)$ & 0 & 0\\
0 & $ 0.09(12)$ & $ 0.09(10)$ & $-0.027(40)$ & $ 1.597(33)$ & 0 & 0\\
0 & 0 & 0 & 0 & 0 & $ 0.97195(23)$ & $ 0.18510(61)$\\
0 & 0 & 0 & 0 & 0 & $ 0.07468(83)$ & $ 1.6056(32)$\\
\hline
\end{tabular}
\caption{The elements of the $7\times 7$ $\SMOMgg$ (upper) and $\SMOMqq$ (lower) step-scaling matrices $\Lambda(4.006 \mathrm{GeV},1.514 \mathrm{GeV})_{ij}$ between renormalization scales $\mu_1=1.514$ and $\mu_2=4.006$ GeV computed on the 32Ifine ensemble. \label{tab-stepscale1.53_4.0}  }
\end{table}

\begin{table}[tb]
\centering
\begin{tabular}{ccccccc}
\hline
$ 0.40845(42)$ & 0 & 0 & 0 & 0 & 0 & 0\\
0 & $ 0.485(23)$ & $-0.114(20)$ & $-0.012(10)$ & $ 0.0077(63)$ & 0 & 0\\
0 & $-0.0908(93)$ & $ 0.5248(89)$ & $-0.0089(37)$ & $ 0.0061(26)$ & 0 & 0\\
0 & $-0.051(70)$ & $-0.067(58)$ & $ 0.432(30)$ & $-0.003(19)$ & 0 & 0\\
0 & $ 0.021(37)$ & $ 0.025(35)$ & $-0.073(15)$ & $ 0.574(10)$ & 0 & 0\\
0 & 0 & 0 & 0 & 0 & $ 0.47514(49)$ & $-0.01786(21)$\\
0 & 0 & 0 & 0 & 0 & $-0.04460(26)$ & $ 0.55914(99)$\\
\hline
\end{tabular}
\newline
\vspace{1cm}
\newline
\begin{tabular}{ccccccc}
\hline
$ 0.42011(43)$ & 0 & 0 & 0 & 0 & 0 & 0\\
0 & $ 0.422(38)$ & $-0.207(36)$ & $-0.005(13)$ & $ 0.0084(77)$ & 0 & 0\\
0 & $-0.094(24)$ & $ 0.570(24)$ & $-0.0120(83)$ & $ 0.0059(47)$ & 0 & 0\\
0 & $-0.14(14)$ & $-0.15(12)$ & $ 0.424(44)$ & $ 0.013(26)$ & 0 & 0\\
0 & $-0.030(63)$ & $-0.073(66)$ & $-0.106(23)$ & $ 0.620(15)$ & 0 & 0\\
0 & 0 & 0 & 0 & 0 & $ 0.47715(49)$ & $-0.02113(24)$\\
0 & 0 & 0 & 0 & 0 & $-0.05960(55)$ & $ 0.6030(14)$\\
\hline
\end{tabular}
\caption{The elements of the $7\times 7$ $\SMOMgg$ (upper) and $\SMOMqq$ (lower) renormalization matrices $Z(4.006\mathrm{GeV})_{ij}^{{\rm RI}\leftarrow{\rm lat}}$ with renormalization scale $\mu=4.006$ GeV computed by applying the step-scaling matrices in Tab.~\ref{tab-stepscale1.53_4.0} with the renormalization matrices in Tab.~\ref{tab-Zmat1.53}.  This matrix converts the lattice matrix elements computed in this paper to the appropriate RI scheme at $\mu=4.006$ GeV \label{tab-Zmat4.00}  }
\end{table}

\begin{table}[tb]
\centering
\begin{tabular}{ccccccc}
\hline\hline
$ 0.43432(44)$ & 0 & 0 & 0 & 0 & 0 & 0\\
0 & $ 0.487(14)$ & $-0.033(14)$ & $-0.0013(47)$ & $-0.0044(35)$ & 0 & 0\\
0 & $-0.0197(63)$ & $ 0.4949(79)$ & $-0.0029(26)$ & $ 0.0082(22)$ & 0 & 0\\
0 & $-0.006(43)$ & $-0.008(42)$ & $ 0.526(14)$ & $-0.111(10)$ & 0 & 0\\
0 & $ 0.024(19)$ & $ 0.043(22)$ & $-0.0350(73)$ & $ 0.2907(63)$ & 0 & 0\\
0 & 0 & 0 & 0 & 0 & $ 0.49785(50)$ & $-0.10138(10)$\\
0 & 0 & 0 & 0 & 0 & $-0.024002(34)$ & $ 0.27024(28)$\\
\hline
\end{tabular}
\newline
\vspace{1cm}
\newline
\begin{tabular}{ccccccc}
\hline
$ 0.43432(44)$ & 0 & 0 & 0 & 0 & 0 & 0\\
0 & $ 0.488(13)$ & $-0.030(12)$ & $-0.0018(46)$ & $-0.0032(28)$ & 0 & 0\\
0 & $-0.0221(59)$ & $ 0.4874(61)$ & $-0.0015(25)$ & $ 0.0060(16)$ & 0 & 0\\
0 & $-0.005(42)$ & $-0.008(36)$ & $ 0.526(14)$ & $-0.1110(81)$ & 0 & 0\\
0 & $ 0.019(18)$ & $ 0.027(19)$ & $-0.0336(69)$ & $ 0.2872(48)$ & 0 & 0\\
0 & 0 & 0 & 0 & 0 & $ 0.49785(50)$ & $-0.10138(10)$\\
0 & 0 & 0 & 0 & 0 & $-0.024002(34)$ & $ 0.27024(28)$\\
\end{tabular}
\caption{The elements of the $7\times 7$ $\SMOMgg$ renormalization matrix $Z(1.33\mathrm{GeV})_{ij}^{{\rm RI}\leftarrow{\rm lat}}$ with (upper) and without (lower) the effects of the $G_1$ operator included. This matrix converts the lattice matrix elements computed in this paper to the $\SMOMgg$ scheme at $\mu=1.33$ GeV \label{tab-Zmat1.33}  }
\end{table}

\begin{table}[tb]
\centering
\begin{tabular}{ccccccc}
\hline\hline
$ 0.41588(42)$ & 0 & 0 & 0 & 0 & 0 & 0\\
0 & $ 0.500(23)$ & $-0.058(43)$ & $-0.0006(82)$ & $ 0.00000(1300)$ & 0 & 0\\
0 & $-0.055(13)$ & $ 0.507(26)$ & $-0.0055(48)$ & $ 0.0115(79)$ & 0 & 0\\
0 & $ 0.020(68)$ & $-0.01(13)$ & $ 0.496(22)$ & $-0.071(37)$ & 0 & 0\\
0 & $ 0.010(46)$ & $-0.059(93)$ & $-0.032(18)$ & $ 0.392(28)$ & 0 & 0\\
0 & 0 & 0 & 0 & 0 & $ 0.48386(49)$ & $-0.063985(81)$\\
0 & 0 & 0 & 0 & 0 & $-0.035289(72)$ & $ 0.40653(45)$\\
\hline
\end{tabular}
\newline
\vspace{1cm}
\newline
\begin{tabular}{ccccccc}
\hline
$ 0.41588(42)$ & 0 & 0 & 0 & 0 & 0 & 0\\
0 & $ 0.498(15)$ & $-0.063(14)$ & $ 0.0003(53)$ & $-0.0011(33)$ & 0 & 0\\
0 & $-0.0570(72)$ & $ 0.5009(76)$ & $-0.0042(28)$ & $ 0.0088(19)$ & 0 & 0\\
0 & $ 0.024(45)$ & $-0.0010(400)$ & $ 0.494(16)$ & $-0.0672(94)$ & 0 & 0\\
0 & $ 0.051(30)$ & $ 0.040(30)$ & $-0.052(11)$ & $ 0.4245(79)$ & 0 & 0\\
0 & 0 & 0 & 0 & 0 & $ 0.48386(49)$ & $-0.063985(81)$\\
0 & 0 & 0 & 0 & 0 & $-0.035289(72)$ & $ 0.40653(45)$\\
\end{tabular}
\caption{The elements of the $7\times 7$ $\SMOMgg$ renormalization matrix $Z(2.29\mathrm{GeV})_{ij}^{{\rm RI}\leftarrow{\rm lat}}$ with (upper) and without (lower) the effects of the $G_1$ operator included. This matrix converts the lattice matrix elements computed in this paper to the $\SMOMgg$ scheme at $\mu=2.29$ GeV \label{tab-Zmat2.29}  }
\end{table}

\section{Results for $\pmb{ A_0}$ and $\pmb{ \epsilon'}$}
\label{sec:resultsA0epsprime}
In this section we combine our lattice measurements with experimental inputs to obtain ${\rm Re}(\epsilon'/\epsilon)$. The set of Standard Model parameters and other experimental values used for these calculations are listed in Tab.~\ref{tab-expt-params} and their uncertainties are accounted for as a systematic error in the following section. In this table the value of ${\rm Re}(A_2)$ was obtained from the experimental measurement of $K^+\to\pi^+\pi^0$ decays, and the value of ${\rm Re}(A_0)$ from $K_S\to\pi^+\pi^-$ and $K_S\to\pi^0\pi^0$ decays. The relationship between the isospin amplitudes and the experimental branching fractions and decay widths is described in detail in Secs. III.A and III.B of Ref.~\cite{Blum:2012uk}.

As previous mentioned, the Wilson coefficients that incorporate the short distance physics ``integrated out'' from the Standard Model are known in perturbation theory in the 10-operator 
basis to NLO in the $\msbar$ scheme. Partial calculations at NNLO are available in the literature ~\cite{Buras:1999st,Gorbahn:2004my,Brod:2010mj,Cerda-Sevilla:2016yzo,Cerda-Sevilla:2018hjk}, together with a preliminary study on a direct lattice determination ~\cite{Bruno:2017iwk}; in this manuscript we utilize the complete NLO results of Ref.~\cite{Buchalla:1995vs} in the $\msbar$-NDR scheme for our central values, and the LO predictions to assign a systematic error due to the truncation of the perturbative series.

For consistency with the NLO determination of the Wilson coefficients we follow Ref.~\cite{Buchalla:1995vs} in utilizing the 2-loop determination of $\alpha_s$ given in Ref.~\cite{Buchalla:1995vs} (and the 1-loop determination for the LO Wilson coefficients used to estimate the systematic error) despite the fact that a 4-loop calculation is available~\cite{vanRitbergen:1997va}. In order to fix the parameters of the 2-loop (1-loop) calculation, a value of $\alpha_s$ at a reference scale is required, and to minimize the perturbative truncation error it is desirable that this scale be close to the typical scale of the physical problem, in our case ${\cal O}(2\ {\rm GeV})$. We therefore utilize the 4-loop calculation of $\alpha_s$ to run the value of $\alpha_s^{N_f=5}(M_Z)$ given in Tab.~\ref{tab-expt-params} down to 1.7 GeV in the 4-flavor theory, and use the result,
\begin{dmath}
\alpha_s^{N_f=4}(1.7\ {\rm GeV})=0.32733
\end{dmath}
as input to our 2-loop (1-loop) calculation. (The reason for choosing this scale will be discussed in Sec.~\ref{sec-syserr-wilsoncoeffs}.)


\begin{table}[tbp]
\begin{tabular}{ccr}
\hline\hline
Quantity & Value\\
\hline
$G_F$ & 1.16638$\times 10^{-5}$ GeV$^{-2}$ \\
$V_{ud}$ & 0.97420 \\
$V_{us}$ & 0.2243 \\
$\phi_\epsilon$ & 0.7596 rad\\
$\tau$ & 0.001558(65) -0.000663(33)i \hspace{0.5cm} & (*)\\
$|\epsilon|$ & 0.002228(11)  & ($\dagger$)\\
$\omega$ & 0.04454(12) & ($\dagger$) \\
${\rm Re}(A_0)_{\rm expt}$ & $3.3201(18)\times 10^{-7}$ GeV & ($\dagger$)\\
${\rm Re}(A_2)_{\rm expt}$ & $1.479(4)\times 10^{-8}$ GeV & ($\dagger$)\\
\hline
$m_c(m_c)$ & 1.27(2) GeV & (*)\\
$m_b(m_b)$ & 4.18(3) GeV & (*)\\
$m_W(m_W)$ & 80.379(12) GeV & (*)\\
$m_Z(m_Z)$ & 91.1876(21) GeV & (*)\\
$m_t(m_t)$ & 160.0(4.8) GeV & (*)\\
$\alpha_s^{N_f=5}(m_Z)$ & 0.1181 \\
$\alpha$ & $1/127.955(10)$ & (*)\\
$\sin^2(\theta_W)$ & 0.23122(3)& (*)\\
\end{tabular}
\caption{Standard Model and other experimental inputs required to determine $A_0$ and ${\rm Re}(\epsilon'/\epsilon)$ from the lattice matrix elements. The parameters given in this table were obtained from the PDG Review of Particle Physics~\cite{PhysRevD.98.030001}, apart from those of ${\rm Re}(A_0)$, ${\rm Re}(A_2)$ and their ratio, $\omega$, which were taken from Ref.~\cite{Bai:2015nea}. Here $\phi_\epsilon$ is the phase of the indirect CP-violation parameter $\epsilon$. The CKM ratio $\tau= -V_{ts}^* V_{td}/V_{us}^* V_{ud}$ is obtained using the Wolfenstein parameterization expanded to eighth order, with parameters taken from the aforementioned review. The impact upon our result of the errors on those quantities marked with a $(*)$ is incorporated as a systematic error in Sec.~\ref{sec-syserr-parametric}. The errors on those quantities marked with $(\dagger)$ are included within the quoted statistical errors on our results. The errors on the remaining quantities are neglected as their contributions to our final error are small in comparison to our statistical error. \label{tab-expt-params} }
\end{table}

\subsection{Lellouch-L\"uscher factor}
\label{sec-llfactor}

The Lellouch-L\"uscher factor $F$~\cite{Lellouch:2000pv} removes the leading power-law finite-volume corrections to the lattice matrix element. It is defined as  
\begin{dmath}
F^2 = \frac{ 4\pi m_K E_{\pi\pi}^2}{k^3} \left( k \frac{d\delta_0}{dk} + q \frac{d\phi}{dq} \right)\,,
\label{eq:LL-factor}
\end{dmath}
where $\delta_0$ is the $I=0$ $\pi\pi$ scattering phase shift and $\phi$ is a known function~\cite{Luscher:1990ux} of $q = \frac{Lk}{2\pi}$, appropriately modified for our antiperiodic pion boundary conditions~\cite{Yamazaki:2004qb}, with $k$ the interacting pion momentum defined via
\begin{dmath}
k^2 = \left(\frac{E_{\pi\pi}}{2}\right)^2 - m_\pi^2\label{eq-llkdef}\,.
\end{dmath}
Note that Eq.~\eqref{eq:LL-factor} differs by a factor of two from the corresponding equation in Ref.~\cite{Lellouch:2000pv} due to our different conventions on the decay amplitude (cf. Ref.~\cite{Blum:2011pu}).

The calculation of the Lellouch-L\"uscher factor requires the derivative of the phase shift with respect to interacting pion momentum, or correspondingly the $\pi\pi$ energy, evaluated at the kaon mass. The determination of this derivative is detailed in Sec.~\ref{sec-phase-shift-deriv} where we present values obtained both directly from the lattice and also from the dispersive prediction. Given the good agreement between our phase shifts and the dispersive predictions~\cite{pipi-paper} we will use the dispersive result given in Eq.~\eqref{eq-delta-deriv-colangelo}. The variation in the results will be incorporated as a systematic error in Sec.~\ref{sec-llfactorsys}.

We find
\begin{dmath}
F = 26.696(52)\,,\label{eq-LLfactor} 
\end{dmath}
where the error arises primarily from the uncertainty in measured $\pi\pi$ energy and its small size results from the small contribution of the $\pi\pi$ scattering phase shift relative to that of the known function $\phi$ in Eq.~\eqref{eq:LL-factor}.

\subsection{Renormalized physical matrix elements}
\label{sec-renorm_phys_matelems}

The infinite-volume matrix elements of the seven chiral-basis operators $Q^{\prime\,R}_j$ in a scheme $R$ at the scale $\mu$ can be expressed without ambiguity in terms of the matrix elements $M^{\prime\,\rm lat}_j = \langle\pi\pi|Q^{\prime\,\rm lat}_j|K\rangle$ of the corresponding lattice operators:
\begin{dmath}
M^{\prime\,R}(\mu) = Z^{R\leftarrow\mathrm{lat}}(\mu) \left(a^{-3}F M^{\prime\,\rm lat}\right)\,,
\label{eq:lat2R}
\end{dmath}
where $a$ is the lattice spacing, $Z^{R\leftarrow\mathrm{lat}}(\mu)$ a $7\times 7$ renormalization matrix and $F$ the Lellouch-L\"uscher factor obtained in Eq.~\eqref{eq-LLfactor}.   

The ten conventional, linearly-dependent operators $Q_i$ are defined in terms of the seven independent operators $Q^\prime_j$ as follows:
\begin{dmath}
Q_i = \sum_i T_{ij} Q^\prime_j,
\label{eq:Q10-def}
\end{dmath}
where $1\le i \le 10$, $j$ runs over the set $\{1,2,3,5,6,7,8\}$ and the matrix $T$ is given by
\begin{dmath}
T = \left(\begin{array}{ccccccc}
1/5   &  1 &  0 & 0 & 0 & 0 & 0\\
1/5   &  0 &  1 & 0 & 0 & 0 & 0\\
0      &  3 &  2 & 0 & 0 & 0 & 0\\
0      &  2 &  3 & 0 & 0 & 0 & 0\\
0      &  0 &  0 & 1 & 0 & 0 & 0\\
0      &  0 &  0 & 0 & 1 & 0 & 0\\
0      &  0 &  0 & 0 & 0 & 1 & 0\\
0      &  0 &  0 & 0 & 0 & 0 & 1\\
3/10 &  0 & -1 & 0 & 0 & 0 & 0\\
3/10 & -1 &  0 & 0 & 0 & 0 & 0\\
\end{array}\right)
\label{eq:T}
\end{dmath}
which can be found as Eqs.~(58) and (59) of Ref.~\cite{Lehner:2011fz}. This relationship applies both to RI scheme and bare lattice operators. 

In our lattice calculation we have evaluated the matrix elements of all ten linearly-dependent operators $Q_i$ as given in Tab.~\ref{tab-fitresultfinal}. This gives us a consistency test of the three Fierz identities: these identities are obeyed to within statistical errors and with an absolute size at the 1\% level, validating our code.  We do not expect the Fierz relations to be obeyed to floating point accuracy since our use of all-to-all propagators introduces a stochastic element into the inversion of the Dirac operator and our use of $\gamma^5$ hermiticity differs between the ten operators introducing statistical noise in different ways into each evaluation.

Since the Fierz identities are not obeyed exactly by the data in Tab.~\ref{tab-fitresultfinal}, we have a choice as to how the ten linearly-dependent matrix elements $M_i^{\rm lat}$ in that table are to be combined to give the seven independent matrix elements $M^{\prime\,{\rm lat}}_i$ needed on the right-hand side of Eq.~\eqref{eq:lat2R}. To this end we choose to treat $M^{\prime\,{\rm lat}}_i$ as fit parameters whose best fit values are obtained by minimizing the correlated $\chi^2$:
\begin{dmath}
\chi^2 = \sum_{ij=1}^{10} \left( M_i^{\rm lat} - \sum_{k=1}^{7} T_{ik}M^{\prime\,{\rm lat}}_k  \right) (C^{-1})_{ij} \left( M_j^{\rm lat} - \sum_{\ell=1}^7 T_{j\ell}M^{\prime\,{\rm lat}}_\ell \right)\,. \label{eq-chiralconvfit-chi2}
\end{dmath}
The result is an optimal combination that provably minimizes the statistical error on the resulting $M^{\prime\,{\rm lat}}_i$. The $10\times 10$ covariance matrix $C_{ij}$ is estimated by studying the variation of the bootstrap means of the matrix elements, and is given in Tab.~\ref{tab-10x10latmatelem_covmat}. Note that we use the same covariance matrix for the fit to each bootstrap sample (a frozen fit) and therefore do not take into account in our errors the fluctuations in the covariance matrix over bootstrap samples. However such effects are expected to be minimal due to our large number of configurations. The results for the bare matrix elements obtained by this procedure, along with those obtained by applying Eq.~\eqref{eq:Q10-def} to convert those results back into the 10-basis, are given in Tab.~\ref{tab-barematelem-chiralconv}. These results are quoted in physical units and incorporate the Lellouch-L\"uscher finite-volume correction.

\begin{table}[tb]
\centering
\resizebox{\textwidth}{!}{
\begin{tabular}{cccccccccc}
\hline
0.001217 & 0.0001759 & 0.001208 & 0.0006908 & 0.001206 & 0.0001964 & 0.0004749 & 7.289$\times 10^{-5}$ & 0.0005008 & -2.695$\times 10^{-5}$ \\
0.0001759 & 0.0008377 & 0.0003157 & 0.001220 & 0.0004747 & 0.0008078 & 0.0004188 & 0.0009140 & 5.226$\times 10^{-5}$ & 0.0003670 \\
0.001208 & 0.0003157 & 0.006443 & 0.003560 & 0.003463 & 0.003764 & -0.0001617 & -0.0007452 & -0.0009426 & -0.001024 \\
0.0006908 & 0.001220 & 0.003560 & 0.008397 & 0.002873 & 0.006152 & 6.055$\times 10^{-6}$ & -0.0002789 & -0.0003660 & -0.001078 \\
0.001206 & 0.0004747 & 0.003463 & 0.002873 & 0.008692 & 0.004380 & -0.0006516 & -0.001387 & -0.0008054 & -0.0003295 \\
0.0001964 & 0.0008078 & 0.003764 & 0.006152 & 0.004380 & 0.02195 & -0.001279 & -0.006099 & -0.0003987 & -0.001377 \\
0.0004749 & 0.0004188 & -0.0001617 & 6.055$\times 10^{-6}$ & -0.0006516 & -0.001279 & 0.002804 & 0.003961 & 0.001241 & 0.0006063 \\
7.289$\times 10^{-5}$ & 0.0009140 & -0.0007452 & -0.0002789 &-0.001387 & -0.006099 & 0.003961 &  0.01150 & 0.0004234 & 0.001589 \\
0.0005008 & 5.226$\times 10^{-5}$ & -0.0009426 & -0.0003660 &-0.0008054 & -0.0003987 & 0.001241 & 0.0004238 & 0.002475 & 0.0003710 \\
-2.695$\times 10^{-5}$ & 0.0003670 & -0.001024 & -0.001078 &-0.0003295 & -0.001377 & 0.0006063 & 0.001589 & 0.0003710 & 0.001571 \\
\hline
\end{tabular}
}
\caption{The $10\times 10$ covariance matrix $C_{ij}$ between the unrenormalized, infinite-volume lattice operators in the conventional basis and physical units of GeV$^3$. \label{tab-10x10latmatelem_covmat} }
\end{table}

\begin{table}[tb]
\centering
\begin{tabular}{l|l|l}
\hline\hline
i & $Q^\prime_i\ ({\rm GeV}^3)$ & $Q_i\ ({\rm GeV}^3)$\\
\hline
1 & $ 0.143(93)$ & $-0.119(32)$\\
2 & $-0.147(24)$ & $ 0.261(27)$\\
3 & $ 0.233(23)$ & $ 0.023(74)$\\
4 & - & $ 0.403(72)$\\
5 & $-0.723(91)$ & $-0.723(91)$\\
6 & $-2.211(144)$ & $-2.211(144)$\\
7 & $ 1.876(52)$ & $ 1.876(52)$\\
8 & $ 5.679(107)$ & $ 5.679(107)$\\
9 & - & $-0.190(39)$\\
10 & - & $ 0.190(35)$\\
\end{tabular}
\caption{The bare lattice matrix elements in the 7-operator chiral basis (second column) that minimize the correlated $\chi^2$ Eq.~\eqref{eq-chiralconvfit-chi2}, and those results converted back into the 10-operator basis by applying Eq.~\eqref{eq:Q10-def} (third column). These results are quoted in physical units and incorporate the Lellouch-L\"uscher finite-volume correction. The errors are statistical, only.\label{tab-barematelem-chiralconv} }
\end{table}

The results for the seven operators converted to the $\SMOMgg$ and $\SMOMqq$ schemes are given in the left two columns of Tab.~\ref{tab-renorm-matrix-elems}. The right two columns of that table show the matrix elements of the ten conventional operators in the $\msbar$ scheme obtained from the left two columns by an application of Eqs.~\eqref{eq:Q10-def} and \eqref{eq:T}. For the convenience of the reader in utilizing these results we also provide the covariance matrices for the $\SMOMqq$ scheme matrix elements, which we will use as our central values in Sec.~\ref{sec:finalresults}, and also the $\msbar$ matrix elements derived from them, in Tabs.~\ref{tab-7x7RIqqmatelem_covmat} and~\ref{tab-10x10MSbarmatelem_covmat}, respectively.

\begin{table}[tb]
\centering
\begin{tabular}{l|l|l||l|l}
\hline\hline
i & SMOM$(\slashed{q},\slashed{q})$ (GeV$^3$) & SMOM$(\gamma^\mu,\gamma^\mu)$ (GeV$^3$) & $\overline{\rm MS}$ via SMOM$(\slashed{q},\slashed{q})$ (GeV$^3$) & $\overline{\rm MS}$ via SMOM$(\gamma^\mu,\gamma^\mu)$ (GeV$^3$)\\
\hline
1 & $ 0.060(39)$ & $ 0.059(38)$ & $-0.107(22)$ & $-0.093(18)$\\
2 & $-0.125(19)$ & $-0.106(16)$ & $ 0.147(15)$ & $ 0.143(14)$\\
3 & $ 0.142(17)$ & $ 0.128(14)$ & $-0.086(61)$ & $-0.053(44)$\\
4 & - & - & $ 0.185(53)$ & $ 0.200(40)$\\
5 & $-0.351(62)$ & $-0.313(48)$ & $-0.348(62)$ & $-0.311(48)$\\
6 & $-1.306(90)$ & $-1.214(82)$ & $-1.308(90)$ & $-1.272(86)$\\
7 & $ 0.775(23)$ & $ 0.790(23)$ & $ 0.769(23)$ & $ 0.784(23)$\\
8 & $ 3.312(63)$ & $ 3.092(58)$ & $ 3.389(64)$ & $ 3.308(63)$\\
9 & - & - & $-0.117(20)$ & $-0.114(19)$\\
10 & - & - & $ 0.137(22)$ & $ 0.123(19)$\\
\end{tabular}
\caption{Physical, infinite-volume matrix elements in the $\SMOMqq$ and $\SMOMgg$ schemes at $\mu=4.006$ GeV given in the 7-operator chiral basis, as well as those converted perturbatively into the $\msbar$ scheme at the same scale in the 10-operator basis. The errors are statistical only.\label{tab-renorm-matrix-elems} }
\end{table}

\begin{table}[tb]
\centering
\begin{tabular}{ccccccc}
\hline
$ 0.001516$ & $ 5.385\times 10^{-5}$ & $-9.167\times 10^{-5}$ & $ 0.0001252$ & $-0.0003965$ & $ 0.0004930$ & $ 0.0007192$\\
$ 5.385\times 10^{-5}$ & $ 0.0003563$ & $-4.099\times 10^{-5}$ & $ 0.0007596$ & $ 0.0002981$ & $ 2.914\times 10^{-5}$ & $-0.0002118$\\
$-9.167\times 10^{-5}$ & $-4.099\times 10^{-5}$ & $ 0.0002808$ & $ 0.0003784$ & $ 0.0004679$ & $-4.656\times 10^{-5}$ & $ 0.0001516$\\
$ 0.0001252$ & $ 0.0007596$ & $ 0.0003784$ & $ 0.003904$ & $ 0.001679$ & $-8.000\times 10^{-5}$ & $-0.0004013$\\
$-0.0003965$ & $ 0.0002981$ & $ 0.0004679$ & $ 0.001679$ & $ 0.008188$ & $-0.0003817$ & $-0.002110$\\
$ 0.0004930$ & $ 2.914\times 10^{-5}$ & $-4.656\times 10^{-5}$ & $-8.000\times 10^{-5}$ & $-0.0003817$ & $ 0.0005395$ & $ 0.0009460$\\
$ 0.0007192$ & $-0.0002118$ & $ 0.0001516$ & $-0.0004013$ & $-0.002110$ & $ 0.0009460$ & $ 0.003937$\\
\hline
\end{tabular}
\caption{The $7\times 7$ covariance matrix between the renormalized, infinite-volume matrix elements in the $\SMOMqq$ scheme in the chiral basis. \label{tab-7x7RIqqmatelem_covmat} }
\end{table}

\begin{table}[tb]
\centering
\resizebox{\textwidth}{!}{
\begin{tabular}{cccccccccc}
\hline
$ 0.0004628$ & $ 8.315\times 10^{-6}$ & $ 0.001058$ & $ 0.0005998$ & $ 0.0008504$ & $ 0.0002622$ & $ 0.0001246$ & $-6.882\times 10^{-5}$ & $ 0.0001651$ & $-0.0002894$\\
$ 8.315\times 10^{-6}$ & $ 0.0002367$ & $ 0.0002796$ & $ 0.0004981$ & $ 0.0002866$ & $ 0.0002532$ & $ 5.669\times 10^{-5}$ & $ 0.0003026$ & $-0.0001273$ & $ 0.0001010$\\
$ 0.001058$ & $ 0.0002796$ & $ 0.003749$ & $ 0.002929$ & $ 0.002999$ & $ 0.001681$ & $-7.629\times 10^{-7}$ & $-0.0003280$ & $-0.0002872$ & $-0.001066$\\
$ 0.0005998$ & $ 0.0004981$ & $ 0.002929$ & $ 0.002784$ & $ 0.002406$ & $ 0.001524$ & $-6.156\times 10^{-5}$ & $ 7.545\times 10^{-5}$ & $-0.0005649$ & $-0.0006666$\\
$ 0.0008504$ & $ 0.0002866$ & $ 0.002999$ & $ 0.002406$ & $ 0.003902$ & $ 0.001607$ & $-7.840\times 10^{-5}$ & $-0.0004062$ & $-0.0002240$ & $-0.0007878$\\
$ 0.0002622$ & $ 0.0002532$ & $ 0.001681$ & $ 0.001524$ & $ 0.001607$ & $ 0.008059$ & $-0.0003739$ & $-0.002158$ & $-0.0004472$ & $-0.0004561$\\
$ 0.0001246$ & $ 5.669\times 10^{-5}$ & $-7.629\times 10^{-7}$ & $-6.156\times 10^{-5}$ & $-7.840\times 10^{-5}$ & $-0.0003739$ & $ 0.0005361$ & $ 0.0009564$ & $ 0.0001873$ & $ 0.0001194$\\
$-6.882\times 10^{-5}$ & $ 0.0003026$ & $-0.0003280$ & $ 7.545\times 10^{-5}$ & $-0.0004062$ & $-0.002158$ & $ 0.0009564$ & $ 0.004120$ & $ 6.076\times 10^{-5}$ & $ 0.0004322$\\
$ 0.0001651$ & $-0.0001273$ & $-0.0002872$ & $-0.0005649$ & $-0.0002240$ & $-0.0004472$ & $ 0.0001873$ & $ 6.076\times 10^{-5}$ & $ 0.0003912$ & $ 9.882\times 10^{-5}$\\
$-0.0002894$ & $ 0.0001010$ & $-0.001066$ & $-0.0006666$ & $-0.0007878$ & $-0.0004561$ & $ 0.0001194$ & $ 0.0004322$ & $ 9.882\times 10^{-5}$ & $ 0.0004892$\\
\hline
\end{tabular}
}
\caption{The $10\times 10$ covariance matrix between the renormalized, infinite-volume matrix elements in the $\msbar$ scheme in the chiral basis obtained using the $\SMOMqq$ intermediate scheme. \label{tab-10x10MSbarmatelem_covmat} }
\end{table}

\subsection{Results for $\pmb{A_0}$}
\label{sec-A0results-staterrsonly}

\begin{table}[tb]
\begin{tabular}{c|cc}
\hline\hline
 $i$ & $y_i$ & $z_i$ \\
\hline
1 & 0 & -0.199111\\
2 & 0 & 1.08976\\
3 & 0.0190166 & -0.00525073\\
4 & -0.0560629 & 0.0244698\\
5 & 0.0132642 & -0.00607434\\
6 & -0.0562033 & 0.0174607\\
7 & -0.000271245 & 0.000134906\\
8 & 0.000521236 & -0.000119628\\
9 & -0.00946862 & 5.60698e-05\\
10 & 0.00186152 & 9.34113e-05\\
\end{tabular}
\caption{The $\msbar$ Wilson coefficients $\vec y$ and $\vec z$ at $\mu=4.006$ GeV computed via NLO QCD+EW perturbation theory. \label{tab-wilson-coeffs} }
\end{table}

We can now obtain $A_0$ from our lattice calculation as follows:
\begin{dmath}
A_0 = \frac{G_F}{\sqrt{2}}V^*_{us}V_{ud} \sum_{i=1}^{10}\left( z_i^\msbar(\mu) + \tau y_i^\msbar(\mu)\right)M^\msbar_i(\mu)\,.
\end{dmath}
The Wilson coefficients have been computed to next-to-leading order in QCD and electroweak perturbation theory in the $\msbar$ scheme~\cite{Buchalla:1995vs}, and at $\mu=4.006$ GeV take the values given in Tab.~\ref{tab-wilson-coeffs}. For the CKM matrix element ratio $\tau$ we use the value given in Tab.~\ref{tab-expt-params}. Combining these with the $\msbar$-renormalized matrix elements obtained in Tab.~\ref{tab-renorm-matrix-elems} we obtain the following for the $\SMOMqq$ intermediate scheme,
\begin{dgroup}
\begin{dmath}
{\rm Re}(A_0) = 2.99(32)\times 10^{-7}\ {\rm GeV}\,, \label{eq-reA0lat-qq}  
\end{dmath}
\begin{dmath}
{\rm Im}(A_0) = -7.15(66)\times 10^{-11}\ {\rm GeV}\,. \label{eq-imA0lat-qq} 
\end{dmath}
\end{dgroup}
and for the $\SMOMgg$ intermediate scheme,
\begin{dgroup}
\begin{dmath}
{\rm Re}(A_0) = 2.86(31)\times 10^{-7}\ {\rm GeV}\,, \label{eq-reA0lat-gg} 
\end{dmath}
\begin{dmath}
{\rm Im}(A_0) = -6.93(64)\times 10^{-11}\ {\rm GeV}\,. \label{eq-imA0lat-gg} 
\end{dmath}
\end{dgroup}

The values of ${\rm Re}(A_0)$ agree to $4.1(4.2)\%$ between the two schemes, and those of ${\rm Im}(A_0)$ to $3.1(3.8)\%$. This excellent agreement suggests that the systematic errors resulting from discretization effects and the truncation of the perturbative series in the non-perturbative renormalization are minimal at our high 4 GeV scale. In the following section a more detailed discussion of these systematic errors is presented.

\begin{table}[tb]
\centering
\begin{tabular}{l|ll|ll}
\hline\hline
  &  \multicolumn{2}{|c|}{Re($A_0$)}                                & \multicolumn{2}{|c}{Im($A_0$)} \\
\hline
i & $(\slashed{q},\slashed{q})$ ($\times 10^{-7}$ GeV) & $(\gamma^\mu,\gamma^\mu)$ ($\times 10^{-7}$ GeV) & $(\slashed{q},\slashed{q})$ ($\times 10^{-11}$ GeV) & $(\gamma^\mu,\gamma^\mu)$ ($\times 10^{-11}$ GeV)\\
\hline
1 & $ 0.383(77)$ 			& $ 0.335(64)$ 			& 0 				& 0\\
2 & $ 2.89(30)$ 			& $ 2.81(28)$ 			& 0 				& 0\\
3 & $ 0.0081(58)$ 			& $ 0.0050(42)$ 		& $ 0.20(14)$ 			& $ 0.12(10)$\\
4 & $ 0.081(23)$ 			& $ 0.088(17)$ 			& $ 1.24(35)$ 			& $ 1.34(27)$\\
5 & $ 0.0380(68)$ 			& $ 0.0339(53)$ 		& $ 0.552(99)$ 			& $ 0.492(77)$\\
6 & $-0.410(28)$ 			& $-0.398(27)$ 			& $-8.78(60)$ 			& $-8.54(57)$\\
7 & $ 0.001863(56)$ 			& $ 0.001900(56)$ 		& $ 0.02491(75)$ 		& $ 0.02540(75)$\\
8 & $-0.00726(14)$ 			& $-0.00708(13)$ 		& $-0.2111(40)$ 		& $-0.2060(39)$\\
9 & $-8.7(1.5)\times 10^{-5}$ 		& $-8.5(1.4)\times 10^{-5}$ 	& $-0.133(22)$ 			& $-0.128(21)$\\
10 & $ 2.37(38)\times 10^{-4}$ 		& $ 2.13(32)\times 10^{-4}$ 	& $-0.0304(49)$ 		& $-0.0273(41)$\\
\hline
Total & $ 2.99(32)$ & $ 2.86(31)$ & $-7.15(66)$ & $-6.93(64)$\\
\end{tabular}
\caption{The contributions of each of the ten four-quark operators to ${\rm Re}(A_0)$ and ${\rm Im}(A_0)$ for the two different RI-SMOM intermediate schemes. The scheme and units are listed in the column headers. The errors are statistical, only.\label{tab-A0contribs} }
\end{table}

The contributions of each of the ten operators to the real and imaginary parts of $A_0$ are given in Tab.~\ref{tab-A0contribs}. The result for ${\rm Im}(A_0)$ is dominated by the $Q_6$ matrix element with a 14(4)\% 
cancelation from $Q_4$, where the errors are statistical only and the value is obtained using the $\SMOMqq$ intermediate scheme to match the scheme used for the previous work. This is in contrast to the 51(29)\%-level cancelation observed in Ref.~\cite{Bai:2015nea} and is largely due to a $5.5\sigma$ 
increase in the $Q_6$ contribution from $-3.57(91)\times 10^{-11}$ GeV to $-8.78(60)\times 10^{-11}$ GeV 
(again using the $\SMOMqq$ intermediate scheme). This change appears to largely result from excited-state contamination in our previous result, as we can see in Fig.~\ref{fig-matelemQ1toQ6} comparing the (larger-statistics) single-operator result at the value of $t'_{\rm min}=4$ used for our previous work to our favored three-operator, two-state result with $t'_{\rm min}=5$. This suggests that the 5\% systematic error we formerly associated with excited-state contamination was significantly underestimated.

\subsection{Incorporating experimental results to improve the determination of Im($\pmb{A_0}$)}

The real and imaginary parts of $A_0$ comprise different linear combinations of the same basis of real lattice matrix elements. As the real part of the amplitude is precisely known from experiment and is not expected to receive significant contributions from new physics, we can use this quantity to replace part of the lattice input and thereby improve the precision of the imaginary part. The appropriate procedure is discussed in Refs.~\cite{Buras:2015yba,Kitahara:2016nld} in the context of the conventional basis of 10 non-independent operators, where the latter authors use it to eliminate the $Q_2$ matrix element. For our purpose it is more convenient to express the method in terms of the unrenormalized matrix elements in the 7-operator basis. We write
\begin{eqnarray}
{\rm Re}(A_0) &=& \frac{G_F}{\sqrt{2}}V^*_{us}V_{ud}\sum_{k=1}^{7} {\rm Re}(w^{\msbar\leftarrow {\rm lat}}_k) M_k^{\prime\,\rm lat}
\label{eq:ReA0} \\
{\rm Im}(A_0) &=& \frac{G_F}{\sqrt{2}}V^*_{us}V_{ud}\sum_{k=1}^{7} {\rm Im}(w^{\msbar\leftarrow {\rm lat}}_k) M_k^{\prime\rm lat}
\label{eq:ImA0}
\end{eqnarray}
where the $M_j^{\prime\,\rm lat} = \langle\pi\pi|Q'_k|K\rangle$ are the matrix elements of the unrenormalized lattice operators in the 7-basis in infinite-volume and physical units, and
\begin{eqnarray}
{\rm Re}(w^{\msbar\leftarrow {\rm lat}}_k) &=& \sum_{i=1}^{10}\sum_{j=1}^7\left(z_i^\msbar + {\rm Re}(\tau) y_i^\msbar\right)T_{ij}Z_{jk}^{\msbar\leftarrow {\rm lat}} \\
{\rm Im}(w^{\msbar\leftarrow {\rm lat}}_k) &=& \sum_{i=1}^{10}\sum_{j=1}^7\left({\rm Im}(\tau) y_i^\msbar\right)T_{ij} Z_{jk}^{\msbar\leftarrow {\rm lat}}
\end{eqnarray}
are the ``lattice Wilson coefficients''. Here $T_{ij}$ is the $10\times 7$ matrix expressing the 10 linearly-dependent operators in terms of the seven independent operators in the chiral basis, given in Eq.~\eqref{eq:T}.  The matrix $Z^{\msbar\leftarrow {\rm lat}}$ is the product of the $7\times 7$ perburbative matrix expressing the seven $\msbar$ operators in terms of the seven RI operators and the non-perturbative $7\times 7$ matrix which determines the RI operators in terms of the lattice operators.  

We can then use Eq.~\eqref{eq:ReA0} to remove the matrix element of the operator $Q'_\ell$ from Im$(A_0)$ if we write
\begin{eqnarray}
{\rm Im}(A_0) &=& \frac{G_F}{\sqrt{2}}V^*_{us}V_{ud}\sum_{k=1}^{7} {\rm Im}(w^{\msbar\leftarrow {\rm lat}}_k) M_k^{\prime\,\rm lat} \nonumber\\ 
&& \hskip 0.8 in 
+ \lambda \left[{\rm Re}(A_0) - \frac{G_F}{\sqrt{2}}V^*_{us}V_{ud}\sum_{k=1}^{7} {\rm Re}(w^{\msbar\leftarrow {\rm lat}}_k) M_k^{\prime\,\rm lat}\right]
\end{eqnarray}
and choose
\begin{dmath}
\lambda = \frac{{\rm Im}(w^{\msbar\leftarrow {\rm lat}}_\ell)}
                       {{\rm Re}(w^{\msbar\leftarrow {\rm lat}}_\ell)}
\end{dmath}

\begin{table}[tp]
\centering
\begin{tabular}{l|ll}
\hline\hline
i & SMOM$(\slashed{q},\slashed{q})$ ($\times 10^{-11}$ GeV) & SMOM$(\gamma^\mu,\gamma^\mu)$ ($\times 10^{-11}$ GeV)\\
\hline
1 & $-7.12(65)$ & $-6.89(63)$\\
2 & $-7.26(72)$ & $-7.23(75)$\\
3 & $-6.98(62)$ & $-6.65(58)$\\
5 & $-5.05(1.98)$ & $-3.72(2.09)$\\
6 & $-0.23(6.16)$ & $ 0.81(4.92)$\\
7 & $-2.09(4.67)$ & $-0.11(4.40)$\\
8 & $ 2.39(9.00)$ & $ 6.07(8.58)$\\
\end{tabular}
\caption{Values of ${\rm Im}(A_0)$ obtained for each of the two intermediate schemes by eliminating lattice data for the matrix element of operator $Q'_\ell$ in favor of experimental value for Re$(A_0)$.\label{tab-imA0-elimvals} }
\end{table}

In Tab.~\ref{tab-imA0-elimvals} we present values for ${\rm Im}(A_0)$ obtained through using this procedure to replace successive lattice matrix elements. The most significant gain in statistical error is achieved by replacing the matrix element $M^{\prime\,\rm lat}_3$, for which we obtain the following for the $\SMOMqq$ intermediate scheme,
\begin{dmath}
{\rm Im}(A_0) = -6.98(62)\times 10^{-11}\ {\rm GeV}\label{eq-imA0-exptinput-qq} 
\label{eq:ImA0-wo-M_3qq}
\end{dmath}
and for the $\SMOMgg$ intermediate scheme,
\begin{dmath}
{\rm Im}(A_0) = -6.65(58)\times 10^{-11}\ {\rm GeV}\label{eq-imA0-exptinput-gg} 
\label{eq:ImA0-wo-M_3gg}
\end{dmath}
which have 6\% smaller statistical errors.

We could instead choose the parameter $\lambda$ to give that result for ${\rm Im}(A_0)$ with the smallest statistical error.  Since the value obtained for $\lambda$ from this procedure is extremely close to that needed to remove the matrix element $M^{\prime\,\rm lat}_3$, we adopt the simpler procedure of eliminating $M^{\prime\,\rm lat}_3$ and the results given in Eqs.~\eqref{eq:ImA0-wo-M_3qq} and \eqref{eq:ImA0-wo-M_3gg}.


\subsection{Determination of $\pmb{\epsilon'}$}

${\rm Re}(\epsilon'/\epsilon)$ can now be obtained via Eq.~\eqref{eq-epsilonprimefromA}. We use the lattice values for the $I=0$ and $I=2$ $\pi\pi$ scattering phase-shifts : $\delta_0$ is given in Eq.~\eqref{eq-delta0value} and for $\delta_2$ we use
\begin{dmath}
\delta_2 = -11.6(2.5)(1.2)^\circ\,,
\end{dmath}
obtained from our continuum result~\cite{Blum:2015ywa}. Here the parentheses list the statistical error and an estimate of the excited-state systematic error, respectively. 

Writing $\epsilon = |\epsilon|e^{i\phi_\epsilon}$, where both $|\epsilon|$ and its phase $\phi_\epsilon$ can be found in Tab.~\ref{tab-expt-params}, the overall complex phase of $\epsilon'/\epsilon$ is
\begin{dmath}
i e^{i(\delta_2-\delta_0)}e^{-i\phi_\epsilon} = e^{i(\delta_2-\delta_0+\pi/2-\phi_\epsilon)}\,.
\end{dmath}
The resulting real part of the complex phase,
\begin{dmath}
\cos(\delta_2-\delta_0+\pi/2-\phi_\epsilon) = 0.999(2)\,,  
\end{dmath}
is in complete agreement with the value of 0.9998(2) obtained by combining PDG inputs~\cite{PhysRevD.98.030001} and the dispersive values for the phase shifts~\cite{Colangelo:2001df}.

For our primary result we use the more precise experimental values of ${\rm Re}(A_0)$ and ${\rm Re}(A_2)$, and use the results for ${\rm Im}(A_0)$ given in Eqs.~\eqref{eq-imA0-exptinput-qq} and~\eqref{eq-imA0-exptinput-gg} that incorporate the experimental value of ${\rm Re}(A_0)$. The continuum, lattice value for ${\rm Im}(A_2)$ is given in Eq.~64 of Ref.~\cite{Blum:2015ywa} and must be corrected for the 20\% change of ${\rm Im}(\tau)=-0.0005558$ used in that work to the value given in Tab.~\ref{sec-syserr-parametric}. We obtain,
\begin{dmath}
{\rm Im}(A_2) = -8.34(1.03)\times 10^{-13}\ {\rm GeV}\label{eq-imA2-lat-corrected}   
\end{dmath}
For the $\SMOMqq$ intermediate scheme we find
\begin{dmath}
{\rm Re}(\epsilon'/\epsilon) = 0.00217(26)\label{eq-ep-div-e-qq-statonly} 
\end{dmath}
and for the $\SMOMgg$ intermediate scheme,
\begin{dmath}
{\rm Re}(\epsilon'/\epsilon) = 0.00203(25)\,,\label{eq-ep-div-e-gg-statonly} 
\end{dmath}
where the error is statistical only.

It is illustrative to break the value of ${\rm Re}(\epsilon'/\epsilon)$ into the so-called ``QCD penguin''
\begin{dmath}
{\rm Re}\left(\frac{\varepsilon'}{\varepsilon}\right)_{\rm QCDP}\ = -\frac{ \omega\cos(\delta_2-\delta_0+\pi/2-\phi_\epsilon) }{\sqrt{2}|\varepsilon|}\frac{\mathrm{Im}A_0}{\mathrm{ReA}_0}
\end{dmath}
and ``electroweak penguin'' 
\begin{dmath}
{\rm Re}\left(\frac{\varepsilon'}{\varepsilon}\right)_{\rm EWP}\ = \frac{ \omega\cos(\delta_2-\delta_0+\pi/2-\phi_\epsilon) }{\sqrt{2}|\varepsilon|}\frac{\mathrm{Im}A_2}{\mathrm{ReA}_2}
\end{dmath}
contributions, the sum of which is equal to ${\rm Re}(\epsilon'/\epsilon)$. These terms have opposite sign such that the sum involves an important cancellation. For the electroweak penguin contribution we find
\begin{dmath}
{\rm Re}\left(\frac{\varepsilon'}{\varepsilon}\right)_{\rm EWP} = -7.96(98)\times 10^{-4}\,.
\end{dmath}
Using the results for $\im(A_0)$ obtained using the $\SMOMqq$ intermediate scheme we find
\begin{dmath}
{\rm Re}\left(\frac{\varepsilon'}{\varepsilon}\right)_{\rm QCDP} = 0.00297(26)\,,
\end{dmath}
and likewise for the $\SMOMgg$ intermediate scheme,
\begin{dmath}
{\rm Re}\left(\frac{\varepsilon'}{\varepsilon}\right)_{\rm QCDP} = 0.00283(25)\,.
\end{dmath}
We observe that the two terms cancel at the 27(4)\% and 28(4)\% level relative to the QCDP contribution for the $\SMOMqq$ and $\SMOMgg$ results, respectively. This degree of cancellation is considerably less than the 71(36)\% observed in our 2015 analysis. Here the errors are statistical only. 




We can also compute a purely lattice value of $\re(\varepsilon'/\varepsilon)$ using ${\rm Re}(A_0)$ from Eqs.~\eqref{eq-reA0lat-qq} and~\eqref{eq-reA0lat-gg}, ${\rm Im}(A_0)$ from Eqs.~\eqref{eq-imA0lat-qq} and~\eqref{eq-imA0lat-gg}, and both ${\rm Re}(A_2)$ and ${\rm Im}(A_2)$ from Eq.~64 of Ref.~\cite{Blum:2015ywa}. Note we do not correct ${\rm Re}(A_2)$ for the change in ${\rm Re}(\tau)$ as its contribution is much smaller than that of the Wilson coefficients $z_i$. For the $\SMOMqq$ intermediate scheme we obtain
\begin{dmath}
{\rm Re}(\epsilon'/\epsilon) = 0.00293(104) 
\end{dmath}
and for the $\SMOMgg$ intermediate scheme,
\begin{dmath}
{\rm Re}(\epsilon'/\epsilon) = 0.00309(112)\,, 
\end{dmath}
where the errors are again statistical. Unfortunately these pure-lattice results have considerably larger statistical errors, which suggests that there is little statistical correlation between the results for ${\rm Im}(A_0)$ and ${\rm Re}(A_0)$ which would be needed to reduce the error in their ratio. Thus, we will use the results given in Eqs.~\eqref{eq-ep-div-e-qq-statonly} and~\eqref{eq-ep-div-e-gg-statonly} for our final results.

\subsection{Origin of the change in $\pmb{\epsilon'}$ compared to our 2015 calculation}
\label{sec:comparison}

In this section we provide further insight into the origin of the significant change between our 2015 result of ${\rm Re}(\epsilon'/\epsilon)=1.38(5.15)(4.59)\times 10^{-4}$ and our results above. Several factors may contribute to this effect:
\begin{enumerate}
 \item The increase in the minimum time separation between the four-quark operator and the sink $\pi\pi$ operator from 4 to 5 in the $K\to\pi\pi$ matrix element fitting.
 \item The change in the procedure for determining the derivative with respect to energy of $\pi\pi$ scattering phase-shift that enters the Lellouch-L\"uscher factor.
 \item The increase in statistics from 216 to 741 configurations.
 \item The addition of the $\pi\pi(311)$ and $\sigma$ sink operators.
 \item The use of step-scaling to raise the renormalization scale from 1.53 GeV to 4.01 GeV.
 \item The change in the value of the experimental inputs, notably that of the CKM ratio $\tau$ from $0.001543 -0.000635i$ to $0.001558 -0.000663i$.
\end{enumerate}

We first note that repeating the $\pi\pi$ two-point function analysis for our larger data set but with a one-state fit to a single operator ($\pi\pi(111)$), and a fit range 6-25 to match that of the 2015 analysis, yields a result (in lattice units),
\begin{dmath}\begin{array}{rl}
A^0_{\pi\pi(111)} & =  0.4028(32)\times \sqrt{1\times 10^{13}} \\
E_0 & = 0.3712(36)
\end{array}\label{eq-pipifit-1op1state-6-25}
\end{dmath}
that is consistent with the results of our 2015 analysis,
\begin{dmath}\begin{array}{rl}
A^0_{\pi\pi(111)} &= 0.3923(60)\times \sqrt{1\times 10^{13}} \\
E_0 &= 0.3606(74) 
\end{array}
\end{dmath}
to $1.5\sigma$ and $1.3\sigma$ for the amplitude and energy, respectively. Furthermore, the p-value of this fit is 0.451 indicating an excellent fit to the one-state model. The ground-state energy is, however, significantly larger than the value of $E_0=0.3479(11)$ found using three operators and two states in Sec.~\ref{sec-pipi2pt}.

We next repeat the analysis of the $K\to\pi\pi$ matrix elements but with only the $\pi\pi(111)$ operator and a one-state fit with $t'_{\rm min}=4$ to match the 2015 analysis, utilizing the $\pi\pi$ fit parameters from Eq.~\eqref{eq-pipifit-1op1state-6-25} above. Recall $t'_{\rm min}$ is the minimum time separation between the four-quark operator and the $\pi\pi$ sink for data included in the fit. We use the same input experimental parameters and other analysis strategies as in the original work, including the approach to obtaining the Lellouch-L\"uscher parameter and the same $\SMOMqq$ non-perturbative renormalization factors with $\mu=1.529$ GeV. We find,
\begin{dmath}
{\rm Re}(\epsilon'/\epsilon) = 2.52(2.12)\times 10^{-4}\,, \label{eq-ep1op1state_741} 
\end{dmath}
where the errors are statistical only. This result is completely consistent with our 2015 result,
\begin{dmath}
{\rm Re}(\epsilon'/\epsilon) = 1.38(5.15)\times 10^{-4} \label{eq-ep1op1state_old} \,,
\end{dmath}
indicating that a 3.4$\times$ increase in statistics is not sufficient to account for the difference.

Repeating the above but with the $K\to\pi\pi$ analysis and input parameters updated to match that of the present work gives,
\begin{dmath}
{\rm Re}(\epsilon'/\epsilon) = 4.20(1.96)\times 10^{-4}\,, 
\end{dmath}
which is slightly larger but still considerably smaller than the results in the previous section. With the step-scaled renormalization factors with $\mu=4.01$ GeV we find,
\begin{dmath}
{\rm Re}(\epsilon'/\epsilon) = 6.50(2.10)\times 10^{-4}\,. \label{eq-ep-d-e-oldfit}  
\end{dmath}
Again we observe a small increase but insufficient to account for the difference.

The result in Eq.~\eqref{eq-ep-d-e-oldfit} differs now from our primary result only in the $\pi\pi$ and $K\to\pi\pi$ fitting strategies. Adopting the final fit ranges determined for the $\pi\pi$ and $K\to\pi\pi$ fits in Secs. ~\ref{sec:TwoPointResults} and~\ref{sec:ThreePointResults}, such that the analysis now differs only in the number of $\pi\pi$ operators, gives
\begin{dmath}
{\rm Re}(\epsilon'/\epsilon) = 12.76(2.71)\times 10^{-4}\,. \label{eq-ep-d-e-newfitranges-1x1}
\end{dmath}
This result is now much closer to our final result. The behavior we observe here is consistent with that displayed in Fig.~\ref{fig-matelemQ1toQ6} where we plot the dependence of the fitted matrix elements on the cut $t'_{\rm min}$ and the number of $\pi\pi$ operators included in the fits to the matrix elements (the $\pi\pi$ two-point function fits remain unchanged between the results displayed in this figure). This figure shows a significant discrepancy between the $Q_6$ matrix element obtained from a one-operator, one-state fit with $t'_{\rm min}=4$ and the plateau observed when further operators are included. With increased statistics the onset of the apparent plateau for the one-operator, one-state fit does not occur until $t'_{\rm min}=5$ (equal to the $t'_{\rm min}=5$ used to obtain the result in Eq.~\eqref{eq-ep-d-e-newfitranges-1x1}) but the resulting value for the $Q_6$ matrix element is still several standard deviations larger than the strong plateau observed in the multi-operator fits. 

We therefore conclude that the difference in ${\rm Re}(\epsilon'/\epsilon)$ between our present and 2015 analysis results can be attributed primarily to unexpectedly large excited-state contamination in our previous analysis masked by the rapid reduction in the signal to noise ratio, and that multiple operators are essential to isolate the ground-state matrix element even with large statistics.

\section{Systematic errors}
\label{sec:syserrs}
In this section we describe the procedure used to estimate the systematic errors on our results. We will quote the values as representative percentage errors on either the matrix elements or on $A_0$ as appropriate. A discussion of the systematic errors in the $\Delta I=3/2$ calculation can be found in Ref.~\cite{Blum:2015ywa}.

\subsection{Excited state contamination}
\label{sec-syserr-excstate}

\begin{table}[tb]
\centering
\begin{tabular}{c|c}
\hline\hline
$i$ & Rel. diff\\
\hline
1 & $-0.04(16)$\\
2 & $ 0.012(39)$\\
3 & $-0.7(6.8)$\\
4 & $-0.08(11)$\\
5 & $ 0.017(38)$\\
6 & $ 0.019(23)$\\
7 & $ 0.0017(95)$\\
8 & $-0.0044(45)$\\
9 & $ 0.093(64)$\\
10 & $-0.032(58)$\\
\end{tabular}
\caption{Relative differences between the ground-state elements obtained by fitting to 3 operators and 3 states with $t'_{\rm min}=4$ and those of our primary fit with 3 operators and 2 states with $t'_{\rm min}=4$. \label{tab-reldiffexcsys}  }
\end{table}

In Sec.~\ref{sec-fitresults} we devoted considerable effort to finding an optimal fit window in which excited state effects are minimal. We were unable to find evidence of such effects arising from excited kaon states, which is to be expected given both the large relative energy of such states and also the fact that the rapid growth of statistical noise as the four-quark insertion is moved away from the $\pi\pi$ operator implies that the data furthest from the kaon operator dominates the fit results. As such we do not assign a systematic error to possible contamination from excited kaon states.

As for the contribution of excited $\pi\pi$ states, we found little evidence for such effects even within the single operator fits to the $\pi\pi(111)$ data, except for the $Q_5$ and $Q_6$ matrix elements where the single-operator fits showed statistically significant deviations from the common plateau region that did not die away until $t'=6$. We observed that by adding further sink operators and allowing for more $\pi\pi$ states substantially reduced the excited-state contamination such that the fits were highly consistent even if we include data at times as low as $t'=3$. Despite this we chose a conservative uniform cut of $t'_{\rm min}=5$ for our fits.

In order to assign a numerical error to the contamination from excited $\pi\pi$ states, we consider the comparison of the 3-operator, 3-state fit with $t'_{\rm min}=4$ and the 3-operator, 2-state fit with $t'_{\rm min}=5$, the latter being our chosen best fit. The former includes a third state and with $t'_{\rm min}=4$ appears capable of describing the data well outside of the fit range, as we observed in Fig.~\ref{fig-Q6_3x2_3x3_tmin6_tpmin4_5-comp} (lower-left panel). We compute relative differences under the bootstrap between the values of the ground-state matrix elements, the results of which are shown in Tab.~\ref{tab-reldiffexcsys}. The only statistically resolvable difference, at $1.5\sigma$, is for the $Q_9$ matrix element, which has only a negligible contribution to ${\rm Im}(A_0)$. For the dominant $Q_4$ and $Q_6$ matrix elements the differences cannot be resolved within our errors. We therefore conclude that the excited state systematic error is likely to be much smaller than our dominant systematic errors and can be neglected.

\subsection{Unphysical kinematics}
\label{sec-syserr-unphyskin}


As our values of $E_{\pi\pi}$ and $m_K$ differ by 2.2(3)\%, the $K\to\pi\pi$ matrix elements are not precisely on shell. As discussed in Sec.~\ref{sec:ThreePointResults}, the primary result of these unphysical kinematics is the rise of a divergent contribution from the pseudoscalar operator $\bar s\gamma^5 d$ that vanishes when on shell by the equations of motion. In order to suppress this error we perform an explicit subtraction of the pseudoscalar operator that leaves behind a finite, regulator-independent term that represents the dominant remaining systematic error from the unequal kaon and $\pi\pi$ energies. As we are close to being on shell we can reasonably assume a linear ansatz for the dependence of our result on the energy difference $E_{\pi\pi}-m_K$. We estimate the associated systematic error by observing the change in the $Q_2$ matrix element as the kaon mass is increased by 4.5\%. The measurement was performed using 69 configurations of our original ensemble~\cite{Bai:2015nea}, with 3 different $K\to\pi$ time separations (10, 12, and 14), and we observed a 6.9\% increase in the matrix element. We scale this increase by the relative difference between our kaon and $\pi\pi$ energies, giving $3\%$. 

\begin{table}[tb]
\centering
\begin{tabular}{c|c}
\hline\hline
$i$ & Rel. diff\\
\hline
1 & $-0.0054(51)$\\
2 & $-0.0086(19)$\\
3 & $-0.06(73)$\\
4 & $-0.0144(75)$\\
5 & $-0.054(12)$\\
6 & $-0.0521(75)$\\
7 & $-0.0053(25)$\\
8 & $-0.0072(21)$\\
9 & $-0.0055(21)$\\
10 & $-0.00234(85)$\\
\end{tabular}
\caption{Relative differences in the unrenormalized lattice matrix elements of $Q_i$ as the pseudoscalar subtraction coefficients $\alpha_i$ are uniformly increased by 5\% \label{tab-alphavary5pc} }
\end{table}

Another means of estimating this systematic error is to vary the subtraction coefficients $\alpha_i$ by an amount consistent with the expected size of the residual contribution of the pseudoscalar operator. Given that the operator is dimension-3, its coefficient is originally ${\cal O}(m_s/a^2)$ where the strange quark mass is in physical units. After the subtraction is performed, the residual term is expected to be of size ${\cal O}(m_s \Lambda_{\rm QCD}^2)$, which has a relative size of ${\sim} a^2\Lambda_{\rm QCD}^2$, or ${\sim}5\%$, of the original contribution, for $\Lambda_{QCD}=300$ MeV. Increasing the subtraction coefficients $\alpha_i$ by this amount gives rise to the differences in the unrenormalized lattice matrix elements given in Tab.~\ref{tab-alphavary5pc}. The observed variations are generally consistent with the above, but to be conservative we assign a relative systematic error of 5\% on the matrix elements resulting from the off-shell difference $E_{\pi\pi} \ne m_K$.

\subsection{Finite lattice spacing}

We use the value provided in Ref.~\cite{Bai:2015nea} as an estimate of the finite lattice spacing systematic error. This was obtained by comparing the values of the $\Delta I=3/2$ matrix elements between the continuum limit~\cite{Blum:2015ywa} and the calculation~\cite{Blum:2012uk} performed on our $32^3\times 64$, $\beta=1.75$ (32ID) lattice. The parameters of the latter ensemble are identical to those used in this work to compute $A_0$, albeit without G-parity boundary conditions and with a larger-than-physical light quark mass giving a unitary pion mass of 170 MeV. The $\msbar$ values for the three continuum matrix elements that contribute to $A_2$ are obtained by combining the continuum values of those matrix elements in the $\SMOMqq$ scheme (Tab. XIV of Ref.~\cite{Blum:2015ywa}) with the RI$\to\msbar$ renormalization matrix computed on the 32ID lattice (Eq. 66 of Ref.~\cite{Blum:2015ywa}). As such this estimate addresses only the discretization errors on the matrix elements and not to those on the renormalization factors (which are expected to be small). We find the values given in Tab.~\ref{tab-syserrfinitea}. Averaging the three relative errors we arrive at an estimate of 12\% discretization errors on the matrix elements.

\begin{table}[tbp]
\begin{tabular}{c|ccc}
\hline\hline
Operator & 32ID & continuum & rel. diff \\
\hline
$(27,1)$ & 0.0461(14) & 0.0502(13) & 8.7(4.1)\% \\
$(8,8)$ & 0.874(49) & 0.993(22) & 13.6(6.1)\% \\
$(8,8)_{\rm mix}$ & 3.96(23) & 4.54(12) & 14.8(6.6)\%
\end{tabular}
\caption{The three $\Delta I=3/2$ matrix elements in the $\msbar$ scheme at $\mu=3.0$ GeV and in units of GeV$^3$ that contribute to $A_2$, calculated on the 32ID ensemble (Ref.~\cite{Blum:2012uk}, Eq.~(31)) and in the continuum limit (Ref~\cite{Blum:2015ywa}, Tab. XIV) along with their relative difference. Only statistical errors are shown.\label{tab-syserrfinitea} }
\end{table}

\subsection{Lellouch-L\"uscher factor}
\label{sec-llfactorsys}
As described in Sec.~\ref{sec-llfactor}, the calculation of the Lellouch-L\"uscher factor, $F$, that accounts for the power-law finite-volume corrections to the matrix element, requires an ansatz for the derivative of the $\pi\pi$ phase shift with respect to energy. In Sec.~\ref{sec-phase-shift-deriv} we present values for this derivative obtained from three methods:
\begin{itemize}
\item The Schenk parameterization~\cite{Schenk:1991xe} of the dispersive energy dependence obtained in Ref.~\cite{Colangelo:2001df} 
\item A linear approximation in the $\pi\pi$ energy above threshold, $\frac{d\delta_0}{dE_{\pi\pi}} = \frac{\delta_0}{E_{\pi\pi} - 2m_\pi}$, which is inspired by the dispersive low-energy dependence found in Ref.~\cite{Colangelo:2001df} and can be related to $d\delta_0/dq$ via Eq.~\eqref{eq-llkdef}.
\item A direct lattice calculation of the phase shift at energies close to and including the kaon mass.
\end{itemize}
Ignoring the noisier of the two lattice determinations, the results varied between $\frac{{\rm d}\delta_0}{{\rm d}q}=1.26$ and 1.41, a 12\% spread. The resulting values of $F$ differ by 1.5\% since the dominant contribution arises from the derivative of the analytic function $\phi$. We therefore assign a 1.5\% systematic error to the matrix elements from this source.

\subsection{Exponentially-suppressed finite volume corrections}

We expect the remaining finite volume corrections to our matrix elements to be dominated by the (exponentially-suppressed) interactions between the final state pions that are not accounted for by the L\"uscher and Lellouch-L\"uscher prescriptions. In Refs.~\cite{Blum:2012uk,Blum:2015ywa} we performed an in-depth analysis of the finite-volume errors on the matrix elements that comprise $A_2$ using SU(3) chiral perturbation theory, in which the mesonic loop integrals are replaced by discrete sums over the allowed momenta. In the earlier work~\cite{Blum:2012uk} we estimated ${\cal O}(6-6.5\%)$ corrections for both classes of operator that enter the calculation of $A_2$ for a lattice volume equal to that used in the present study. However, in the later study~\cite{Blum:2015ywa} we identified a mistake in the literature upon which the earlier analysis was based, leading to a dramatically smaller, ${\cal O}(2.5\%)$ estimate of the finite-volume systematic error, albeit for a somewhat larger $(5.5{\rm fm})^3$ volume than the $(4.6{\rm fm})^3$ of our present study. Due to the absence of correspondingly accurate finite volume effect estimates for the $I=0$ final state, we retain a 7\% systematic error estimate for the matrix elements of $A_0$. It is useful to bear in mind that while this assigned 7\% error is itself uncertain, like our other small sources of systematic error, it is unlikely to make an appreciable difference to our total ${\cal O}(35\%)$ systematic error.

\subsection{Neglecting the contribution of the $\pmb{G_1}$ operator}
\label{sec-G1syserrs}

\begin{table}[tb]
\centering
\begin{tabular}{l|l|l}
\hline\hline
i & Relative difference\\
\hline
1 & $-0.038(36)$\\
2 & $-0.022(12)$\\
3 & $ 0.070(576)$\\
4 & $-0.018(31)$\\
5 & $ 0.003(41)$\\
6 & $ 0.006(6)$\\
7 & $0(0)$\\
8 & $0(0)$\\
9 & $-0.031(17)$\\
10 & $-0.023(21)$\\
\end{tabular}
\caption{The relative difference in $\overline{MS}$ matrix elements at $\mu=1.33$ GeV obtained through the $\SMOMgg$ intermediate scheme due to including the $G_1$ operator. \label{tab-G1effect-MSbarmatelem}  }
\end{table}

In the calculation of our step-scaled non-perturbative renormalization factors with scale $\mu=4.01$ GeV we have not incorporated the effects of the $G_1$ operator. A previous lattice study~\cite{McGlynnThesis}, performed in the $\SMOMgg$ scheme and utilizing step-scaling from a low-scale of $\mu=1.33$ GeV on our 32ID ensemble to a high scale of $2.29$ GeV on a finer lattice, revealed the effects on $A_0$ of including this operator to be on the order of a few percent when combined with the matrix elements measured in our 2015 work~\cite{Bai:2015nea}. Unfortunately the statistical errors on the differences in the renormalized matrix elements at $\mu=2.29$ GeV with and without $G_1$ included were found to be too large to resolve the effect with any precision, and we find that this also applies to the matrix elements obtained in the present work. (The renormalization matrices with and without $G_1$ at $\mu=2.29$ GeV can be found in Tab.~\ref{tab-Zmat2.29}.)

As discussed in Ref.~\cite{McGlynnThesis}, the increase in the relative error on the bootstrap differences is associated largely with the step-scaling matrix $\Lambda^{\rm RI}$ that describes the running between the low and high energy scales. However it is reasonable to expect that the largest effects of neglecting $G_1$ appear at the low energy scale in the step-scaling where the QCD coupling is larger. We therefore compare the matrix elements renormalized at the low scale in the $\msbar$ scheme in order to estimate the size of this systematic error with greater precision. We perform this comparison using the $\SMOMgg$ intermediate scheme with $\mu=1.33$ GeV, the renormalization matrices of which are given in Tab.~\ref{tab-Zmat1.33}. The relative differences of the resulting $\msbar$ matrix elements are given in Tab.~\ref{tab-G1effect-MSbarmatelem}. While the observed differences are still poorly resolved, the typical size of the effect appears to be ${\cal O}(3\%)$, and we therefore assign a 3\% systematic error to the effect of neglecting $G_1$. (This estimate is quite conservative given the tiny difference in the dominant, $Q_6$ operator observed in the table.)


\subsection{Sytematic errors in $\pmb{\msbar}$ operator renormalization}
\label{sec-renorm-err}

The most important systematic errors in determining the renormalization matrix $Z^{\msbar\leftarrow\mathrm{lat}}$ arise from three sources: i)  The omission of dimension-6, quark bilinear operators which vanish on shell such as $G_1$ discussed above. ii) Finite lattice spacing errors that result from our choice of a large RI renormalization scale $\mu$. iii)  The perturbative truncation error introduced when one-loop QCD perturbation theory is used to relate the RI-SMOM and $\msbar$ schemes.  In order to estimate these systematic errors, we examine the difference between the results in the $\msbar$ scheme obtained from our two different intermediate RI-SMOM schemes. Rather than examining the matrix elements themselves, which can be statistically noisy and vary significantly in size and importance, it is convenient to study instead the differences between the elements of the $7\times$7 lattice$\to\msbar$ renormalization matrix
\begin{equation}
R^{\msbar\leftarrow {\rm RI} \leftarrow {\rm lat}}_{1\hbox{-}{\rm loop}\,j\ell}(\mu) =  H^{\msbar\leftarrow{\rm RI}}_{1\hbox{-}{\rm loop}\,jk}(\mu) R_{k\ell}^{{\rm RI}\leftarrow{\rm lat}}(\mu)\,,
\end{equation}
where $H$ is the perturbative matching matrix. In the absence of systematic errors the matrix $R^{\msbar\leftarrow RI \leftarrow {\rm lat}}$ is independent of the intermediate RI scheme.  We can then study this systematic error by examining the matrix
\begin{equation}
\Xi \equiv \left|{\mathbb I} - \left[R_{1\hbox{-}{\rm loop}}^{\msbar\leftarrow\SMOMqq\leftarrow {\rm lat}}\right]^{-1}
R_{1\hbox{-}{\rm loop}}^{\msbar\leftarrow\SMOMgg\leftarrow{\rm lat}}\right|\,,
\end{equation}
where ${\mathbb I}$ is the $7\times 7$ unit matrix and $|.|$ implies that the absolute value of each element is taken. The ratio of $R$-matrices in this equation converts from the lattice scheme to $\msbar$ through one intermediate scheme, and converts back to the lattice scheme via the other scheme, and hence becomes the unit matrix if no systematic errors exist. The difference from the unit matrix is therefore a measure of the size of the systematic error: Under the reasonable assumption that the systematic errors in the two schemes are comparable in size, we expect the elements of $\Xi$ to vary between zero and approximately twice the size of the systematic error present in each. We therefore assign a percentage systematic error that is one half of the largest observed element of $\Xi$ at a scale $\mu$.

\begin{table}[tbp]
\centering
\begin{tabular}{c|cccc}
\hline\hline
Element $(i,j)$ & 1.33 GeV & 1.53 GeV & 2.29 GeV & 4.01 GeV\\
\hline
(1,1) & $0.07406(36)$ & $0.062571(56)$ & $0.04936(42)$ & $0.01686(36)$\\
(2,2) & $ 0.182(34)$ & $ 0.173(15)$ & $ 0.044(54)$ & $ 0.128(83)$\\
(2,3) & $ 0.313(38)$ & $ 0.282(16)$ & $ 0.132(58)$ & $ 0.135(83)$\\
(2,5) & $ 0.006(11)$ & $ 0.0036(50)$ & $0.013(16)$ & $0.009(31)$\\
(2,6) & $ 0.0005(95)$ & $0.0030(42)$ & $ 0.0099(100)$ & $0.005(13)$\\
(3,2) & $0.276(33)$ & $0.256(14)$ & $0.119(33)$ & $ 0.058(42)$\\
(3,3) & $0.417(38)$ & $0.399(16)$ & $0.197(37)$ & $ 0.047(43)$\\
(3,5) & $0.006(10)$ & $0.0076(47)$ & $0.0084(94)$ & $ 0.005(13)$\\
(3,6) & $ 0.0420(96)$ & $ 0.0212(40)$ & $ 0.0315(68)$ & $ 0.0020(59)$\\
(5,2) & $0.00(14)$ & $ 0.042(59)$ & $0.18(18)$ & $ 0.22(27)$\\
(5,3) & $0.04(15)$ & $0.001(60)$ & $0.20(19)$ & $ 0.21(26)$\\
(5,5) & $0.004(39)$ & $0.012(18)$ & $0.034(50)$ & $ 0.022(97)$\\
(5,6) & $ 0.037(34)$ & $0.007(15)$ & $ 0.044(31)$ & $0.032(38)$\\
(6,2) & $ 0.139(65)$ & $ 0.173(27)$ & $0.010(110)$ & $ 0.16(13)$\\
(6,3) & $ 0.321(74)$ & $ 0.291(33)$ & $ 0.14(12)$ & $ 0.23(14)$\\
(6,5) & $ 0.027(20)$ & $ 0.0104(75)$ & $ 0.024(34)$ & $ 0.055(46)$\\
(6,6) & $0.110(22)$ & $0.0752(89)$ & $0.052(26)$ & $0.031(24)$\\
(7,7) & $0.01424(34)$ & $0.008152(35)$ & $0.01096(40)$ & $0.00360(25)$\\
(7,8) & $ 0.003429(46)$ & $ 0.002120(29)$ & $ 0.002029(51)$ & $ 0.00548(19)$\\
(8,7) & $ 0.026523(94)$ & $ 0.024917(63)$ & $ 0.02364(24)$ & $ 0.00710(92)$\\
(8,8) & $0.14784(44)$ & $0.12752(14)$ & $0.09866(58)$ & $0.0263(10)$\\
\end{tabular}
\caption{The non-zero elements of the matrix $\Xi$ computed using the renormalization matrices obtained at $\mu=1.33$ GeV and 1.53 GeV on the 32ID ensemble, as well as the step-scaled renormalization matrices with $\mu=2.29$ GeV and 4.01 GeV. We do not include the $G_1$ operator here, and its absence is treated as a separate systematic error in Sec.~\ref{sec-G1syserrs}.\label{tab-onemZrat} }
\end{table}

In Tab.~\ref{tab-onemZrat} we tabulate the non-zero elements of $\Xi$ for various $\msbar$ scales and step-scaling procedures. Once again we observe that the effects of including or discounting the $G_1$ operator, while harder to statistically resolve after passing through the step-scaling procedure, are at the percent scale.

As expected there is a general trend towards smaller values as we increase the scale that appears consistent with the factor of three decrease in $\alpha_s^2$ between 1.33 GeV and 4.01 GeV that is expected to describe the scaling of the missing NNLO terms. Unfortunately the statistical errors on the results at 4.01 GeV are too large to resolve the residual systematic effects. Nevertheless, considering the results of this table and also the 3-4\% differences observed in $\re A_0$ and $\im A_0$ between the schemes in Sec.~\ref{sec-A0results-staterrsonly}, we assign a 4\% systematic error to the non-perturbative renormalization factors.


\subsection{Parametric errors}
\label{sec-syserr-parametric}


We propagate the parametric uncertainties shown in Tab.~\ref{tab-expt-params} to $\re A_0$ and $\im A_0$. For $\re A_0$ the largest such uncertainty is the charm-mass dependence, which, however, is only a $0.3\%$ effect. For $\im A_0$, the largest uncertainty is $5\%$ from the $\tau$ parameter, $3\%$ from $\alpha_s$, and less than $1\%$ from the charm and top quark masses. The other uncertainties have been estimated but are negligible compared to those quoted. We therefore estimate a total parametric uncertainty of $6\%$ for $\im A_0$ and $0.3\%$ for $\re A_0$.


\subsection{Wilson coefficients}
\label{sec-syserr-wilsoncoeffs}

As mentioned previously we compare the NLO and LO determinations of the Wilson coefficients in order to estimate the systematic error arising due to missing higher-order terms. More specifically, we compare $\im(A_0)$ obtained from LO and NLO Wilson coefficients, computed using the 1-loop and 2-loop determinations of $\alpha_s$, respectively, while keeping fixed the renormalized matrix elements in the $\msbar$ scheme at 4.01 GeV obtained using the $\SMOMqq$ intermediate scheme, given in Tab.~\ref{tab-renorm-matrix-elems}, together with the various input parameters, such as the quark masses and the QCD coupling constant. For the latter we use the solution of the 4-loop $\beta$ function~\cite{vanRitbergen:1997va} to compute $\alpha_s^{N_f=4}(\hat \mu)$ in the 4-flavor theory, starting from the value of $\alpha_s(m_Z)$ in Tab.~\ref{tab-expt-params}, and we study the dependence of the LO prediction of $\im(A_0)$ as a function of $\hat \mu$, relative to the NLO result. (As expected, the NLO shows a mild dependence simply due to the mismatch between the running of $\alpha_s$ from the $Z$ pole (4 loops) and the running used in the calculation of the Wilson Coefficients (2 loops).) Starting at 8\% at $\hat \mu\approx m_c$, it increases up to 16\% at $\hat \mu\approx m_b$; hence for our systematic error estimate on the Wilson coefficients, we choose the intermediate point $\hat \mu=1.7~\mathrm{GeV}$ for which the NLO and LO difference is 12\%. We have verified that fixing the value of $\Lambda^{N_f=4}$ leads to similar conclusions.

Additionally we consider the same difference of LO vs NLO predictions for $\im(A_0)$, as a function of the RI intermediate schemes and the scale of the RI to $\msbar$ conversion, while keeping fixed all parameters, $\alpha_s^{N_f=4}(\hat \mu)$ included. We find that, despite varying the renormalization scale by almost a factor of two and the use of different intermediate RI schemes, the differences in the values of $\im(A_0)$ are quite consistent, in the range 11-15\%. This suggests that the bulk of the observed difference arises from the perturbative 3-to-4 flavor matching and running above the charm threshold, which is common to all of these determinations, and that improved theory input for the 3-to-4 flavor matching could significantly reduce it. (Note that in our calculation we take the matching scale across a flavor threshold equal to the corresponding quark mass in order to avoid large logarithms. Additional insights could be gained by studying the dependence on this  matching scale as in Ref.~\cite{Cerda-Sevilla:2018hjk}.)

In conclusion we assign a 12\% systematic error on both $\re A_0$ and $\im A_0$ associated with the NLO determination of the Wilson coefficients.

\subsection{Error budget}

\begin{table}[tb]
\begin{tabular}{c|c}
\hline\hline
Error source & Value \\
\hline
Excited state & - \\
Unphysical kinematics & 5\% \\
Finite lattice spacing & 12\% \\
Lellouch-L\"uscher factor & 1.5\% \\
Finite-volume corrections & 7\% \\
Missing $G_1$ operator & 3\% \\
Renormalization & 4\% \\
\hline 
Total & 15.7\% 
\end{tabular}
\caption{Relative systematic errors on the infinite-volume matrix elements of the $\msbar$-renormalized four-quark operators $Q_j'$. \label{tab-syserrQ}}
\end{table}

\begin{table}[tb]
\begin{tabular}{c|c|c}
\hline\hline
Error source & \multicolumn{2}{|c}{Value} \\
\hline
                               & ${\rm Re}(A_0)$ & ${\rm Im}(A_0)$ \\
\hline 
Matrix elements      & 15.7\%  & 15.7\% \\
Parametric errors   & 0.3\%  & 6\%  \\
Wilson coefficients & 12\%  & 12\% \\
\hline 
Total                        & 19.8\%  & 20.7\% 
\end{tabular}
\caption{Relative systematic errors on ${\rm Re}(A_0)$ and ${\rm Im}(A_0)$. \label{tab-syserrImA0}}
\end{table}

We divide the systematic errors into those that affect the calculation of the matrix elements of the $\msbar$ weak operators $Q'_j$ and those that enter when these matrix elements are combined to produce the complex, physical decay amplitude $A_0$. The former are collected in Tab.~\ref{tab-syserrQ}. In order to obtain the final systematic error on ${\rm Im}(A_0)$ arising from these matrix elements we note that the result is dominated by the $Q_6$ operator with only a 20\% cancellation from $Q_4$.  In this circumstance it is reasonable simply to apply the same flat percentage error to ${\rm Im}(A_0)$ as to $Q_6$. Since ${\rm Re}(A_0)$ is similarly dominated by $Q_2$, we apply the same strategy. For $A_0$ we then arrive at the error budget given in Tab.~\ref{tab-syserrImA0} which includes this error arising from the uncertainties in the matrix elements as well as those arising from the use of perturbation theory when computing the $\msbar$ Wilson coefficients and the values of the needed Standard Model input parameters.

\section{Final results and discussion}
\label{sec:finalresults}
In this section we collect our final results including systematic errors and discuss the implications of our results. For consistency with our previous work we will use the $\SMOMqq$ intermediate scheme for our central value.

\subsection{Matrix elements}

The renormalized, infinite-volume matrix elements in the RI and $\msbar$ schemes are given in Tab.~\ref{tab-renorm-matrix-elems}, where the errors are statistical only. The corresponding relative systematic errors can be found in Tab.~\ref{tab-syserrQ}. For the convenience of the reader we have reproduced the matrix elements in the $\SMOMqq$ scheme including their systematic errors in Tab.~\ref{tab-renorm-matrix-elems-qq-full}. In order to allow the reader to compute derivative quantities from these matrix elements, the covariance matrices for the renormalized matrix elements in the $\SMOMqq$ and $\msbar$ schemes at 4.01 GeV can be found in Tabs.~\ref{tab-7x7RIqqmatelem_covmat} and~\ref{tab-10x10MSbarmatelem_covmat}, respectively. 

\begin{table}[tb]
\centering
\begin{tabular}{l|r}
\hline\hline
i & SMOM$(\slashed{q},\slashed{q})$ (GeV$^3$)\\
\hline
1 & $ 0.060(39)(9)$\\
2 & $-0.125(19)(20)$\\
3 & $ 0.142(17)(22)$\\
5 & $-0.351(62)(55)$\\
6 & $-1.306(90)(205)$\\
7 & $ 0.775(23)(122)$\\
8 & $ 3.312(63)(520)$\\
\end{tabular}
\caption{Physical, infinite-volume matrix elements in the $\SMOMqq$ scheme at $\mu=4.006$ GeV given in the 7-operator chiral basis. The errors are statistical and systematic respectively. Note that our 4\% estimate of the renormalization systematic error includes both lattice systematic errors and those associated with the truncation of the perturbative series in the RI$\to\msbar$ matching. While the latter are inappropriate to apply to matrix elements in the non-perturbative schemes, due to our estimation procedure we are at present unable to isolate these two effects and as such apply the full 4\% systematic error also to these RI matrix elements. \label{tab-renorm-matrix-elems-qq-full} }
\end{table}

\subsection{Decay amplitude}

For the real part of the decay amplitude we take the value from Eq.~\eqref{eq-reA0lat-qq} and apply the systematic errors given in Tab.~\ref{tab-syserrImA0} to obtain
\begin{dmath}
{\rm Re}(A_0) = 2.99(0.32)(0.59)\times 10^{-7}\ {\rm GeV}\,,~\label{eq-reA0-final}  
\end{dmath}
where the errors are statistical and systematic, respectively. The imaginary part is obtained likewise from Eq.~\eqref{eq-imA0-exptinput-qq}, giving
\begin{dmath}
{\rm Im}(A_0) = -6.98(0.62)(1.44)\times 10^{-11}\ {\rm GeV}\,.
\end{dmath}
The breakdown of the contributions of each of the 10 operators to these amplitudes can be found in Tab.~\ref{tab-A0contribs}. We observe that, at the scale at which we are working, the dominant contribution to $\re(A_0)$ (97\%) originates from the tree operator $Q_2$, while $Q_1$ has a contribution of about 13\% that is largely cancelled by that of the penguin operator~\cite{Shifman:1975tn, Gilman:1979bc} $Q_6$. Likewise, the dominant contribution to $\im(A_0)$ is from the QCD penguin~\cite{Shifman:1975tn, Gilman:1979bc} operator, $Q_6$, with a 14\% cancellation from $Q_4$.

\subsection{A comment on the $\pmb{\Delta I=1/2}$ rule}

The ``$\Delta I=1/2$ rule''  refers to the enhancement by almost a factor of 450 of the $I=0$ $K\to\pi\pi$ decay rate relative to that of the $I=2$ decay, corresponding to the experimentally-determined ratio ${\rm Re}(A_0)/{\rm Re}(A_2)=22.45(6)$.  A factor of two contribution to this ratio arises from the perturbative Wilson coefficients~\cite{Gaillard:1974nj, Altarelli:1974exa, Gilman:1978wm}. While the remaining factor of ten has been viewed for some time as a consequence of the strong dynamics of QCD, the origin of this large factor has remained something of a mystery with no widely-accepted dynamical explanation. 

In the past~\cite{Boyle:2012ys, Blum:2011ng, Blum:2012uk}, and most recently in Ref.~\cite{Blum:2015ywa}, when simulating with physical pion masses we have observed a sizeable cancellation between the two Wick contractions of the dominant $(27,1)$ operator contributing to the $\Delta I = 3/2$ decay amplitude, leading to a significant suppression of Re\,$(A_2)$. In these calculations we reproduced the experimental value of Re\,$(A_2)$ and concluded that this cancellation was likely to be a very significant element in the $\Delta I=1/2$ rule. We stress that the cancellation between the two leading contributions to Re\,$(A_2)$ depends sensitively on the light quark mass and becomes much less significant as the light quark mass is increased above its physical value. Note also that such a cancellation is not consistent with na\"ive factorization, which predicts that both contributions have the same sign and differ in size by a factor of three due to color suppression.


In order to obtain a quantitative, first-principles result for Re\,$(A_0)/$Re\,$(A_2)$, we also require knowledge of Re\,$(A_0)$ which we provide in Eq.~\eqref{eq-reA0-final} of the present paper. Combining this with our earlier result for $A_2$~\cite{Blum:2015ywa}, we obtain
%
%
\begin{equation}
\frac{\mathrm{Re}(A_0)}{\mathrm{Re}(A_2)} = 19.9\,(2.3)\,(4.4)\,,   \label{eq-deltI1/2rule-result}
\end{equation}
where the errors are statistical and systematic respectively. The value in Eq.~\eqref{eq-deltI1/2rule-result} agrees very well with the experimental result, demonstrating quantitatively that, within the uncertainties, the $\Delta I=1/2$ rule is indeed a consequence of QCD and thus providing an answer to an important long-standing puzzle. 

For earlier theoretical papers on the $\Delta I=1/2$ rule and the real parts of the individual amplitudes $A_0$ and $A_2$, as well as some recent work, see Refs.~\cite{Donoghue:1979mu, Bardeen:1986vz, Pich:1995qp, Bertolini:1998vd, Buras:2014maa, Donini:2020qfu}.

\subsection{Result for Re($\pmb{\varepsilon'/\varepsilon})$}
\label{sec:epsilonp} 

For $\varepsilon'/\varepsilon$ we use Eq.~\eqref{eq-epsilonprimefromA}, combining the lattice values for the imaginary parts of the decay amplitudes with the experimental measurements of the real parts. The systematic error for ${\rm Im}(A_0)$ is taken from Tab.~\ref{tab-syserrImA0} and that of ${\rm Im}(A_2)$ from Eq. 64 of Ref.~\cite{Blum:2015ywa}.   The statistical and systematic errors on ${\rm Im}(A_0)$ and ${\rm Im}(A_2)$ are combined in quadrature and are therefore  enhanced by the cancellation between the two terms in Eq.~\eqref{eq-epsilonprimefromA}.  However, one further important systematic error should be addressed: that arising from the effects of electromagnetism and the isospin-breaking difference, $m_d-m_u$, between the down and up quark masses.

While for most quantities these corrections enter at the 1\% level or below, for $\epsilon'$ this familiar situation does not hold.  As can be seen from the formula used to compute $\epsilon'$ in the Standard Model given in Eq.~\eqref{eq-epsilonprimefromA}, the $I=0$ and $I=2$ amplitudes $A_0$ and $A_2$ enter with equal weight.  However, as is summarized by the $\Delta I=1/2$ rule, the amplitude $A_2$ is 22.5 times smaller than $A_0$.  Thus, a 1\% correction to $A_0$ can introduce an ${\cal O}(20\%)$ correction to $A_2$ and a potential correction to $\epsilon'$ of  20\% or more.  

The effects on $\epsilon'$ of electromagnetism and $m_d-m_u$ have been the subject of active research for some time~\cite{Buras:1987wc,Cirigliano:2003nn,Cirigliano:2019cpi}.  The most recent results are those of Cirigliano {\it et al.}~\cite{Cirigliano:2019cpi}. They provide a correction that is appropriate for our calculation in which the contribution of the electro-weak penguin operators $Q_7$ and $Q_8$ has been included.  Their result is parametrized by $\hat\Omega_{\rm eff}$ which is introduced into a version of 
Eq.~\eqref{eq-epsilonprimefromA} which incorporates these effects:
\begin{equation}
\frac{\varepsilon'}{\varepsilon}\ = \frac{i\omega_+ e^{i(\delta_2-\delta_0)}}{\sqrt{2}\varepsilon}\left[ \frac{\im(A^{\rm emp}_2)}{\re(A^{(0)}_2)}-\frac{\im(A^{(0)}_0)}{\re(A^{(0)}_0)}\left(1-\hat\Omega_{\rm eff}\right) \right] \label{eq-epsilonprime-E&M}.
\end{equation}
and find $\hat\Omega_{\rm eff}=(17.0 {+9.1 \atop -9.0})\times 10^{-2}$.  Here we are reproducing Eqs. (54) and (60) from Ref.~\cite{Cirigliano:2019cpi}, where $\re(A_{0,2}^{(0)})$ refer to the real amplitudes in the absence of isospin breaking, $\im(A_2^{\rm emp})$ represents the dominant contribution to $\im(A_2)$ and arises from the electroweak penguin operators $Q_{7,8}$, and $\im(A_0^{(0)})$ additionally includes the effects of QCD penguin operators. At the present level of accuracy, our use of the experimental rates for the real amplitudes, together with small differences from the definition of the isosymmetric limit in Ref.~\cite{Cirigliano:2019cpi}, do not affect the applicability of Eq.~\eqref{eq-epsilonprime-E&M} to our calculation. (For a review of earlier work on this topic see Ref.~\cite{Cirigliano:2011ny}.) Note also that $\omega_+ = \re(A_2^+)/\re(A_0)$, where the plus (+) indicates the amplitude obtained from charged kaon decay, is equal to the value of $\omega$ used to represent the isospin-symmetric ratio in this work and given in Tab.~\ref{tab-expt-params}.

Since a careful discussion of these corrections is beyond the scope of this paper we choose to treat these effects of isospin breaking as a systematic error whose size is given by the effect of including $\hat\Omega_{\rm eff}$ in Eq.~\eqref{eq-epsilonprime-E&M}.  We find
\begin{dmath}
%
%
{\rm Re}(\epsilon'/\epsilon) = 0.00217(26)(62)(50) \label{eq-ep-div-e-qq-full}\,,
\end{dmath}
where the errors are statistical and systematic, with the systematic error separated as  isospin-conserving and isospin-breaking, respectively. We note that if we were to apply this negative correction directly to our result for ${\rm Re}(\epsilon'/\epsilon)$, the central value obtained, 0.00167, would nearly coincide with the experimental value, albeit with appreciable errors.

\begin{figure}[tb]
\centering
\includegraphics[width=0.85\textwidth]{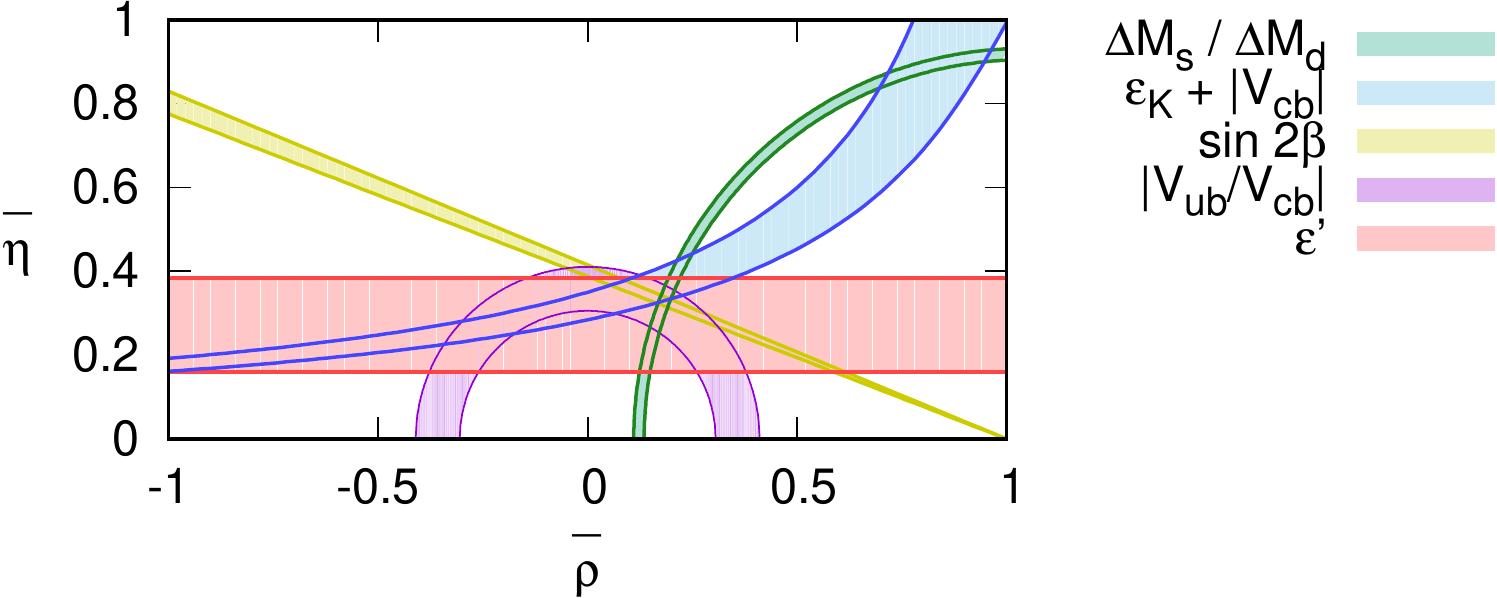}
\caption{The horizontal-band constraint on the CKM matrix unitarity triangle in the $\bar\rho-\bar\eta$ plane obtained from our calculation of $\epsilon'$, along with constraints obtained from other inputs~\cite{PDG2019,PhysRevD.98.030001,Boyle:2018knm}. The error bands represent the statistical and systematic errors combined in quadrature. Note that the band labeled $\epsilon'$ is historically (e.g. in Ref.~\cite{Lehner:2015jga}) labeled as $\epsilon'/\epsilon$, where $\epsilon$ is taken from experiment. \label{fig-utplot} }
\end{figure}

Our first-principles calculation of $\epsilon'/\epsilon$ also allows us to place a new, horizontal-band constraint on the CKM matrix unitarity triangle in the $\bar\rho-\bar\eta$ plane.  In Fig.~\ref{fig-utplot} we overlay this band with constraints arising from other sources.  We find that our result is consistent with the other constraints and does not at present suggest any violation of the CKM paradigm. For more information on how this band was obtained, as well as the corresponding plot obtained using our 2015 results, we refer the reader to Ref.~\cite{Lehner:2015jga}.

\section{Conclusions}
\label{sec:conclusions}
We have described in detail a calculation which substantially enhances our 2015 lattice calculation of $\epsilon'$~\cite{Bai:2015nea}.  Both the 2015 and the current calculation were performed on a single, $32^3\times 64$ M\"obius domain wall ensemble with the Iwasaki+DSDR gauge action, with an inverse lattice spacing of 1.378(7) GeV and physical pion masses. G-parity boundary conditions are used in the three spatial directions which induces non-zero momentum for the ground-state pions so that the energy of the lightest two-pion state matches the kaon mass to around 2\%, thereby ensuring a physical, energy-conserving decay.

The new calculation reported here is based on an increase by a factor of 3.4 in the number of Monte Carlo samples and includes two additional $\pi\pi$ interpolating operators, which have dramatically improved our control over contamination arising from excited $\pi\pi$ states. The greater resolution among the excited finite-volume $\pi\pi$ states provided by our now three interpolating operators has allowed us to resolve the approximately $2\sigma$ discrepancy between our earlier lattice result for the $I=0$ $\pi\pi$ scattering phase shift and the dispersive prediction, as will be detailed in Ref.~\cite{pipi-paper}.   These improved techniques result in a significant, 70\% ($2.6\sigma$ if $\sigma$ is determined from only the statistical error) relative increase in the size of the unrenormalized lattice value of $Q_6$, suggesting that our excited-state systematic error was previously underestimated. A  detailed comparison of our old and new result can be found in Sec.~\ref{sec:comparison}. 

We have also included in this new calculation, an improved renormalization technique. As discussed in Sec.~\ref{sec:NPR}, the lattice matrix operators must be renormalized in the $\msbar$ scheme in which the Wilson coefficients that parameterize the high-energy weak interactions have been evaluated. This is accomplished by performing an intermediate non-perturbative conversion into two RI-SMOM schemes, each of which can be matched perturbatively to $\msbar$ at some high energy scale.  As we use a somewhat coarse, $a^{-1}=1.38$ GeV ensemble, our renormalization scale was formerly limited by this cutoff and $\mu=1.53$ GeV was chosen as the momemtum scale at which our RI-SMOM schemes were converted to $\msbar$. In the new calculation reported here we have applied the step-scaling procedure to bypass the limitation imposed by the lattice cutoff and raise our renormalization scale to $4.006$ GeV, thereby improving our control over the systematic error resulting from the perturbative matching to $\msbar$.  This improved method results in a reduced discrepancy between the results obtained from the two different RI-SMOM intermediate schemes and a reduction in the renormalization systematic error.  In the future we expect to improve this systematic error by further raising the renormalization scale. 

Unfortunately raising the renormalization scale does not result in a similar improvement for the Wilson coefficients as their error is dominated by the use of perturbation theory at the scale of the charm quark mass to match the effective weak interaction theory between 3 and 4 flavors. We are presently working~\cite{Tomii:2019esd} to circumvent this issue by computing the 3- to 4-flavors matching non-perturbatively using a position-space NPR technique~\cite{Tomii:2018zix}.

Finally in the current calculation we have adopted a new bootstrap method~\cite{Kelly:2019wfj} to determine the $\chi^2$ distributions appropriate for our calculation in which the data is both correlated and non-Gaussian.  The resulting improved $p$-values provide better guidance in our choice of fitting ranges and multi-state fitting functions. 

Finite lattice spacing effects remain a significant source of systematic error as at present we have computed $\epsilon'$ at a single, somewhat coarse lattice spacing. In the future we intend to follow the procedure used in our $A_2$ calculation~\cite{Blum:2015ywa} to compute $A_0$ at two different lattice spacings, allowing us to perform a full continuum limit. This is hampered by the need to generate new ensembles with GPBC, which alongside the high computational cost of the measurements and the need for large statistics requires significantly more computing power than is presently available.  

A second important systematic error, which we plan to reduce in future work, comes from the effects of electromagnetic and light quark mass isospin breaking.  As discussed in Sec.~\ref{sec:epsilonp}, the small size of the amplitude $A_2$ relative to $A_0$ gives a potential twenty times enhancement of such effects which are normally at the 1\% level.  The effects of electromagnetism and the quark mass difference $m_d-m_u$ have been studied in considerable detail using chiral perturbation theory and large $N_c$ arguments, most recently in Ref.~\cite{Cirigliano:2019cpi}.  We take the size of their correction as an important systematic error for our present result and are exploring possible methods to also use lattice techniques to determine these effects~\cite{Christ:2017pze, Cai:2018why}.

For our final result we obtain 
\begin{dmath}
{\rm Re}(\epsilon'/\epsilon)_{\rm lattice} = 0.00217(26)(62)(50)\,.  
\label{eq:epsilon-prime-lattice}
\end{dmath}
The third error here is the systematic error associated with isospin breaking and electromagnetic effects, and the first and second errors are the statistical error and the remaining systematic error.
This result can be compared to the experimental value 
\begin{dmath}
{\rm Re}(\epsilon'/\epsilon)_{\rm expt.} = 0.00166(23)\,.
\label{eq:epsilon-prime-expt}
\end{dmath}
These values are consistent within the quoted errors.

We believe that $\epsilon'$ continues to offer a very important test of the Standard Model with exciting opportunities for the discovery of new physics.  For this promise to be realized substantially more accurate Standard Model predictions are needed.  Important improvements can be expected from a simple extension of the work presented here, studying a sequence of ensembles with decreasing lattice spacing so that a continuum limit can be evaluated. In addition, we are developing a second, complementary approach to the study of $K\to\pi\pi$ decay which is based on periodic boundary conditions.  This avoids the complexity of the G-parity boundary conditions used in the present work but requires that higher excited $\pi\pi$ states be used as the decay final state~\cite{Hoying:2019ejx}. More challenging is the problem posed by the inclusion of electromagnetism where new methods~\cite{Christ:2017pze, Cai:2018why} are needed to combine the finite-volume methods of L\"uscher~\cite{Luscher:1990ux} and Lellouch and L\"uscher~\cite{Lellouch:2000pv} with the long-range character of electromagnetism.

\section*{Acknowledgments}

We would like to thank our RBC and UKQCD collaborators for their ideas and
support.  We are also pleased to acknowledge Vincenzo Cirigliano for
helpful discussion and explanation of isospin breaking effects.  The
generation of the gauge configurations used in this work was primarily
performed using the IBM BlueGene/Q (BG/Q) installation at BNL (supported
by the RIKEN BNL Research Center and BNL),  the Mira computer at the ALCF
(as part of the Incite program), Japan's KEKSC 1540 computer and the STFC
DiRAC machine at the University of Edinburgh (STFC grants ST/R00238X/1, ST/S002537/1, ST/R001006/1), 
with additional generation
performed using the NCSA Blue Waters machine at the University of
Illinois. The majority of the measurements and analysis, including the NPR
calculations, were performed using the Cori supercomputer at NERSC, with
contributions also from the Hokusai machine at ACCC RIKEN and the BG/Q
machines at BNL.  

T.B. and M.T. were supported by U.S. DOE grant \#DE-SC0010339, N.H.C., R.D.M., M.T. and 
T.W. by U.S. DOE grant \#DE-SC0011941. D.H. was supported by U.S. DOE grant
\#DE-SC0010339.  The work of C.J. and A.S. was supported in part by the
U.S. DOE contract \#DE-SC0012704.  C.K. was supported by the Intel
Corporation.  C.L. was in part supported by the U.S. DOE contract
\#DE-SC0012704.  D.J.M. was supported in part by the U.S. DOE grant
\#DE-SC0011090.  C.T.S. was partially supported by STFC (UK) grant No.
ST/P000711/1 and by an Emeritus Fellowship from the Leverhulme Trust. P.B is a Wolfson Fellow 
WM/60035 and is supported by STFC grants ST/L000458/1 and ST/P000630/1.

\appendix

\section{Wick contractions for the $\pmb{ K\to\pi\pi}$ three-point function with the $\pmb{ \sigma}$ operator}
\label{sec:appendix-sigmacon}
In this appendix we provide the expressions for the Wick contraction  required to compute the $K\to\pi\pi$ three-point function with the $\sigma$ operator. The corresponding diagrams for the $\pi\pi(\ldots)$ operators can be found in in Appendix B.1 and B.2 of Ref.~\cite{ZhangThesis}.

For this appendix we will utilize the notation described in Sec.~\ref{sec-GPnotation} whereby the quark field operators are placed in two-component ``flavor'' vectors $\psi_l$ and $\psi_h$ for the light and heavy quarks, respectively, and the corresponding propagators are matrices also in this flavor index. In this notation the creation operator for the G-parity even neutral kaon analog has the form,
\begin{dmath}
{\cal O}_{\tilde K^0} = \frac{i}{\sqrt{2}}(\bar d \gamma^5 s + \bar s' \gamma^5 u) = \frac{i}{\sqrt{2}}\bar\psi_l \gamma^5 \psi_h\,,
\end{dmath}
where the physical component corresponds to the usual neutral kaon operator (cf. Sec. VI.A of Ref.~\cite{Christ:2019sah}). The $\sigma$ creation operator has the form,
\begin{dmath}
{\cal O}_{\sigma} = \frac{1}{\sqrt{2}}(\bar u u + \bar d d) = \frac{1}{\sqrt{2}}\bar\psi_l \psi_l\,.
\end{dmath}
For convenience we will treat the meson bilinears as point operators in which both quarks reside on the same lattice site.  (In our actual lattice calculation we use more elaborate source and sink operators but those details are not needed to specify how we evaluate the Wick contractions.)   The ten effective four-quark operators $Q_i$ for $i\in\{1\ldots 10\}$ written in the above notation are given in Sec. 3.2.2 of Ref.~\cite{ZhangThesis}. While the exact forms are not important for this discussion, we highlight the fact that the operators are written in terms of a common set of matrices,
\begin{dmath}
{ M^\mu_{0,V\pm A} = F_0\gamma^\mu(1\pm \gamma^5)\,, }\\
{ M^\mu_{1,V\pm A} = -F_1\gamma^\mu(1\pm \gamma^5)\,, }
\end{dmath}
where $F_i$ are diagonal flavor matrices that pick out either the upper (0) or lower (1) element of the vector upon which they act:
\begin{dmath}
{ F_0 = \left(\begin{array}{cc}1 & 0\\0 & 0\end{array}\right)\,,  \hspace{2cm}  F_1 = \left(\begin{array}{cc}0 & 0\\0 & 1\end{array}\right)\,.  } \label{eq-F0F1def}
\end{dmath}
The matrices $M^\mu_{i,V\pm A}$ appear inside products of two bilinear operators and the space-time index $\mu$ is summed over implicitly. Following the notation of Ref.~\cite{ZhangThesis} we will suppress this index.

The Wick contractions of the $K\to\pi\pi$ three-point function with the $\sigma$ operator,
\begin{dmath}
{\cal A}_i = \langle 0 | {\cal O}^\dagger_{\sigma}(z) \hat Q_i(y) {\cal O}_{\tilde K^0}(x) | 0\rangle\,,
\end{dmath}
where $\hat Q_i$ are the unsubtracted four-quark operators, are divided into three classes by their topology that we label with indices 1, 3 and 4 in homage to the conventional labeling of the $\pi\pi(\ldots)$ contractions. The {\it type3} and {\it type4} diagrams are those that contain a quark loop at the location of the four-quark operator, with {\it type4} corresponding to that subset of those diagrams that are disconnected (i.e. for which the $\sigma$ operator self-contracts). 
For the $\pi\pi(\ldots)$ operators the remaining, connected, contractions can be subdivided based on whether the two pion bilinear operators are directly connected by a quark line ({\it type2}) or not ({\it type1}); no such distinction exists of course for the $\sigma$ sink operator.   Hence we classify all remaining diagrams as {\it type1}.

As in Ref.~\cite{ZhangThesis} it is convenient to write the ten expressions ${\cal A}_i$ in terms of a common basis of, in this case 23, functions $D_\alpha(\Gamma_1, \Gamma_2)$ where the subscript indexes the function and $\Gamma_{1,2}$ are spin-flavor matrices. 

We will first write down the expressions for the correlation functions ${\cal A}_i$ in terms of these functions and will conclude the section with their definition. We list the contributions for each of the three types separately. The {\it type1} contributions are as follows:
\begin{dgroup}
\begin{dmath}
{\cal A}_{1}^{\rm type1} = \frac{1}{2} D_{6}(M_{0,V-A}, M_{1,V+A}) -\frac{1}{2} D_{1}(M_{0,V-A}, M_{1,V+A})
\end{dmath}
\begin{dmath}
    {\cal A}_{2}^{\rm type1} = \frac{1}{2} D_{11}(M_{0,V-A}, M_{1,V+A}) -\frac{1}{2} D_{8}(M_{0,V-A}, M_{1,V+A})
\end{dmath}
\begin{dmath} 
     {\cal A}_{3}^{\rm type1} = \frac{1}{2} D_{6}(M_{0,V-A}, M_{1,V+A}) + \frac{1}{2} D_{6}(M_{0,V-A}, M_{0,V-A}) -\frac{1}{2} D_{1}(M_{0,V-A}, M_{1,V+A}) -\frac{1}{2} D_{1}(M_{0,V-A}, M_{0,V-A})
\end{dmath}
\begin{dmath} 
     {\cal A}_{4}^{\rm type1} = D_{11}(M_{0,V-A}, M_{1,V+A}) -\frac{1}{2} D_{8}(M_{0,V-A}, M_{1,V+A}) -\frac{1}{2} D_{19}(M_{0,V-A}, M_{0,V-A})
\end{dmath}
\begin{dmath} 
     {\cal A}_{5}^{\rm type1} = D_{6}(M_{0,V-A}, M_{1,V-A}) -\frac{1}{2} D_{1}(M_{0,V-A}, M_{1,V-A}) -\frac{1}{2} D_{1}(M_{0,V-A}, M_{0,V+A})
\end{dmath}
\begin{dmath} 
     {\cal A}_{6}^{\rm type1} = D_{11}(M_{0,V-A}, M_{1,V-A}) -\frac{1}{2} D_{8}(M_{0,V-A}, M_{1,V-A}) -\frac{1}{2} D_{19}(M_{0,V-A}, M_{0,V+A})
\end{dmath}
\begin{dmath} 
     {\cal A}_{7}^{\rm type1} = \frac{1}{4} D_{6}(M_{0,V-A}, M_{1,V-A}) -\frac{1}{2} D_{1}(M_{0,V-A}, M_{1,V-A}) + \frac{1}{4} D_{1}(M_{0,V-A}, M_{0,V+A})
\end{dmath}
\begin{dmath} 
     {\cal A}_{8}^{\rm type1} = \frac{1}{4} D_{11}(M_{0,V-A}, M_{1,V-A}) -\frac{1}{2} D_{8}(M_{0,V-A}, M_{1,V-A}) + \frac{1}{4} D_{19}(M_{0,V-A}, M_{0,V+A})
\end{dmath}
\begin{dmath} 
     {\cal A}_{9}^{\rm type1} = \frac{1}{4} D_{6}(M_{0,V-A}, M_{1,V+A}) -\frac{1}{2} D_{1}(M_{0,V-A}, M_{1,V+A})+ \frac{1}{4} D_{1}(M_{0,V-A}, M_{0,V-A})
\end{dmath}
\begin{dmath} 
     {\cal A}_{10}^{\rm type1} = \frac{1}{4} D_{11}(M_{0,V-A}, M_{1,V+A}) -\frac{1}{2} D_{8}(M_{0,V-A}, M_{1,V+A}) + \frac{1}{4} D_{19}(M_{0,V-A}, M_{0,V-A})\,,
\end{dmath}
\end{dgroup}
the {\it type3} contributions are:
\begin{dgroup}
\begin{dmath}
    {\cal A}_{1}^{\rm type3} = \frac{1}{2} D_{2}(M_{0,V-A}, M_{1,V+A}) -\frac{1}{2} D_{3}(M_{0,V-A}, M_{1,V+A})
\end{dmath}
\begin{dmath} 
    {\cal A}_{2}^{\rm type3} = \frac{1}{2} D_{10}(M_{0,V-A}, M_{1,V+A}) -\frac{1}{2} D_{7}(M_{0,V-A}, M_{1,V+A})
\end{dmath}
\begin{dmath} 
    {\cal A}_{3}^{\rm type3} = \frac{1}{2} D_{2}(M_{0,V-A}, M_{1,V+A}) + \frac{1}{2} D_{2}(M_{0,V-A}, M_{0,V-A}) -\frac{1}{2} D_{3}(M_{0,V-A}, M_{1,V+A}) 
      -\frac{1}{2} D_{3}(M_{0,V-A}, M_{0,V-A}) + \frac{1}{2} D_{14}(M_{0,V-A}, M_{0,V-A}) -\frac{1}{2} D_{16}(M_{0,V-A}, M_{0,V-A})
\end{dmath}
\begin{dmath}    
    {\cal A}_{4}^{\rm type3} = D_{10}(M_{0,V-A}, M_{1,V+A}) -\frac{1}{2} D_{7}(M_{0,V-A}, M_{1,V+A}) -\frac{1}{2} D_{18}(M_{0,V-A}, M_{0,V-A}) 
      + \frac{1}{2} D_{21}(M_{0,V-A}, M_{0,V-A}) -\frac{1}{2} D_{23}(M_{0,V-A}, M_{0,V-A})
\end{dmath}
\begin{dmath}    
    {\cal A}_{5}^{\rm type3} = D_{2}(M_{0,V-A}, M_{1,V-A}) -\frac{1}{2} D_{3}(M_{0,V-A}, M_{1,V-A}) -\frac{1}{2} D_{3}(M_{0,V-A}, M_{0,V+A}) 
      + \frac{1}{2} D_{14}(M_{0,V-A}, M_{0,V+A}) -\frac{1}{2} D_{16}(M_{0,V-A}, M_{0,V+A})
\end{dmath}
\begin{dmath}     
    {\cal A}_{6}^{\rm type3} = D_{10}(M_{0,V-A}, M_{1,V-A}) -\frac{1}{2} D_{7}(M_{0,V-A}, M_{1,V-A}) -\frac{1}{2} D_{18}(M_{0,V-A}, M_{0,V+A})
      + \frac{1}{2} D_{21}(M_{0,V-A}, M_{0,V+A}) -\frac{1}{2} D_{23}(M_{0,V-A}, M_{0,V+A})
\end{dmath}
\begin{dmath}     
    {\cal A}_{7}^{\rm type3} = \frac{1}{4} D_{2}(M_{0,V-A}, M_{1,V-A}) -\frac{1}{2} D_{3}(M_{0,V-A}, M_{1,V-A}) + \frac{1}{4} D_{3}(M_{0,V-A}, M_{0,V+A}) 
      -\frac{1}{4} D_{14}(M_{0,V-A}, M_{0,V+A}) + \frac{1}{4} D_{16}(M_{0,V-A}, M_{0,V+A})
\end{dmath}
\begin{dmath}     
    {\cal A}_{8}^{\rm type3} = \frac{1}{4} D_{10}(M_{0,V-A}, M_{1,V-A}) -\frac{1}{2} D_{7}(M_{0,V-A}, M_{1,V-A}) + \frac{1}{4} D_{18}(M_{0,V-A}, M_{0,V+A})
      -\frac{1}{4} D_{21}(M_{0,V-A}, M_{0,V+A}) + \frac{1}{4} D_{23}(M_{0,V-A}, M_{0,V+A})
\end{dmath}
\begin{dmath}    
    {\cal A}_{9}^{\rm type3} = \frac{1}{4} D_{2}(M_{0,V-A}, M_{1,V+A}) -\frac{1}{2} D_{3}(M_{0,V-A}, M_{1,V+A}) + \frac{1}{4} D_{3}(M_{0,V-A}, M_{0,V-A})
      -\frac{1}{4} D_{14}(M_{0,V-A}, M_{0,V-A}) + \frac{1}{4} D_{16}(M_{0,V-A}, M_{0,V-A})
\end{dmath}
\begin{dmath}    
    {\cal A}_{10}^{\rm type3} = \frac{1}{4} D_{10}(M_{0,V-A}, M_{1,V+A}) -\frac{1}{2} D_{7}(M_{0,V-A}, M_{1,V+A}) + \frac{1}{4} D_{18}(M_{0,V-A}, M_{0,V-A})
      -\frac{1}{4} D_{21}(M_{0,V-A}, M_{0,V-A}) + \frac{1}{4} D_{23}(M_{0,V-A}, M_{0,V-A})\,,
\end{dmath}
\end{dgroup}
and the {\it type4}:
\begin{dgroup}
\begin{dmath}
    {\cal A}_{1}^{\rm type4} = -\frac{1}{2} D_{5}(M_{0,V-A}, M_{1,V+A}) + \frac{1}{2} D_{4}(M_{0,V-A}, M_{1,V+A})
\end{dmath}
\begin{dmath}    
    {\cal A}_{2}^{\rm type4} = -\frac{1}{2} D_{12}(M_{0,V-A}, M_{1,V+A}) + \frac{1}{2} D_{9}(M_{0,V-A}, M_{1,V+A})
\end{dmath}
\begin{dmath}    
    {\cal A}_{3}^{\rm type4} = -\frac{1}{2} D_{5}(M_{0,V-A}, M_{1,V+A}) -\frac{1}{2} D_{5}(M_{0,V-A}, M_{0,V-A}) + \frac{1}{2} D_{4}(M_{0,V-A}, M_{1,V+A})
      + \frac{1}{2} D_{4}(M_{0,V-A}, M_{0,V-A}) -\frac{1}{2} D_{13}(M_{0,V-A}, M_{0,V-A}) + \frac{1}{2} D_{15}(M_{0,V-A}, M_{0,V-A})
\end{dmath}
\begin{dmath}        
    {\cal A}_{4}^{\rm type4} = - D_{12}(M_{0,V-A}, M_{1,V+A}) + \frac{1}{2} D_{9}(M_{0,V-A}, M_{1,V+A}) + \frac{1}{2} D_{17}(M_{0,V-A}, M_{0,V-A})
      -\frac{1}{2} D_{20}(M_{0,V-A}, M_{0,V-A}) + \frac{1}{2} D_{22}(M_{0,V-A}, M_{0,V-A})
\end{dmath}
\begin{dmath}        
    {\cal A}_{5}^{\rm type4} = - D_{5}(M_{0,V-A}, M_{1,V-A}) + \frac{1}{2} D_{4}(M_{0,V-A}, M_{1,V-A}) + \frac{1}{2} D_{4}(M_{0,V-A}, M_{0,V+A})
      -\frac{1}{2} D_{13}(M_{0,V-A}, M_{0,V+A}) + \frac{1}{2} D_{15}(M_{0,V-A}, M_{0,V+A})
\end{dmath}
\begin{dmath}        
    {\cal A}_{6}^{\rm type4} = - D_{12}(M_{0,V-A}, M_{1,V-A}) + \frac{1}{2} D_{9}(M_{0,V-A}, M_{1,V-A}) + \frac{1}{2} D_{17}(M_{0,V-A}, M_{0,V+A})
      -\frac{1}{2} D_{20}(M_{0,V-A}, M_{0,V+A}) + \frac{1}{2} D_{22}(M_{0,V-A}, M_{0,V+A})
\end{dmath}
\begin{dmath}        
    {\cal A}_{7}^{\rm type4} = -\frac{1}{4} D_{5}(M_{0,V-A}, M_{1,V-A}) + \frac{1}{2} D_{4}(M_{0,V-A}, M_{1,V-A}) -\frac{1}{4} D_{4}(M_{0,V-A}, M_{0,V+A})
      + \frac{1}{4} D_{13}(M_{0,V-A}, M_{0,V+A}) -\frac{1}{4} D_{15}(M_{0,V-A}, M_{0,V+A})
\end{dmath}
\begin{dmath}        
    {\cal A}_{8}^{\rm type4} = -\frac{1}{4} D_{12}(M_{0,V-A}, M_{1,V-A}) + \frac{1}{2} D_{9}(M_{0,V-A}, M_{1,V-A}) -\frac{1}{4} D_{17}(M_{0,V-A}, M_{0,V+A})
      + \frac{1}{4} D_{20}(M_{0,V-A}, M_{0,V+A}) -\frac{1}{4} D_{22}(M_{0,V-A}, M_{0,V+A})
\end{dmath}
\begin{dmath}        
    {\cal A}_{9}^{\rm type4} = -\frac{1}{4} D_{5}(M_{0,V-A}, M_{1,V+A}) + \frac{1}{2} D_{4}(M_{0,V-A}, M_{1,V+A}) -\frac{1}{4} D_{4}(M_{0,V-A}, M_{0,V-A})
      + \frac{1}{4} D_{13}(M_{0,V-A}, M_{0,V-A}) -\frac{1}{4} D_{15}(M_{0,V-A}, M_{0,V-A})
\end{dmath}
\begin{dmath}        
    {\cal A}_{10}^{\rm type4} = -\frac{1}{4} D_{12}(M_{0,V-A}, M_{1,V+A}) + \frac{1}{2} D_{9}(M_{0,V-A}, M_{1,V+A}) -\frac{1}{4} D_{17}(M_{0,V-A}, M_{0,V-A})
      + \frac{1}{4} D_{20}(M_{0,V-A}, M_{0,V-A}) -\frac{1}{4} D_{22}(M_{0,V-A}, M_{0,V-A})\,.
\end{dmath}      
\end{dgroup}

The {\it type1} contractions are:
\begin{dgroup}
\begin{dmath}
D_{1}(\Gamma_1,\Gamma_2) =   {\rm tr}\left( \Gamma_2 \prop^l_{y,x} \gamma^5 \prop^h_{x,y} \Gamma_1 \prop^l_{y,z} \prop^l_{z,y} \right)   
\end{dmath}
\begin{dmath}
D_{6}(\Gamma_1,\Gamma_2) =   {\rm tr}\left( \prop^h_{x,y} \Gamma_1 \prop^l_{y,x} \gamma^5\right)     {\rm tr}\left( \prop^l_{z,y} \Gamma_2 \prop^l_{y,z}\right)   
\end{dmath}
\begin{dmath}
D_{8}(\Gamma_1,\Gamma_2) =   {\rm tr}_{sf}\left( \left[\Gamma_1 \prop^l_{y,z} \prop^l_{z,y}\right]_{\alpha\beta} \left[\Gamma_2 \prop^l_{y,x} \gamma^5 \prop^h_{x,y} \right]_{\alpha\beta}\right)  
\end{dmath}
\begin{dmath}
D_{11}(\Gamma_1,\Gamma_2) =   {\rm tr}_{sf}\left( \prop^l_{y,x} \gamma^5 \prop^h_{x,y} \Gamma_1\right)_{\alpha\beta}     {\rm tr}_{sf}\left(\Gamma_2 \prop^l_{y,z} \prop^l_{z,y}\right)_{\alpha\beta} 
\end{dmath}
\begin{dmath}
D_{19}(\Gamma_1,\Gamma_2) =   {\rm tr}_{sf}\left( {\rm tr}_{c}\left[ \prop^l_{y,x}\gamma^5 \prop^h_{x,y}\Gamma_1 \right] {\rm tr}_{c}\left[ \prop^l_{y,z} \prop^l_{z,y} \Gamma_2\right] \right)\,,
\end{dmath}
\end{dgroup}
and the {\it type3} are:
\begin{dgroup}
\begin{dmath} 
D_{2}(\Gamma_1,\Gamma_2) =   {\rm tr}\left( \gamma^5 \prop^h_{x,y} \Gamma_1 \prop^l_{y,z} \prop^l_{z,x} \right) {\rm tr}\left( \prop^l_{y,y} \Gamma_2\right)   
\end{dmath}
\begin{dmath} 
D_{3}(\Gamma_1,\Gamma_2) =   {\rm tr}\left( \prop^l_{y,z} \prop^l_{z,x} \gamma^5 \prop^h_{x,y} \Gamma_1 \prop^l_{y,y} \Gamma_2\right)
\end{dmath}
\begin{dmath}
D_{7}(\Gamma_1,\Gamma_2) =   {\rm tr}_{sf}\left( \left[\Gamma_2 \prop^l_{y,z} \prop^l_{z,x} \gamma^5 \prop^h_{x,y}\right]_{\alpha\beta} \left[\Gamma_1 \prop^l_{y,y} \right]_{\alpha\beta}\right)  
\end{dmath}
\begin{dmath}
D_{10}(\Gamma_1,\Gamma_2) =   {\rm tr}_{sf}\left( \Gamma_2 \prop^l_{y,y} \right)_{\alpha\beta}     {\rm tr}_{sf}\left( \Gamma_1 \prop^l_{y,z} \prop^l_{z,x} \gamma^5 \prop^h_{x,y}\right)_{\alpha\beta}  
\end{dmath}
\begin{dmath}
D_{14}(\Gamma_1,\Gamma_2) =   {\rm tr}\left( \prop^l_{y,z} \prop^l_{z,x} \gamma^5 \prop^h_{x,y} \Gamma_1\right)     {\rm tr}\left( \prop^h_{y,y} \Gamma_2\right)
\end{dmath}
\begin{dmath}
D_{16}(\Gamma_1,\Gamma_2) =   {\rm tr}\left( \prop^h_{y,y} \Gamma_1 \prop^l_{y,z} \prop^l_{z,x} \gamma^5 \prop^h_{x,y} \Gamma_2\right)
\end{dmath}
\begin{dmath}
D_{18}(\Gamma_1,\Gamma_2) =   {\rm tr}_{sf}\left( {\rm tr}_{c}\left[ \prop^l_{y,y} \right]{\rm tr}_{c}\left[ \Gamma_2 \prop^l_{y,z} \prop^l_{z,x} \gamma^5 \prop^h_{x,y} \Gamma_1\right] \right)
\end{dmath}
\begin{dmath}
D_{21}(\Gamma_1,\Gamma_2) =   {\rm tr}_c\left({\rm tr}_{sf}\left[ \prop^h_{y,y} \Gamma_2\right]     {\rm tr}_{sf}\left[ \Gamma_1 \prop^l_{y,z} \prop^l_{z,x} \gamma^5 \prop^h_{x,y}\right]   \right)
\end{dmath}
\begin{dmath}
D_{23}(\Gamma_1,\Gamma_2) =   {\rm tr}_{sf}\left( {\rm tr}_{c}\left[ \prop^h_{y,y} \right] {\rm tr}_{c}\left[ \Gamma_1 \prop^l_{y,z} \prop^l_{z,x} \gamma^5 \prop^h_{x,y} \Gamma_2\right]  \right)\,,
\end{dmath}
\end{dgroup}
where $\prop^l$ and $\prop^h$ are light and strange quark propagators, respectively, and $\alpha,\beta$ are color indices. We indicate spin and flavor traces as ${\rm tr}_{sf}$ and color traces as ${\rm tr}_c$; traces over all three indices (spin, color and flavor) are denoted as ${\rm tr}$ without a subscript.

For simplicity, in Eqs.~\eqref{eq-ktosigma-type4-con} given below for the {\it type4} diagrams we do not include the disconnected $\sigma$ ``bubble'',
\begin{dmath}
B_\sigma = {\rm tr}\left( \prop^l_{z,z}\right)\,.\label{eq-def-Bsigma}
\end{dmath}
In computing the expectation values of these diagrams it is also necessary to perform a vacuum subtraction.  Thus, the expressions $D_i^*$ given in Eqs.~\eqref{eq-ktosigma-type4-con} can be used to obtain the complete contributions of the corresponding diagrams to the {\it type4} amplitudes as follows:
\begin{dmath}
\langle D_{i}(\Gamma_1,\Gamma_2) \rangle = \langle D^*_{i}(\Gamma_1,\Gamma_2) B_\sigma \rangle - \langle D^*_{i}(\Gamma_1,\Gamma_2)\rangle\times\langle B_\sigma \rangle\,,
\end{dmath}
where $D^*$ are defined as:
\begin{dgroup}
\begin{dmath} 
D^*_{4}(\Gamma_1,\Gamma_2) =   {\rm tr}\left( \prop^l_{y,x} \gamma^5 \prop^h_{x,y} \Gamma_1 \prop^l_{y,y} \Gamma_2\right)     
\end{dmath}
\begin{dmath} 
D^*_{5}(\Gamma_1,\Gamma_2) =   {\rm tr}\left( \prop^h_{x,y} \Gamma_1 \prop^l_{y,x} \gamma^5\right)     {\rm tr}\left( \prop^l_{y,y} \Gamma_2\right)     
\end{dmath}
\begin{dmath}
D^*_{9}(\Gamma_1,\Gamma_2) =   {\rm tr}_{sf}\left( \left[\Gamma_1 \prop^l_{y,y}\right]_{\alpha\beta} \left[\Gamma_2 \prop^l_{y,x}\gamma^5 \prop^h_{x,y}\right]_{\alpha\beta}  \right) 
\end{dmath}
\begin{dmath}
D^*_{12}(\Gamma_1,\Gamma_2) =       {\rm tr}_{sf}\left( \prop^l_{y,x}\gamma^5 \prop^h_{x,y} \Gamma_1\right)_{\alpha\beta}     {\rm tr}_{sf}\left( \prop^l_{y,y} \Gamma_2\right)_{\alpha\beta}   
\end{dmath}
\begin{dmath}
D^*_{13}(\Gamma_1,\Gamma_2) =        {\rm tr}\left( \gamma^5 \prop^h_{x,y} \Gamma_1 \prop^l_{y,x}\right)     {\rm tr}\left( \Gamma_2 \prop^h_{y,y}\right)   
\end{dmath}
\begin{dmath}
D^*_{15}(\Gamma_1,\Gamma_2) =    {\rm tr}\left( \Gamma_1 \prop^l_{y,x} \gamma^5 \prop^h_{x,y} \Gamma_2 \prop^h_{y,y}\right)   
\end{dmath}
\begin{dmath}
D^*_{17}(\Gamma_1,\Gamma_2) =        {\rm tr}_{sf}\left( {\rm tr}_{c}\left[\prop^l_{y,y}\right] {\rm tr}_{c}\left[ \Gamma_2 \prop^l_{y,x} \gamma^5 \prop^h_{x,y} \Gamma_1\right] \right) 
\end{dmath}
\begin{dmath}
D^*_{20}(\Gamma_1,\Gamma_2) =        {\rm tr}_c\left(  {\rm tr}_{sf}\left[ \prop^l_{y,x} \gamma^5 \prop^h_{x,y} \Gamma_1 \right] {\rm tr}_{sf}\left[ \prop^h_{y,y} \Gamma_2\right] \right)   
\end{dmath}
\begin{dmath}
D^*_{22}(\Gamma_1,\Gamma_2) =        {\rm tr}_{sf}\left( {\rm tr}_{c}\left[ \prop^h_{y,y}\right] {\rm tr}_{c}\left[ \Gamma_1 \prop^l_{y,x} \gamma^5 \prop^h_{x,y} \Gamma_2\right]  \right)\,.  
\end{dmath}
\label{eq-ktosigma-type4-con}
\end{dgroup}

\section{Wick contractions for matrix elements required for subtraction of the vacuum and pseudoscalar operator contributions}
\label{sec:appendix-Pop}
As described in Sec.~\ref{sec:ThreePointResults} it is necessary to subtract a pseudoscalar operator $P=\bar s\gamma^5 d$ from the unsubtracted weak effective four-quark operators $\hat Q_i$ in order to remove a divergent contribution for off-shell terms. The subtraction and the evaluation of the corresponding coeffients, $\alpha_i$, require the measurement of $\langle {\cal O}_{\pi\pi}^\dagger P \tilde {\cal O}_{\tilde K^0}\rangle$, $\langle P {\cal O}_{\tilde K^0}\rangle$ and $\langle \hat Q_i {\cal O}_{\tilde K^0}\rangle$ correlation functions. The vacuum subtraction of the {\it type4} diagrams also requires evaluating the $\langle \hat Q_i {\cal O}_{\tilde K^0}\rangle$ correlation functions. Here and below we use the shorthand $\langle A B C\ldots \rangle$ to denote the n-point Green's functions of the operators $A$, $B$, $C$, and so on, in descending time order.

It is easy to see that the ${\cal A}^{\rm vac}_i = \langle \hat Q_i {\cal O}_{\tilde K^0}\rangle$ are directly proportional to the {\it type4}, disconnected contributions to $\langle {\cal O}_{\pi\pi}^\dagger \hat Q_i {\cal O}_{\tilde K^0}\rangle$ with the $\pi\pi$ ``bubble'' removed. The results are
\begin{dgroup}
\begin{dmath}
{\cal A}^{\rm vac}_1 = \frac{1}{\sqrt{2}}\left( C_{23}(M_{0,V-A}, M_{1,V+A}) - C_{26}(M_{1,V+A}, M_{0,V-A})\right)
\end{dmath}
\begin{dmath}
{\cal A}^{\rm vac}_2 = \frac{1}{\sqrt{2}}\left( C_{24}(M_{0,V-A}, M_{1,V+A}) - C_{27}(M_{1,V+A}, M_{0,V-A})\right)
\end{dmath}
\begin{dmath}
{\cal A}^{\rm vac}_3 = \frac{1}{\sqrt{2}}\left( C_{23}(M_{0,V-A}, M_{1,V+A}) +  C_{23}(M_{0,V-A}, M_{0,V-A}) - C_{26}(M_{1,V+A}, M_{0,V-A}) - C_{26}(M_{0,V-A}, M_{0,V-A}) + C_{29}(M_{0,V-A}, M_{0,V-A} ) - C_{31}(M_{0,V-A} , M_{0,V-A})   \right)
\end{dmath}
\begin{dmath}
{\cal A}^{\rm vac}_4 = \frac{1}{\sqrt{2}}\left(  
C_{24}(M_{0,V-A} , M_{1,V+A})
+ C_{25}(M_{0,V-A} , M_{0,V-A} ) - C_{27}(M_{1,V+A}, M_{0,V-A} ) - C_{28}(M_{0,V-A} , M_{0,V-A})
+ C_{30}(M_{0,V-A} , M_{0,V-A} ) - C_{32}(M_{0,V-A}, M_{0,V-A} )
\right)
\end{dmath}
\begin{dmath}
{\cal A}^{\rm vac}_5 = \frac{1}{\sqrt{2}}\left(  
C_{23}(M_{0,V-A} , M_{1,V-A} ) + C_{23}(M_{0,V-A}, M_{0,V +A}) - C_{26}(M_{1,V-A} , M_{0,V-A})
- C_{26}(M_{0,V +A} , M_{0,V-A}) + C_{29}(M_{0,V-A}, M_{0,V +A} ) - C_{31}(M_{0,V-A} , M_{0,V +A})
\right)
\end{dmath}
\begin{dmath}
{\cal A}^{\rm vac}_6 = \frac{1}{\sqrt{2}}\left(  
   C_{24}(M_{0,V -A} , M_{1,V -A})
+  C_{25}(M_{0,V -A} , M_{0,V +A} ) -   C_{27}(M_{1,V -A}, M_{0,V -A} ) -   C_{28}(M_{0,V +A} , M_{0,V -A})
+  C_{30}(M_{0,V -A} , M_{0,V +A} ) -   C_{32}(M_{0,V -A}, M_{0,V +A} )
\right)
\end{dmath}
\begin{dmath}
{\cal A}^{\rm vac}_7 = \frac{1}{\sqrt{2}}\left(  
 C_{23}(M_{0,V -A}, M_{1,V -A })
-\frac{1}{2} C_{23}(M_{0,V -A} , M_{0,V +A} )
- C_{26}(M_{1,V -A} , M_{0,V -A}) 
+ \frac{1}{2} C_{26}(M_{0,V +A} , M_{0,V -A}) 
-\frac{1}{2}  C_{29}(M_{0,V -A} , M_{0,V +A})
+ \frac{1}{2} C_{31}(M_{0,V -A} , M_{0,V +A})
\right)
\end{dmath}
\begin{dmath}
{\cal A}^{\rm vac}_8 = \frac{1}{\sqrt{2}}\left(  
C_{24}(M_{0,V -A}, M_{1,V -A} )
-\frac{1}{2} C_{25}(M_{0,V -A} , M_{0,V +A} )
- C_{27}(M_{1,V -A} , M_{0,V -A})
+\frac{1}{2}C_{28}(M_{0,V +A} , M_{0,V -A})
-\frac{1}{2}C_{30}(M_{0,V -A} , M_{0,V +A})
+\frac{1}{2}C_{32}(M_{0,V -A} , M_{0,V +A})
\right)
\end{dmath}
\begin{dmath}
{\cal A}^{\rm vac}_9 = \frac{1}{\sqrt{2}}\left(  
 C_{23}(M_{0,V -A}, M_{1,V +A} )
-\frac{1}{2}  C_{23}(M_{0,V -A} , M_{0,V -A} )
- C_{26}(M_{1,V +A} , M_{0,V -A})
+\frac{1}{2}  C_{26}(M_{0,V -A} , M_{0,V -A})
-\frac{1}{2}  C_{29}(M_{0,V -A}, M_{0,V -A} )
+\frac{1}{2}  C_{31}(M_{0,V -A} , M_{0,V -A})
\right)
\end{dmath}

\begin{dmath}
{\cal A}^{\rm vac}_9 = \frac{1}{\sqrt{2}}\left(  
C_{24}(M_{0,V -A}, M_{1,V +A} )
-\frac{1}{2} C_{25}(M_{0,V -A} , M_{0,V -A} )
- C_{27}(M_{1,V +A} , M_{0,V -A})
+\frac{1}{2} C_{28}(M_{0,V -A} , M_{0,V -A})
-\frac{1}{2} C_{30}(M_{0,V -A} , M_{0,V -A})
+\frac{1}{2} C_{32}(M_{0,V -A} , M_{0,V -A})
\right)\,.
\end{dmath}

\end{dgroup}
These results can be obtained by isolating the $C_{23}-C_{32}$ {\it type4} contributions from the expressions in Sec. 3.2.2 of Ref.~\cite{ZhangThesis} and multiplying the result by a factor of $1/\sqrt{3}$. Equivalent results can also be obtained from the {\it type4} contributions given in Eq.~\eqref{eq-ktosigma-type4-con} by multiplying the result by a factor of $\sqrt{2}$. When measured with A2A propagators the results computed in these two bases are not exactly equal due to differing choices of where to employ $\gamma^5$-hermiticity, a symmetry that is broken by the stochastic ``high-mode'' approximation and restored only in the large ensemble-size limit (or the large-hit limit on a single configuration). This gives rise to the small differences observed in Sec.~\ref{sec-alphai-comput}.

In our notation the pseudoscalar operator becomes
\begin{dmath}
{ P = \bar s\gamma^5 d = \bar\psi_h  \gamma^5 F_0\psi_l }\,,
\end{dmath}
where $F_0$ is defined in Eq.~\eqref{eq-F0F1def}.

The $\langle P {\cal O}_{\tilde K^0}\rangle$ and $\langle {\cal O}_{\pi\pi}^\dagger  P {\cal O}_{\tilde K^0}\rangle$ correlation functions with the $\pi\pi(\ldots)$ and $\sigma$ operators can be written in terms of three diagrams:
\begin{dgroup}
\begin{dmath}
{\rm mix3} = {\rm tr}\left( \prop^l_{z_2,x} \gamma^5 \prop^h_{x,y} \gamma^5 F_0 \prop^l_{y,z_1} \gamma^5 \sigma_3 \prop^l_{z_1,z_2} \gamma^5 \sigma_3 \right)
\end{dmath}
\begin{dmath}
{\rm mix3}_\sigma = {\rm tr}\left( \prop^l_{z,x} \gamma^5 \prop^h_{x,y} \gamma^5 F_0 \prop^l_{y,z} \right)
\end{dmath}
\begin{dmath}
{\rm mix4} = {\rm tr}\left(  \prop^h_{x,y} \gamma^5 F_0 \prop^l_{y,x}\gamma^5 \right)\,.
\end{dmath}
\end{dgroup}
where $x$ and $y$ are the locations of the kaon source and the operator insertion, respectively. The $\sigma$ sink operator is located at $z$, and the coordinates of the two pion bilinear operators in the $\pi\pi(\ldots)$ operators are $z_1$ and $z_2$.

The result for ${\cal A}^{\rm vac, P} = \langle  P {\cal O}_{\tilde K^0}\rangle$ is
\begin{dmath}
{\cal A}^{\rm vac, P} = -\frac{1}{\sqrt{2}}\ {\rm mix4}\,.
\end{dmath}
The amplitudes ${\cal A}^{\pi\pi(\ldots), P} = \langle {\cal O}_{\pi\pi}^\dagger  P {\cal O}_{\tilde K^0}\rangle$ for the $\pi\pi(\ldots)$ operators are computed as
\begin{dmath}
{\cal A}^{\pi\pi(\ldots), P} = -\frac{3}{\sqrt{6}}\left( B\ {\rm mix4} + {\rm mix3} \right)
\end{dmath}
where 
\begin{dmath}
B = -\frac{1}{2}{\rm tr}\left(  \prop^l_{z_1,z_2}\gamma^5 \sigma_3 \prop^l_{z_2,z_1} \gamma^5\sigma_3 \right)
\end{dmath}
is the $\pi\pi$ self-contraction ``bubble'' introduced in Sec. B.2 of Ref.~\cite{ZhangThesis}. The corresponding result for the $\sigma$ sink operator is
\begin{dmath}
{\cal A}^{\sigma, P} = \frac{1}{2}\left( B_\sigma\ {\rm mix4} - {\rm mix3}_\sigma \right)
\end{dmath}
where $B_\sigma$ is defined in Eq.~\eqref{eq-def-Bsigma}.





\FloatBarrier
\bibliography{paper}

\end{document}